\documentclass[aps,prl,twocolumn,superscriptaddress]{revtex4-2}

\usepackage{graphics,graphicx,array,afterpage}
\usepackage{grffile}
\usepackage[export]{adjustbox}
\usepackage{lineno}
\usepackage{xcolor}
\usepackage{longtable}
\usepackage{svg}
\usepackage{braket}
\usepackage{multirow}
\usepackage{enumerate}
\usepackage{amsmath}
\usepackage{amssymb}
\usepackage{amsthm}
\usepackage{mathrsfs}
\usepackage{pifont}

\usepackage[caption=false]{subfig}
\usepackage[colorlinks, citecolor=blue, urlcolor=blue]{hyperref}

\newcommand{\tr}{\mbox{tr}}

\newcommand{\bdry}{\partial}
\newcommand{\cobdry}{\delta}
\newcommand{\BBc}[1]{\ensuremath{\mathrm{BB}(#1)}}
\newcommand{\im}{\mathop{\mathrm{im}}}
\newcommand{\B}[2]{\mathcal{B}(#1, #2)}

\ifdefined\pdfminorversion
\fi
\ifdefined\pdfsuppresswarningpagegroup
\fi

\newtheorem{theorem}{Theorem}
\newtheorem{corollary}{Corollary}
\newtheorem{lemma}{Lemma}
\newtheorem{definition}{Definition}
\newtheorem{remark}{Remark}
\newtheorem{proposition}{Proposition}

\makeatletter
\newcommand{\bibnotoc}{\let\nabbACL\addcontentsline\let\addcontentsline\@gobblethree}
\newcommand{\bibtoc}{\let\addcontentsline\nabbACL}
\makeatother

\begin{document}

\title{Non-Abelian Twist Defects in Bivariate Bicycle Codes}
\author{Rui Zeng}
\affiliation{Theoretical Physics Division, Chern Institute of Mathematics and LPMC,
Nankai University, Tianjin 300071, China}
\author{Li-Wei Yu}
\email{yulw@nankai.edu.cn}
\affiliation{Theoretical Physics Division, Chern Institute of Mathematics and LPMC,
Nankai University, Tianjin 300071, China}
\begin{abstract}
Bivariate bicycle quantum low-density parity-check codes combine efficient quantum storage with a rich Abelian anyon structure, providing a compelling platform for exploring the interplay between quantum error correction and topological order. A central challenge is how to realize domain walls that host non-Abelian twist defects at their endpoints. Here we explicitly construct electric-magnetic duality walls for arbitrary bivariate bicycle codes with finite anyon content using composite, interblock, and patch checks. These checks remove the unwanted local modes while preserving charge conversion, ensuring that topologically contractible closed ribbons act trivially on encoded states.  For each fixed code, check weights and interaction ranges remain bounded independently of wall length. The wall endpoints have quantum dimension $2^{Q/2}$, where $Q$ counts the independent generators in each electric or magnetic charge space. As an application, we illustrate measurement-only braiding for $Q=3$ under the correspondence between measurements and defect transport. Our construction provides a valuable basis for exploring non-Abelian defect physics and developing logical operations in quantum low-density parity-check codes with Abelian topological order in future.
\end{abstract}
\maketitle

Quantum low-density parity-check (qLDPC) codes have opened a new route toward scalable quantum computing, allowing high encoding rates and growing code distance to coexist with sparse parity checks~\cite{mackaySparse2004,kovalevQuantum2013,tillichQuantum2014, breuckmannBalanced2021,panteleevAsymptotically2022,breuckmannQuantum2021}. As experiments advance toward reliable logical processors~\cite{krinnerRealizing2022,googleQuantum2025,bluvsteinLogical2024}, this combination offers a path to protecting large logical registers with substantially reduced physical-qubit overhead. Among these codes, bivariate bicycle (BB) codes combine sparse stabilizer checks with high encoding rates in finite instances~\cite{bravyiHighthreshold2024,vossMultivariate2024,wangCoprime2026, linQuantum2024,wangAbelian2023,wangDistance2022,shawLowering2025,liangGeneralized2025,Wang2026Demonstration}. They can store more logical qubits than surface codes~\cite{dennisTopological2002,kitaevFaulttolerant2003,fowlerSurface2012} at comparable distance~\cite{bravyiHighthreshold2024}, motivating a number of proposals for their implementation and logical operations~\cite{berthusenToward2025,yoderTourdegross2025,breuckmannFoldtransversal2024,liangSelfdual2026,eberhardtLogicalTIT2025}.Turning these memories into computational architectures requires logical operations that manipulate and entangle logical qubits while preserving the advantages of sparse error correction~\cite{eastinRestrictions2009,bravyiClassification2013}. Existing strategies include transversal gates~\cite{quintavallePartitioning2023}, measurement-based schemes~\cite{horsmanSurface2012,litinskiGame2019,cohenLowoverhead2022,yangPlanar2026,williamsonLowoverhead2026}, and so on. The Abelian topological order~\cite{wenTopological1990,kitaevAnyons2006,haahCommuting2013} associated with BB codes~\cite{chenAnyon2025} offers an alternative perspective, where the stabilizer algebra encodes emergent charge sectors with well-defined fusion and braiding. For BB codes with multiple independent electric and magnetic charge generators, this richer charge structure provides a natural basis for engineering defect degrees of freedom beyond those of a single toric-code layer. A natural question is how non-Abelian twist defects and coherent operations on their fusion spaces can be realized within the original Pauli stabilizer framework of BB codes.

\begin{figure}[tb]
 \centering
 \makebox[.455\columnwidth][l]{\footnotesize\textbf{(a)}}%
 \makebox[.545\columnwidth][l]{\footnotesize\textbf{(b)}}\par\vspace{2pt}
 \includegraphics[width=\columnwidth]{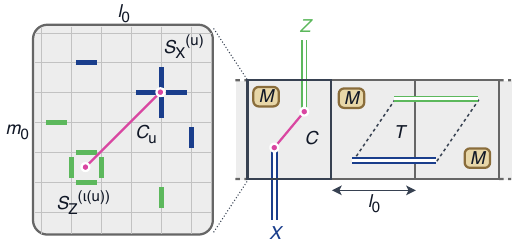}
 \caption{Electric-magnetic duality wall constructed from mutually commuting Pauli checks. Blue and green lines denote $X$ and $Z$ support, respectively. (a) Within an $l_0\times m_0$ block, original checks at reflected positions form the composite check $C_u=S_X^{(u)}S_Z^{(\iota(u))}$. (b) The wall consists of a strip of these blocks. Double lines represent ribbons. A crossing ribbon changes Pauli type at the wall, and an interblock check $T$ combines electric and magnetic ribbons between adjacent blocks. Patch checks $M$ remove the remaining charge-neutral local logical modes. The operator supports are schematic.
}
 \label{fig:defect}
\end{figure}

Non-Abelian twist defects arise at the endpoints of domain walls that permute topological charges, allowing an Abelian topological medium to support collective fusion spaces~\cite{bombinTopological2010}. Quantum information is encoded non-locally in these spaces, while braiding implements noncommuting logical transformations whose action depends on the topology of the exchange paths~\cite{nayakNonAbelian2008,freedmanModular2002,freedmanTopological2003,frankoExtraspecial2006}. In the toric code, an electric-magnetic wall exchanges electric and magnetic charges and terminates at defects with Ising fusion rules~\cite{bombinTopological2010,kitaevModels2012,brownPoking2017}, which enables the logical-gate implementations~\cite{litinskiLattice2018,yoderSurface2017,bombinClifford2011,hiraiSpacetime2026}. On the other hand, measurement-only protocols based on pair-charge projections can also reproduce non-Abelian braiding transformations without moving the defects~\cite{bondersonMeasurementOnly2008,bondersonMeasurementTunable2013}. This connection between defects, fusion, and logical operations has motivated further theoretical explorations on topological phases and quantum codes~\cite{barkeshliTwist2013,barkeshliTopological2012,barkeshliTheory2013,kesselringBoundaries2018,youProjective2012,teoUnconventional2014,zhengDemonstrating2015,barkeshliSymmetry2019,teoTheory2015,tarantinoSymmetry2016,scrubyHierarchy2020,zhuNonAbelian2026,zhuSipserSpielman2026,christosNonAbelian2026,liangOperator2024,jiaGeneralized2026,mousaComposite2025,wangAlgebra2026,yoshidaGapped2017}, as well as experimental studies of non-Abelian operations~\cite{andersenNonAbelian2023,xuNonAbelian2024,iqbalNonAbelian2024}. For BB codes, however, an anyon permutation does not itself specify a wall in the sparse stabilizer Hamiltonian. One needs to construct mutually commuting checks that enforce the prescribed conversion of every charge type while removing accidental local modes, so that the surviving degeneracy belongs to the intended defect fusion space. Furthermore, to implement the non-Abelian braiding for logical operations, one should demonstrate how the protocol reproduces the braiding transformation while preserving the coherence of the encoded quantum information.

In this paper, we develop a general framework for constructing non-Abelian twist defects in arbitrary BB code with finite anyon content.  We build electric-magnetic walls through defining composite checks, interblock checks, and patch checks. This converts the $Q$ independent generators in each charge space while removing all accidental local modes. For sufficiently long and well-separated walls, we show that each additional twist pair encodes $Q$ logical qubits and obtain a twist quantum dimension of $2^{Q/2}$. For $Q=3$, we construct three families of pair-charge measurements and derive their logical action directly from microscopic Pauli operators. Under the correspondence between microscopic measurements and defect transport, the protocol realizes measurement-only braiding.

{\it Charge algebra and the duality wall}---We consider the BB code $\BBc{\alpha,\beta,a,b}$ with $f=1+x+x^{-\alpha}y^b$, $g=1+y+x^ay^{-\beta}$ over $R=\mathbb F_2[x^{\pm1},y^{\pm1}]$~\cite{bravyiHighthreshold2024}. Electric (magnetic) excitations violate $X$ ($Z$) checks, with $x^iy^j$ recording a violation at $(i,j)$. Excitation patterns related by finite-support Pauli operations carry the same topological charge. Identifying such patterns gives the electric and magnetic charge spaces
\begin{equation}\label{eq:charges}
 \begin{gathered}
 V_e=R/\langle f^\star,g^\star\rangle,\qquad
 V_m=R/\langle f,g\rangle,\\
 \dim_{\mathbb{F}_2} V_e=\dim_{\mathbb{F}_2}  V_m=Q,
 \end{gathered}
\end{equation}
where $h^\star(x,y)=h(x^{-1},y^{-1})$ and $0<Q<\infty$. Fusion adds charge classes. Each space has $Q$ independent generators and $2^Q$ charge types, including the vacuum. The bulk corresponds to $Q$ toric-code copies in the local classification~\cite{bombinUniversal2012,haahClassification2021,wangDecoupling2026}. Finite period ribbons obtained by polynomial division~\cite{coxIdeals2015} transport these charges. The commutation phases of intersecting magnetic and electric period ribbons define the nondegenerate mutual braiding pairing $\mathcal B:V_m\times V_e\to\{\pm1\}$~\cite{supplementalMaterial}.

We construct the wall along a strip of blocks of size $l_0\times m_0$. The block dimensions are chosen so that translating an excitation by $l_0$ lattice spacings along $x$ or $m_0$ along $y$ leaves its topological charge unchanged. These common charge periods satisfy $1+x^{l_0},1+y^{m_0}\in\langle f,g\rangle$. Within each block, reflection about its center pairs the position $u=(i,j)$ with $\iota(u)=(l_0-1-i,m_0-1-j)$. For an excitation pattern represented by $r$, the reflected pattern is $r^\iota=\kappa r^\star$, where $\kappa=x^{l_0-1}y^{m_0-1}$ accounts for the position of the block center. One can then define the  electric-magnetic charge duality 
\begin{equation}\label{eq:duality}
 \mathcal D(e_r)=m_{r^\iota},\qquad
 \mathcal D(m_s)=e_{s^\iota},
\end{equation}
which preserves fusion, mutual braiding, and topological spin. Microscopically, this duality is realized by a Hadamard-type fold symmetry that exchanges $X$ and $Z$ checks at reflected positions~\cite{eberhardtLogicalTIT2025}.

In the wall, the original checks are replaced by the three types of checks: the  composite checks $C$, the interblock checks $T$, and the patch checks $M$ (Fig.~\ref{fig:defect}). The composite check on the position $u$ takes the form $C_u=S_X^{(u)}S_Z^{(\iota(u))}$, where $S_X^{(u)}$ and $S_Z^{(\iota(u))}$ denote the original checks on the position $u$ and $\iota(u)$  respectively. Consequently, the eigenvalues of $(S_X^{(u)}, S_Z^{(\iota(u))})$ can flip together, while leaving $C_u$ unchanged. 

Interblock checks combine a $Z$-type electric period ribbon with its $X$-type magnetic dual shifted forward by one block~\cite{supplementalMaterial}. Let $\mathcal R^e_{j,v}$ and $\mathcal R^m_{j,v}$ denote the electric and magnetic ribbons connecting position $v$ in block $j$ to the corresponding position in block $j+1$, respectively. We define $T_{j,u}=\mathcal R^e_{j,\iota(u)}\mathcal R^m_{j,u}$. This ribbon--dual construction yields Hermitian, mutually commuting interblock checks. The only original checks that anticommute with $T_{j,u}$ are $S_X^{(\iota(u))}$ and $S_Z^{(u)}$ in each of the two blocks. The two minus signs cancel in their composite $C_{\iota(u)}$, so $T_{j,u}$ commutes with all composite checks and all original checks retained in the bulk. 

A contractible closed ribbon crossing the wall twice without leaving excitations should preserve the encoded information. It must therefore be a product of stabilizers up to an overall phase. However, the composite and interblock checks alone leave local logical modes along the wall. To characterize these modes, we  define the quantity $\ell(P)=\sum_jw_j\pmod 2$, where $P$ is a Pauli operator, $w_j$ denotes the pattern of removed original $X$ checks flipped by $P$ in block $j$. Multiplication by an interblock stabilizer preserves $P$'s action on the code space but flips corresponding bits in two neighboring blocks. These changes leave  $\ell(P)$ unchanged. Aligning corresponding positions across blocks gives a single $l_0\times m_0$ pattern in $\mathsf A=R/\langle1+x^{l_0},1+y^{m_0}\rangle$. For any finite-support $P$ commuting with the retained bulk, composite, and interblock checks, $\ell(P)$ lies in the subspace $\mathcal I_e=(f^\star,g^\star)\mathsf A$. The quotient $\mathsf A/\mathcal I_e\simeq V_e$ records the total topological charge with $Q$ binary coordinates, leaving $l_0m_0-Q$ independent charge-neutral patterns.

We remove these modes with patch checks. For each charge-neutral basis pattern $h\in\mathcal I_e$, we construct a finite-support Pauli that flips removed $X$ and $Z$ checks in reflected pairs $(h,h^\iota)$. The two sign changes cancel in each affected composite, and no retained bulk check is flipped. Multiplying by suitable removed original checks then cancels any interblock-check flips without altering these paired patterns. This yields the patch check $M(h)$. For a fixed code, their weights and interaction ranges remain bounded independently of wall length. With the patch checks added, every finite-support Pauli that leaves no excitations on the infinite wall is a product of stabilizers up to an overall phase~\cite{supplementalMaterial}. Thus, a contractible closed ribbon crossing the wall twice cannot change the encoded information if it leaves no excitations.

\begin{figure}[tb]
 \centering
 \subfloat[\label{fig:loops-a}]{%
   \includegraphics[trim=1 16 4 6,clip,height=78bp]{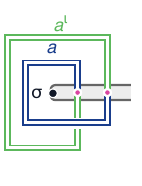}}\hfill
 \subfloat[\label{fig:loops-b}]{%
   \includegraphics[trim=0 1 1 1,clip,height=78bp]{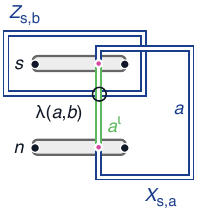}}\hfill
 \subfloat[\label{fig:loops-c}]{%
   \includegraphics[trim=1 3 1 12,clip,height=78bp]{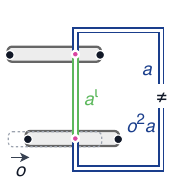}}
 \caption{Ribbon operators for twist defects. Black dots indicate twist endpoints. (a) A ribbon winds twice around a twist before closing, alternating between charges $m_a$ and $e_{a^\iota}$ that form the invariant composite $\epsilon_a=m_a e_{a^\iota}$~\cite{bombinTopological2010}. (b) $\mathsf Z_{s,b}$ reads pair $s$'s charge, and $\mathsf X_{s,a}$ transfers charge to reference pair $n$ at fixed total charge. Their commutation phase is $\lambda(a,b)$. (c) Displacing the lower wall by $o$ changes the returning charge to $m_{o^2a}$, so closure requires $(1+o^2)a=0$.}
 \label{fig:loops}
\end{figure}

Ribbon operators can transport any charge across the infinite wall in either direction without exciting the wall. For such a crossing, commutation with the interblock checks requires the electric charge on one side and the electric dual of the magnetic charge on the other to have identical braiding phases with every magnetic charge. Since these phases distinguish all electric charge types, they fix the conversion from magnetic to electric charge. At the same time, the composite checks require the electric and magnetic syndromes on the removed original checks to form reflected pairs. Charge conservation, including both the bulk endpoint charges and these paired wall syndromes, then fixes the conversion from electric to magnetic charge, establishing the duality in Eq.~\eqref{eq:duality}. Furthermore, we show that crossing-wall ribbons with identical bulk endpoint syndromes differ only by a product of stabilizers and an overall phase [SM, Theorem~\ref{thm:transmission}]. These conclusions also hold for finite walls when the crossing region lies sufficiently far from both endpoints [SM, Corollary~\ref{cor:finite-transmission}]. In the toric-code limit, our wall implements the same electric-magnetic exchange $e\leftrightarrow m$ as the dislocation line~\cite{bombinTopological2010,kitaevModels2012}.

{\it Endpoint sectors and fusion}---Terminating the duality wall at block boundaries creates a pair of non-Abelian twist defects of type $\sigma$ in the Abelian bulk~\cite{barkeshliTwist2013}. We retain the composite checks within the finite strip and the interblock checks between its blocks, with patch checks selected according to the natural termination rule~\cite{supplementalMaterial}. For sufficiently long walls, any Pauli operator confined to a prescribed endpoint neighborhood either excites a check or leaves every encoded state unchanged up to an overall phase [SM, Theorem~\ref{thm:endpoint-exactness}]. An individual endpoint therefore carries no local logical degree of freedom.

Quantum information is encoded collectively in the joint charge states of separated endpoint pairs. To describe this space, we consider $n$ long, well-separated parallel walls on a torus whose side lengths are integer multiples of the corresponding charge periods. A translation represented by the monomial $o$ moves each block's reflection center. The reflected image of a fixed point then shifts by twice the displacement, changing $\kappa$ in Eq.~(\ref{eq:duality}) to $o^2\kappa$. We therefore choose relative wall displacements of whole charge periods so that all walls implement the same electric-magnetic duality. In this arrangement, loops enclosing a pair read its charge, while ribbons connecting two pairs change their charges without altering the total topological charge. At fixed total topological charge, we find that the defect-space dimension and the quantum dimension $d_\sigma$ of each twist are
\begin{equation}\label{eq:dimension}
 \dim\mathcal H_{\rm def}=2^{(n-1)Q},\qquad d_\sigma=2^{Q/2}.
\end{equation}
Thus the $2n$ twists encode $(n-1)Q$ logical qubits, in addition to those of the original BB code on the torus.

To establish the above dimension, we construct charge-readout loops and charge-transfer ribbons, then analyze their algebra. We choose the $n$-th wall  as the reference, whose pair charge is fixed by the other $n-1$ pair charges once the total topological charge is specified. For each $s<n$, a magnetic loop $\mathsf Z_{s,b}$ carrying charge $m_b$ encloses the  $s$-th pair and reads its charge (Fig.~\ref{fig:loops}(b)). A ribbon operator $\mathsf X_{s,a}$ transfers charge between this pair and the reference, preserving the total. Here $a,b\in V_m$, and the magnetic and electric segments of the transfer ribbon carry $m_a$ and its dual $e_{a^\iota}$, respectively. The symbols $\mathsf Z$ and $\mathsf X$ describe their logical roles. The braiding phase $\lambda(a,b)=\mathcal B(a,b^\iota)$ determines whether a transfer flips a readout eigenvalue, giving $\mathsf Z_{s,b}\mathsf X_{t,a} =\lambda(a,b)^{\delta_{st}}\mathsf X_{t,a}\mathsf Z_{s,b}$.
The pairing $\lambda$ is nondegenerate, meaning that every nonzero charge transfer flips the eigenvalue of at least one readout loop.  The loop and transfer operators therefore generate $Q$ logical Pauli $X$--$Z$ pairs for each $s<n$, supporting at least $(n-1)Q$ defect qubits. An independent stabilizer-rank count gives exactly this number of additional qubits. Thus, these operators account for the entire defect logical space, as shown in Eq.~\eqref{eq:dimension}.

The pair-charge sectors are labeled by the Abelian composites $\epsilon_a=m_a e_{a^\iota}$, $a\in V_m$, which remain unchanged under the wall's electric-magnetic duality (Fig.~\ref{fig:loops}(a)). For four twists at fixed total charge, pair loops distinguish $2^Q$ such sectors. Under the assumed correspondence between charge sectors and fusion channels~\cite{supplementalMaterial}, these sectors describe the possible fusion outcomes of a twist pair, each with a one-dimensional fusion space. A twist $\sigma$ can also absorb any $\epsilon_a$ without changing its type. Together with Abelian charge addition, one can realize 
\begin{equation}\label{eq:fusion}
 \sigma\times\sigma=\sum_{a\in V_m}\epsilon_a,\quad
 \epsilon_a\times\epsilon_b=\epsilon_{a+b},\quad
 \sigma\times\epsilon_a=\sigma.
\end{equation}
These are Tambara--Yamagami fusion rules~\cite{tambaraTensor1998, siehlerBraided2000,gelakiCenters2009}. Each $\epsilon_a$ has quantum dimension $d_{\epsilon_a}=1$, and the corresponding topological spin $\theta_{\epsilon_a}$ equals the mutual braiding phase of its electric and magnetic constituents, as $\theta_{\epsilon_a}=\lambda(a,a)$. The $2^Q$ invariant charges $\{\epsilon_a\mid a\in V_m\}$ are either all bosons or equally divided between bosons and fermions.

\begin{figure}[t]
 \centering
 \subfloat[\label{fig:q3-routes-a}]{%
   \includegraphics[trim=0 1 13 0,clip,height=99bp]{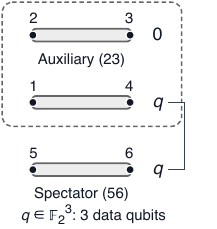}}\hfill
 \subfloat[\label{fig:q3-routes-b}]{%
   \includegraphics[trim=1 1 4 4,clip,height=99bp]{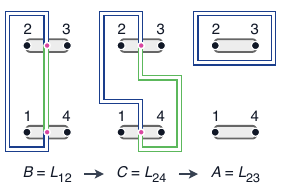}}
 \caption{Six-twist setup for $Q=3$. (a) With total charge and auxiliary pair $(23)$ fixed to vacuum, pairs $(14)$ and $(56)$ share $q\in\mathbb F_2^3$, encoding three logical qubits. The spectator $(56)$ remains outside the dashed active region. (b) $B\to C\to A$ targets vacuum outcomes on $(12)$, $(24)$, and $(23)$, restoring the auxiliary vacuum. Each pair measurement uses three commuting Wilson observables.}
 \label{fig:q3-routes}
\end{figure}

{\it Measurement-only braiding for $\BBc{1,1,3,1}$}---The pair-charge loops provide the measurements required by the measurement-only braiding protocol~\cite{bondersonMeasurementOnly2008,bondersonMeasurementTunable2013,karzigScalable2017,vijayTeleportationbased2016}. We implement this protocol using six twists in $\BBc{1,1,3,1}$ (Fig.~\ref{fig:q3-routes}(a)). At fixed total vacuum charge, the auxiliary pair $(23)$ is initialized in vacuum. Pairs $(14)$ and $(56)$ then share a charge label $q\in\mathbb F_2^3$, whose eight values encode three data qubits. We measure pair charges based on Wilson loops $A_a=L_{23,a}$, $B_a=L_{12,a}$, and $C_a=L_{24,a}$. Here $a=(a_0,a_1,a_2)\in\mathbb F_2^3$ labels the ribbon charge in the magnetic basis $(1,x,x^2)$. Each pair charge is determined by three commuting basis observables~\cite{bondersonInterferometry2008}. We  define the projector associated with pair-charge outcome $t$ as
$P_F^t=\frac18\sum_{a\in\mathbb F_2^3}\lambda(t,a)F_a$, $F=A,B,C$.
Here $\lambda(t,a)=\mathcal B(t,a^{\iota})$ denotes the braiding phase between $m_t$ and $e_{a^{\iota}}$.

Now we implement the braiding of twists based on projective measurements. We initialize the auxiliary pair $(23)$ in the vacuum channel. The protocol targets vacuum outcomes for pairs $(12)$, $(24)$, and $(23)$ in the sequence $B\to C\to A$ (Fig.~\ref{fig:q3-routes}(b)). The first two projections couple an auxiliary twist to each data twist in turn, and the last restores the auxiliary pair to vacuum. Multiplying the vacuum projectors gives~\cite{supplementalMaterial}
\begin{equation}
\begin{gathered}
P_A^0P_C^0P_B^0P_A^0
\propto |0\rangle\langle0|_{\rm aux}\otimes U_0,\\
U_0|q\rangle=i^{q_0+q_2}(-1)^{q_1q_2}|q\rangle, \quad U_0=S_0S_2\mathrm{CZ}_{12},
\end{gathered}
\end{equation}
where $|0\rangle_{\rm aux}$ denotes the vacuum charge state of the auxiliary pair $(23)$,  $|q\rangle=|q_0q_1q_2\rangle$ denotes the charge-$q$ state of pair $(14)$, $S=\operatorname{diag}(1,i)$ and $\mathrm{CZ}=\operatorname{diag}(1,1,1,-1)$ gates act on the sites indicated by their subscripts. Here one can adopt the forced measurements to obtain the target vacuum outcomes \cite{bondersonMeasurementOnly2008}. Under the correspondence between microscopic measurements and defect transport, we implement the braiding operation $U_0$ between the two twists $1$ and $4$. The Clifford gates and Pauli measurements used here remain within the stabilizer framework~\cite{gottesmanHeisenberg1998,aaronsonImproved2004}. Universal computation would require additional non-stabilizer resources, such as magic-state injection and distillation~\cite{bravyiUniversal2005,bravyiMagic2012}.


We would like to stress that BB wall configurations can have different collective encoding spaces despite identical bulk order and individual twist quantum dimensions. In BB codes, translating a wall can change its electric-magnetic matching. At fixed total charge, charge changes at different endpoint pairs must compensate one another, so the allowed joint states depend on this matching. We rigorously establish the resulting displacement dependence through stabilizer-rank counting and show that individual endpoints carry no local logical degrees of freedom~\cite{supplementalMaterial}.  For the $Q=3$ example, it shows that two long and well-separated walls on a torus compatible with the charge periods encode three additional qubits, whereas a configuration with one wall shifted by one lattice spacing has no additional defect qubits (Fig.~\ref{fig:loops}(c)). Our results establish how lattice symmetry determines the compatibility of charges carried by different endpoint pairs, making the relative placement of walls a physical parameter that controls their collective encoding space.

{\it Discussion}---Our findings would suggest further research on exploring the control and protection of quantum information in BB codes. One is to determine how quantum information can be preserved when wall displacements change the dimension of the defect encoding space~\cite{vuillotCode2019}. This would clarify whether the position dependence of defect encoding can be used for controlled logical operations. Another direction is to adapt existing qLDPC surgery methods~\cite{crossImproved2025} for information transfer between the defect register and selected logical qubits of the original BB code. Such an interface could bring defect operations to stored quantum information, while requiring assessing error protection throughout the transfer, together with the number of auxiliary qubits and measurement rounds needed. For stationary walls, a further question is how the defect code distance and the splitting of fusion states under perturbations depend on wall length and defect separation~\cite{bravyiTopologicalStability2010,bravyiShort2011}. Establishing these scaling relations would identify the geometric scales that govern the robustness of the encoded information.

In conclusion, we construct electric-magnetic duality walls in BB codes with finite anyon content using mutually commuting Pauli checks. These checks remove local logical modes along the wall, while termination produces twist endpoints that encode quantum information collectively. We determine the resulting encoding space and its dependence on the bulk charge structure and relative wall placement. In an explicit example, we implement the braiding operation of those non-Abelian twist defects through the microscopic pair-charge measurements. Our results shed light on how bulk topological order and lattice translations jointly shape the quantum information that defects can encode collectively, providing a  foundation for future studies of non-Abelian operations in qLDPC codes.

\emph{Acknowledgments---} We thank Liang Chang, Jiucai Wang, Weikang Li for helpful discussions. This work was supported by the National Natural Science Foundation of China (Grant No. 12375060), and the Fundamental Research Funds for the Central Universities (Grants No. 63253097, No. 63263090).

\bibnotoc
\bibliography{references}
\bibtoc

\makeatletter
\g@addto@macro\normalsize{%
  \setlength{\abovedisplayskip}{6pt plus 2pt minus 4pt}%
  \setlength{\belowdisplayskip}{6pt plus 2pt minus 4pt}%
  \setlength{\abovedisplayshortskip}{0pt plus 1pt}%
  \setlength{\belowdisplayshortskip}{3pt plus 1pt minus 2pt}}
\makeatother
\normalsize

\makeatletter
\def\section{\@startsection{section}{1}{\z@}{16\p@ \@plus3\p@ \@minus2\p@}{7\p@}%
  {\normalfont\small\bfseries\centering}}
\def\subsection{\@startsection{subsection}{2}{\z@}{14\p@ \@plus3\p@ \@minus2\p@}{6\p@}%
  {\normalfont\small\bfseries\centering}}
\def\subsubsection{\@startsection{subsubsection}{3}{\z@}{12\p@ \@plus2\p@ \@minus2\p@}{5\p@}%
  {\normalfont\small\itshape\centering}}
\def\thm@space@setup{\thm@preskip=6\p@ \@plus2\p@ \@minus2\p@
  \thm@postskip=\thm@preskip}
\renewenvironment{proof}[1][\proofname]{\par
  \pushQED{\qed}\normalfont \topsep5\p@\@plus2\p@\@minus2\p@\relax
  \trivlist\item[\hskip\labelsep\itshape #1\@addpunct{.}]\ignorespaces
}{\popQED\endtrivlist\@endpefalse}
\makeatother

\clearpage
\onecolumngrid
\makeatletter
\setcounter{figure}{0}
\setcounter{equation}{0}
\setcounter{theorem}{0}
\setcounter{secnumdepth}{3}
\renewcommand{\thefigure}{S\@arabic\c@figure}
\renewcommand \theequation{S\@arabic\c@equation}
\renewcommand \thetable{S\@arabic\c@table}
\renewcommand{\thetheorem}{S\arabic{theorem}}
\renewcommand{\thelemma}{S\arabic{lemma}}
\renewcommand{\thedefinition}{S\arabic{definition}}
\renewcommand{\thecorollary}{S\arabic{corollary}}
\renewcommand{\theproposition}{S\arabic{proposition}}
\renewcommand{\theremark}{S\arabic{remark}}
\renewcommand{\theHfigure}{S\arabic{figure}}
\renewcommand{\theHequation}{S\arabic{equation}}
\renewcommand{\theHtable}{S\arabic{table}}
\renewcommand{\theHtheorem}{S\arabic{theorem}}
\renewcommand{\theHlemma}{S\arabic{lemma}}
\renewcommand{\theHdefinition}{S\arabic{definition}}
\renewcommand{\theHcorollary}{S\arabic{corollary}}
\renewcommand{\theHproposition}{S\arabic{proposition}}
\renewcommand{\theHremark}{S\arabic{remark}}
\makeatother
\newpage

\begin{center}
	{\large \bf Supplemental Material for:\\
    ``Non-Abelian Twist Defects in Bivariate Bicycle Codes''} \\
    \vspace{0.8em}
    Rui Zeng and Li-Wei Yu\textsuperscript{*}\\
    \textit{Theoretical Physics Division, Chern Institute of Mathematics and LPMC,\\
    Nankai University, Tianjin 300071, China}\\
    \textsuperscript{*}\href{mailto:yulw@nankai.edu.cn}{yulw@nankai.edu.cn}
\end{center}

\renewcommand\thesection{S\Roman{section}}
{
  \hypersetup{linkcolor=blue}
  \makeatletter
  \let\l@subsubsection\@gobble@tw@
  \let\l@paragraph\@gobble@tw@
  \makeatother
  \tableofcontents
}
\vspace{2em}

\noindent\textit{Overview.}---This Supplemental Material develops a route from the defining polynomials $(f,g)$ of a
bivariate-bicycle (BB) code to twist-defect logic.  Its destination is the
Letter's central claim: one defect line, which starts from the code's own checks merged in mirror
pairs and is completed by interblock and patch layers,
whose endpoints inherit the code's full logical multiplicity --- $Q$ logical
qubits for every twist pair beyond the first in the common-frame, commensurate
configurations of Theorem~\ref{thm:exhaustion}.  The arc is \emph{algebra first, geometry
second, defects third}.
Sec.~\ref{sec:twists} uses the crossing law of Sec.~\ref{sec:transmission} only to
make its pair-charge crossings admissible.  The code $\BBc{1,1,3,1}$, with $Q=3$, is the running example
throughout.

\section{Algebraic Foundations of Abelian Braiding}\label{sec:algebra}

Everything in this work is computed from two Laurent polynomials.  This
section hands the rest of the paper four objects: the chain complex that turns
stabilizer configurations, Pauli operators and syndromes into modules over a
Laurent polynomial ring (Sec.~\ref{sec:chain}); the label spaces $V_e$ and
$V_m$, whose sum is the entire anyon content; the topological index $Q$ that
counts them; and the anyon periods $l_0,m_0$, which fix the block size of the
defect construction (Sec.~\ref{sec:anyontypes}).

\subsection{The BB Code as a Chain Complex}\label{sec:chain}

Bivariate bicycle (BB) codes are a promising class of high-rate quantum
low-density parity-check (qLDPC) codes: translation-invariant CSS
codes~\cite{calderbankQuantum1998}
with two qubits and two checks per unit cell of a two-dimensional lattice.
Translation invariance is the structural feature we exploit throughout: it
lets us encode an entire lattice pattern in a single Laurent polynomial,
with the monomial $x^iy^j$ recording an insertion at cell $(i,j)$. To track
topological excitations and stabilizer interactions, we map the physical
stabilizer operators onto a chain complex over this polynomial ring.

Throughout, $\BBc{\alpha,\beta,a,b}$ denotes the bivariate bicycle
code~\cite{bravyiHighthreshold2024, chenAnyon2025} defined by the polynomials
\begin{equation*}
    f = 1 + x + x^{-\alpha} y^{b}, \qquad g = 1 + y + x^{a} y^{-\beta}
\end{equation*}
over $R = \mathbb{F}_2[x, y, x^{-1}, y^{-1}]$ (signs are immaterial in
characteristic 2). For $h = \sum_{i,j} c_{ij}\, x^{i} y^{j} \in R$ we write
$h^{\star} = h(x^{-1}, y^{-1})$ for its antipode; for a matrix $M$ over $R$,
$M^{\dagger}$ denotes the transpose composed with the entrywise antipode.
The map $\tr : R \to \mathbb{F}_2$ extracts the coefficient of the identity
monomial, $\tr(h) = c_{00}$.

Three $R$-modules organize the code. Physically, $G = R^{2}$ indexes the two
stabilizer types per unit cell, $P = R^{4} \cong R^{2} \oplus R^{2}$ encodes
Pauli operators on the two qubits per cell in symplectic form, and
$E = R^{2}$ indexes the two syndrome types. Throughout this SM we order the
Pauli module as
\begin{equation}\label{eq:pauli-component-order}
    P=P_Z\oplus P_X,
    \qquad
    p=(u_1,v_1\mid u_2,v_2),
\end{equation}
so the first two entries are the $Z$ support and the last two are the $X$
support. With this convention the two components of $\cobdry p$ below are
the electric and magnetic syndrome polynomials, respectively.

Unless a coherent phase section is explicitly fixed, every statement that two
Pauli operators are equal, or that one lies in a stabilizer group, is meant in
the binary symplectic module, modulo the scalar $\{\pm1,\pm i\}$; all rank,
commutator and exactness arguments below are computed there, with Hermitian
representatives chosen where a Hamiltonian term is required.

The BB code can be described by a chain
complex~\cite{haahCommuting2013,chenAnyon2025}:
\begin{align}
    G \xrightarrow{\bdry} P \xrightarrow{\cobdry} E .
\end{align}
The boundary map $\bdry$ sends a formal combination of checks to the Pauli
operator obtained by multiplying them:
\begin{align*}
    G &\xrightarrow{\bdry} P \\
    \begin{pmatrix} p \\ v \end{pmatrix} &\mapsto
    \begin{pmatrix} g^{\star} & 0 \\ f^{\star} & 0 \\ 0 & f \\ 0 & g \end{pmatrix}
    \begin{pmatrix} p \\ v \end{pmatrix} =
    \begin{pmatrix} g^{\star}p \\ f^{\star}p \\ fv \\ gv \end{pmatrix}.
\end{align*}
Similarly, the co-boundary map $\cobdry$ sends a Pauli configuration to its
syndrome:
\begin{align*}
    P &\xrightarrow{\cobdry} E \\
    \begin{pmatrix} u_1 \\ v_1 \\ u_2 \\ v_2 \end{pmatrix} &\mapsto
    \begin{pmatrix} f^{\star} & g^{\star} & 0 & 0 \\ 0 & 0 & g & f \end{pmatrix}
    \begin{pmatrix} u_1 \\ v_1 \\ u_2 \\ v_2 \end{pmatrix} =
\begin{pmatrix} f^{\star}u_1 + g^{\star}v_1 \\ gu_2 + fv_2 \end{pmatrix}.
\end{align*}
One checks directly that $\cobdry \circ \bdry = 0$, since
$f^{\star}(g^{\star}p) + g^{\star}(f^{\star}p) = 0$ and $g(fv) + f(gv) = 0$
over $\mathbb{F}_2$, so the sequence is a chain complex. The identity
$\cobdry\circ\bdry=0$ is simply the statement that products of stabilizers
create no syndrome; the interesting physics lives in the failure of the
sequence to be exact at $E$, which we quantify next.

\subsection{Anyon Types and the Topological Index}\label{sec:anyontypes}

An excitation pattern $e\in E$ is \emph{trivial} if some local Pauli
operator creates it, i.e., if $e\in\im\cobdry$. Topological charge is what
survives modulo such trivial patterns, so all anyon types in a BB code are
labeled by the syndrome quotient~\cite{chenAnyon2025}. With the component
ordering fixed above, the first syndrome component is vertex/electric and
the second is plaquette/magnetic. We therefore write
\[
    V_e=\frac{R}{\langle f^\star,g^\star\rangle},
    \qquad
    V_m=\frac{R}{\langle f,g\rangle},
\]
so that
\begin{equation}\label{eq:C}
    \frac{E}{\im\cobdry}\cong V_e\oplus V_m .
\end{equation}

Anyon labels are therefore polynomials, expanded in the standard monomials
of the quotient with coefficients in $\mathbb{F}_2$: a basic excitation at
lattice site $(i,j)$ is the monomial $x^iy^j$, and fusion of labels $p_1$
and $p_2$ is the sum $p_1+p_2$ taken modulo the relevant ideal. The additive
structure is the Abelian fusion rule of the code, and characteristic $2$
makes every anyon its own inverse.  The standard monomials depend on a chosen
presentation of the localized quotient and a compatible order; the quotient,
its dimension and its additive structure do not.

The size of the anyon content is measured by the \textbf{topological index}
\[
    Q=\dim_{\mathbb F_2} R/\langle f,g\rangle ,
\]
the $\mathbb{F}_2$-dimension of the quotient module $R/\langle f,g\rangle$,
which measures the space of irreducible excitations and sets the topological
order: the Abelian sector has $2^{2Q}$ anyon types, and on a pristine torus
whose side lengths are multiples of the anyon periods below the code encodes exactly $2Q$
logical qubits (the rank computation in the proof of
Lemma~\ref{lem:routed-logical-basis}). For the running example $\BBc{1,1,3,1}$, the set
of standard monomials is $\{ 1, x, x^2 \}$, so $Q=3$ and the code supports
$2^{6} = 64$ different kinds of anyon types.

We also use the \textbf{anyon periods} $l_0,m_0$, the minimal positive
integers with
\(1+x^{l_0},\;1+y^{m_0}\in \langle f,g\rangle\)
respectively.  These minima exist whenever $Q<\infty$, and every result below
assumes $0<Q<\infty$, a finite and non-trivial anyon theory;
Lemma~\ref{lem:common-periods} below supplies them and shows that the dual ideal $\langle f^\star,g^\star\rangle$ has
exactly the same periods. Physically,
these are the translation lengths after which every anyon returns to its
original type; they set the natural block size of the defect construction in
Sec.~\ref{sec:scaffold}.

\begin{lemma}[Common periods]\label{lem:common-periods}
If $Q<\infty$, positive $x$- and $y$-periods exist.  For every positive
integer $n$,
\begin{equation}
  1+x^n\in\langle f,g\rangle
  \quad\Longleftrightarrow\quad
  1+x^n\in\langle f^\star,g^\star\rangle,
\end{equation}
and likewise for $y$.  Hence the primal and dual anyon periods coincide.
\end{lemma}
\begin{proof}
Multiplication by $x$ and by $y$ is invertible on the finite space
$R/\langle f,g\rangle$, hence of finite order.  The antipode carries
$\langle f,g\rangle$ onto $\langle f^\star,g^\star\rangle$ and $1+x^n$ to the
unit multiple $x^{-n}(1+x^n)$, which gives both directions.
\end{proof}

An $L_1\times L_2$ torus is \emph{commensurate} when $l_0\mid L_1$ and
$m_0\mid L_2$.  On a torus that is not, transport around a cycle returns an anyon
multiplied by $x^{L_1}$ or $y^{L_2}$.  Put
\[
 J_L=(1+x^{L_1})V_m+(1+y^{L_2})V_m,
 \qquad
 Q_L=\dim_{\mathbb F_2}V_m/J_L ,
\]
and $J_{e,L}$ for the electric counterpart in $V_e$.  The pristine code on that
torus encodes $2Q_L$ qubits: each CSS check map has rank $L_1L_2-Q_L$ by the
annihilator--quotient duality of Sec.~\ref{sec:kernel}, applied to the
Frobenius torus ring $R/(1+x^{L_1},1+y^{L_2})$.  Here $Q_L=Q$ exactly when
the torus is commensurate; otherwise fewer logical qubits survive, the
topological frustration of Ref.~\cite{chenAnyon2025}.

\section{Geometric Abelian Braiding}\label{sec:geometry}

The algebra of Sec.~\ref{sec:algebra} tells us \emph{which} anyons exist;
this section explains how they \emph{move} and \emph{braid}.  Three geometric
objects carry the physics: ribbon operators transport anyons
(Sec.~\ref{sec:ribbons}); closed loops decide commutation through their
well-defined \emph{interiors} (Sec.~\ref{sec:commutation}); and the braiding
pairing $\mathcal B$, a commutator of period ribbons computed in closed form
from $(f,g)$, records the resulting statistics (Sec.~\ref{sec:braidingstats}).
Section~\ref{sec:kernel} proves that pairing perfect.

\subsection{Ribbon Operators and Polynomial Division}\label{sec:ribbons}

Given an arbitrary localized vertex-type syndrome $e \in R$, the
multivariate division algorithm with respect to a Gr\"obner basis of the
ideal $\langle f^{\star}, g^{\star} \rangle$~\cite{coxIdeals2015} gives a
Pauli configuration $(u_1, v_1)$ that reduces $e$ to its normal form, and
creates (or corrects) $e$ exactly when that normal form vanishes:
\begin{align}\label{eq:excitation_division}
    e = f^{\star}u_1 + g^{\star}v_1 + r,
\end{align}
where $r$ is the remainder. The finite Pauli $(u_1,v_1\mid 0,0)$ has
syndrome $f^{\star}u_1+g^{\star}v_1=e+r$.  The remainder
$r$, which resides in the quotient space
$R/\langle f^{\star}, g^{\star} \rangle$, is the
irreducible topological charge (the anyon) at that location; if $e$ carries
no topological charge then $r = 0$ and that finite Pauli creates $e$ exactly.
These operators are not unique: for any $h \in R$, the pair
$(u_1 + h g^{\star},\; v_1 + h f^{\star})$ produces the same syndrome.

In the toric code one draws a string operator along a chosen lattice path and
the geometry is an \emph{input}; in a BB code the support of a ribbon is the
\emph{output} of the division algorithm, dictated by the ideal
$\langle f^{\star},g^{\star}\rangle$ and the chosen monomial order.  Any two ribbons with
the same endpoint syndromes differ by local stabilizer configurations, so
modulo stabilizers a ribbon's shape is pure gauge while its endpoints carry the
physics; its commutator with another excitation can still depend on the shape
(Sec.~\ref{sec:commutation}).

\begin{figure}[htbp]
    \centering
    \includegraphics[width=0.5\linewidth]{\detokenize{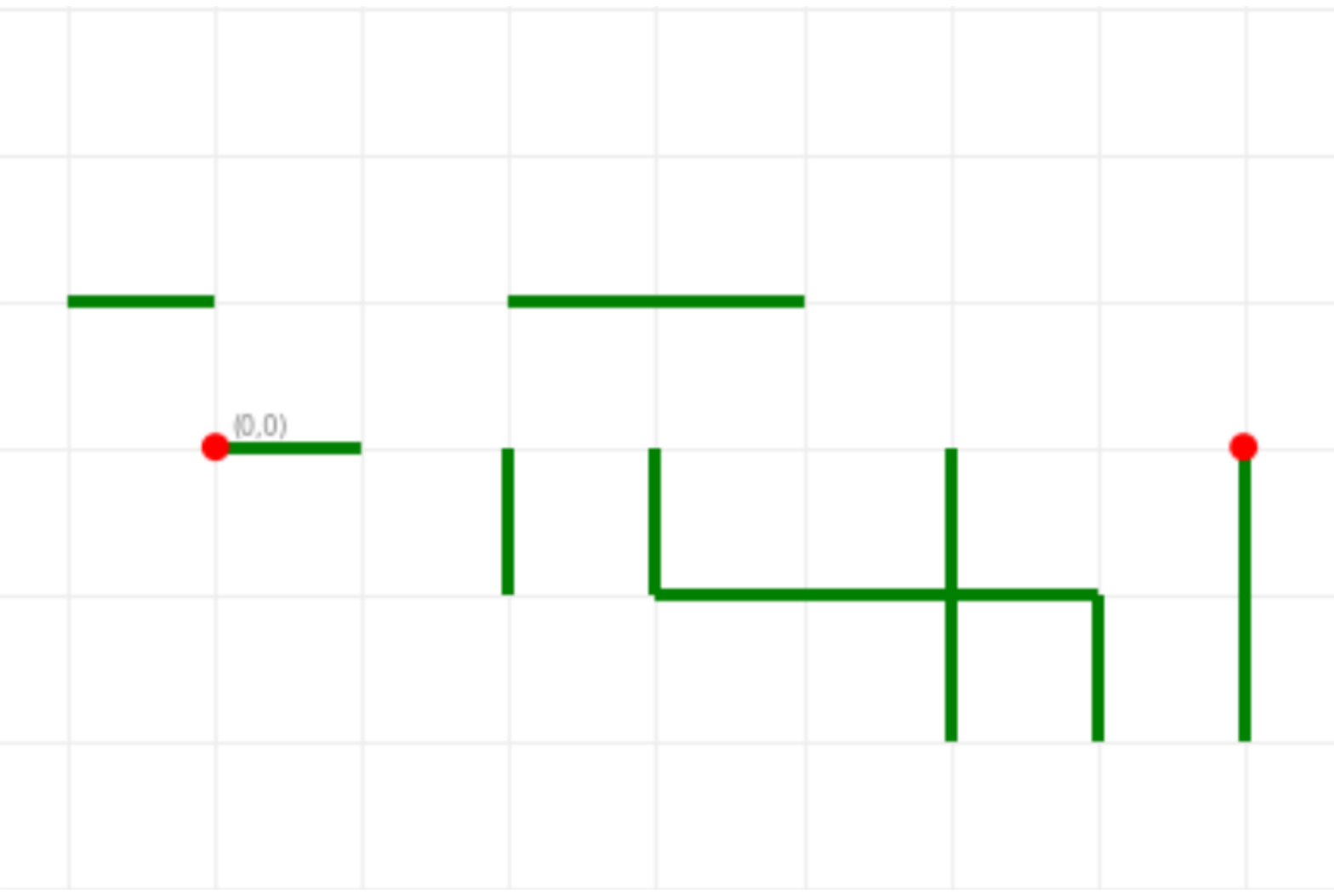}}
    \caption{A basis ribbon of $\BBc{1,1,3,1}$ creating the excitation
    $1+x^{7}$. Green edges represent $Z$ operators, and red points mark
    vertex excitations.}
    \label{fig:BB1131_ribbon_1px7}
\end{figure}

One possible basis ribbon of $\BBc{1,1,3,1}$ creating an excitation of
$1+x^7$ is shown in Fig.~\ref{fig:BB1131_ribbon_1px7}. Using these basis
ribbons, we can propagate anyons via discrete ``jumps''.

\subsection{Closed Loops, Interiors, and Commutation Relations}\label{sec:commutation}

If an anyon moves along a path and fuses with itself, it traces out a closed
loop. To determine whether an arbitrary localized excitation commutes or
anti-commutes with this loop, it is useful to define the interior of the
loop as a set of stabilizers rather than a region of the plane, since
commutation then reduces to a count: the parity of the excitation's overlap
with this interior, read componentwise (Lemma~\ref{lem:closed-loop-interior}).

\begin{definition}
    The \textbf{interior} of a closed loop $p_{\mathrm{c}}$ is the stabilizer
    configuration $a \in G$ such that $\bdry a = p_{\mathrm{c}}$.
\end{definition}

On the infinite plane, and for $0<Q<\infty$, the interior exists and is
unique: existence is exactness of the complex at $P$, equivalent for a
topologically ordered translation-invariant Pauli code to the absence of
locally generated logical operators~\cite{haahCommuting2013}, and uniqueness is
injectivity of $\bdry$, since $\bdry(p,v)=0$ forces
$g^{\star}p=f^{\star}p=fv=gv=0$ over an integral domain.  The compact form,
with its support bound, is Lemma~\ref{lem:compact-filling}.

\begin{figure*}[t]
    \centering
    \includegraphics[width=\textwidth]{\detokenize{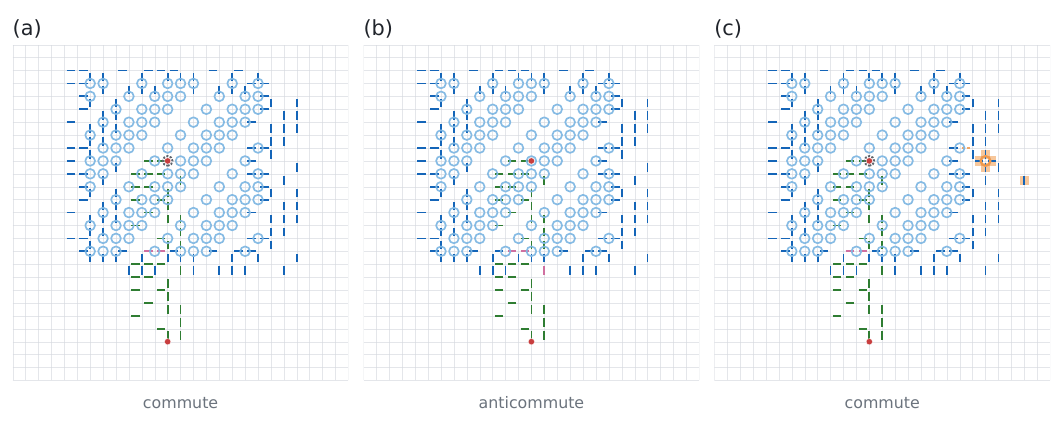}}
    \caption{A closed loop, its interior, and a gauge deformation in
    $\BBc{1,1,3,1}$. Blue edges are $X$, green edges are $Z$, and pink edges
    are both; red points denote vertex excitations; light-blue circles mark
    the interior $a$ of the loop. In every panel the lattice squares that
    meet the loop's support surround the endpoint $e_1$ of the green string,
    and not $e_2$. (a) Yet the vertex check at $e_1$ (grey dashed circle) is
    not in $a$, and loop and string commute. (b) The same string translated by one lattice step puts $e_1$ on
    a check of $a$, and they anticommute. (c) The loop of (a) multiplied by
    one vertex check that shares a qubit with it: orange marks the six qubits
    of that check, which are not all adjacent (dashed where one cancels an
    edge of the loop), and the check added to $a$. This gauge
    transformation leaves the excitations and the commutator unchanged; an
    excitation sitting on the added check would have flipped the commutator,
    which reflects the ribbon choice, not a change of the pairing
    $\mathcal B$. All supports are exact, computed on a $42\times42$ torus.}
    \label{fig:closed-loop-BB1131}
\end{figure*}

An example is shown in Fig.~\ref{fig:closed-loop-BB1131}.

We use the symplectic form to determine whether two operators commute:
\begin{definition}
    The \textbf{commutation value} between two elements $a, b \in P$ is given by:
    \[ \langle a, b \rangle = \tr( a^\dagger \lambda_P b ), \]
    where
    $\lambda_P = \begin{pmatrix} 0 & I_2 \\ -I_2 & 0 \end{pmatrix}$
    is the symplectic form on the four-component BB Pauli module $P=R^4$.
\end{definition}
For $a, b \in P$, the value is zero if and only if the corresponding Pauli
operators commute~\cite{haahCommuting2013}.

The central fact is that pairing a closed loop against any Pauli
reads only the syndrome on the checks of its interior.

\begin{lemma}[Closed loops see only the exact syndrome]
\label{lem:closed-loop-interior}
Let $Z=\bdry A$ be a compact closed pristine Pauli loop with finite pristine
filling $A$, let $p$ be any finite Pauli, and write
$S=\left(\begin{smallmatrix}0&1\\1&0\end{smallmatrix}\right)$ for the exchange
of the two syndrome components.  Then
\begin{equation}\label{eq:adjoint-swap}
 \bdry^{\dagger}\lambda_P=S\,\cobdry ,
\end{equation}
and
\begin{equation}
 \left\langle Z,p\right\rangle
 =\left\langle\bdry A,p\right\rangle
 =\tr\!\left(A^{\dagger}S\,\cobdry p\right).
\label{eq:closed-loop-syndrome-evaluation}
\end{equation}
The pairing sees only the exact pristine syndrome $\cobdry p$, never the thick
support of the ribbons used to build $p$, and the identity holds also when $A$
contains checks later removed from the completed Hamiltonian, being an
identity in the pristine Pauli complex.  In particular, an open string whose
syndrome is two elementary excitations $e_1+e_2$ anticommutes with $Z$ if and
only if exactly one of them lies in $A$, an electric $e_k$ being tested against
the second component of $A$ and a magnetic one against the first.
\end{lemma}
\begin{proof}
Eq.~\eqref{eq:adjoint-swap} is read off the matrices displayed above, the
exchange arising because $E$ is ordered electric--magnetic while $G^{\ast}$ is
ordered dual to the two columns of $\bdry$; $S$ thus pairs each component of the
filling against the syndrome its own check type detects.  Bilinearity of the
pairing gives Eq.~\eqref{eq:closed-loop-syndrome-evaluation}, and
$\tr(A^{\dagger}Se_k)=1$ exactly when $e_k$ lies in $A$.
\end{proof}

\subsection{The Braiding Pairing}\label{sec:braidingstats}

When anyon $a$ is braided around anyon $b$ in a full closed loop, the
wavefunction acquires a phase $\pm1$.  Following standard practice for Pauli
stabilizer models, we define this mutual braiding phase as a commutator of
ribbon operators, as for the toric-code mutual
statistics~\cite{kitaevFaulttolerant2003} and the algebraic braiding pairing
of Ref.~\cite{rubaHomological2024}, whose Proposition~49 identifies that
pairing with the commutation form of the extended operators; monodromy and
mutual statistics always mean this commutator below, and topological spin is
defined through the standard CSS formula of Lemma~\ref{lem:label-duality}, with
pure electric and pure magnetic charges bosonic.  A planar loop does not serve
directly: multiplication by local stabilizers deforms a closed loop and moves
checks into or out of its interior (Fig.~\ref{fig:closed-loop-BB1131}(c)),
which changes its commutator with a string ending there.  We use instead the
two transverse period ribbons of a torus, where every choice is explicit.

Let $z$ and $w$ be the two period translations $x^{l_0}$ and $y^{m_0}$; their
order is fixed with the wall orientation in Sec.~\ref{sec:junctions}, and
nothing in this section depends on it.  Since
$1+z,1+w\in\langle f^\star,g^\star\rangle$ (Lemma~\ref{lem:common-periods}),
there are finite Laurent solutions $(h,v)$ and $(c,d)$ of
\begin{equation}\label{eq:parallel-transverse-ribbons}
 f^\star h+g^\star v=1+z,
 \qquad
 f^\star c+g^\star d=1+w,
\end{equation}
the electric \emph{period ribbons}: $Z[(h,v)]$ and $Z[(c,d)]$ transport an
electric charge through one period in the two directions.  Their
\emph{comparison polynomial} is
\begin{equation}\label{eq:comparison-determinant}
 \Delta=cv+dh.
\end{equation}
For a finite Laurent polynomial $F$ write
\begin{equation}\label{eq:double-trace}
    \operatorname{Tr}_{z,w}(F)=\sum_{a,b\in\mathbb Z}\tr(z^aw^bF),
\end{equation}
a finite sum: the coefficient of the identity monomial of the image of $F$ in
$R/(1+z,1+w)$, the ring of the $l_0\times m_0$ torus.

\begin{definition}\label{def:braiding-phase}
For a code with $0<Q<\infty$, the \emph{mutual braiding phase}
$\mathcal{B}: V_m\times V_e\to\{+1,-1\}$ and its binary exponent $\phi$ are
\begin{equation}\label{eq:direct-braiding-form}
 \mathcal B(\xi,\eta)=(-1)^{\phi(\xi,\eta)},
 \qquad
 \phi(\xi,\eta)
 =\operatorname{Tr}_{z,w}\!\left(\widehat\xi^{\,\star}\widehat\eta\,\Delta\right)
 \in\mathbb F_2,
\end{equation}
for any finite representatives $\widehat\xi,\widehat\eta\in R$ of the two
labels.  The first argument is $m$-type and the second $e$-type; every
commutator computation below is carried out on $\phi$.
\end{definition}

This is a ribbon commutator.  The CSS dual $(v^\star,h^\star)$ of the first
period ribbon has magnetic syndrome
$gv^\star+fh^\star=(f^\star h+g^\star v)^\star=1+z^{-1}$, so on the
$l_0\times m_0$ torus $X[\widehat\xi\,z(v^\star,h^\star)]$ and
$Z[\widehat\eta\,(c,d)]$ are closed ribbons, carrying $m_\xi$ and $e_\eta$ once
around the two different cycles.  With
$\langle Z[p],X[r]\rangle=\tr(p_1r_1^\star+p_2r_2^\star)$ their commutator is
$\operatorname{Tr}_{z,w}(\widehat\xi^{\,\star}\widehat\eta\,z^{-1}\Delta)
=\phi(\xi,\eta)$, as $z=1$ on that torus.  The same value results with the two
cycles exchanged, and for the period ribbons concatenated around the cycles of
any commensurate torus (the computation is in the proof of
Lemma~\ref{lem:routed-logical-basis}).  The right-hand sides of
Eq.~\eqref{eq:parallel-transverse-ribbons} fix the relative anchoring of the
two ribbons: translating one closed ribbon against the other by a monomial $t$
replaces $\phi(\xi,\eta)$ by $\phi(\xi,t^{\pm1}\eta)$, a different pairing
unless $t$ acts trivially on $V_e$, so the number of transverse crossings does
not by itself fix the pairing.

\begin{lemma}[Well-definedness, bilinearity and translation invariance]
\label{lemma:bilinear}
The exponent $\phi$ of Eq.~\eqref{eq:direct-braiding-form} depends neither on
the representatives of the labels nor on the choice of period ribbons in
Eq.~\eqref{eq:parallel-transverse-ribbons}.  For $\xi,\xi'\in V_m$,
$\eta,\eta'\in V_e$ and any Laurent monomial $t$,
\begin{align}
    \B{\xi}{\eta+\eta'} &= \B{\xi}{\eta}\B{\xi}{\eta'},
    &\B{\xi+\xi'}{\eta} &= \B{\xi}{\eta}\B{\xi'}{\eta},
    &\B{t\xi}{t\eta} &= \B{\xi}{\eta},
\end{align}
and, with $\xi^\star\eta$ the product formed in $R$ and then reduced in $V_e$,
\begin{equation}\label{eq:phi-reduction}
 \B{\xi}{\eta}=\B{1}{\xi^\star\eta}.
\end{equation}
That is, $\mathcal B$ is a bicharacter invariant under simultaneous
translation of its two arguments, and the whole table of mutual statistics is
the one-argument functional $\B{1}{\cdot}$.
\end{lemma}
\begin{proof}
As $Q<\infty$, the polynomials $f^\star$ and $g^\star$ are coprime in the
unique-factorization domain $R$, so two solutions of a period equation differ
by a multiple of the syzygy $(g^\star,f^\star)$.  Replacing $(h,v)$ by
$(h+g^\star s,v+f^\star s)$ changes $\Delta$ by $s(f^\star c+g^\star d)=s(1+w)$,
and likewise a change of $(c,d)$ changes it by a multiple of $1+z$; neither
changes $\operatorname{Tr}_{z,w}$, which is invariant under translation by $z$
and by $w$.  Moreover
\begin{equation}\label{eq:delta-annihilates}
 f^\star\Delta=(1+w)v+(1+z)d,
 \qquad
 g^\star\Delta=(1+w)h+(1+z)c,
\end{equation}
so $\operatorname{Tr}_{z,w}(F\Delta)=0$ for every
$F\in\langle f^\star,g^\star\rangle$; since the antipode carries
$\langle f,g\rangle$ onto $\langle f^\star,g^\star\rangle$, $\phi$ is unchanged
when $\widehat\xi$ moves by an element of $\langle f,g\rangle$ or $\widehat\eta$
by one of $\langle f^\star,g^\star\rangle$.  The trace is
$\mathbb F_2$-linear, $t^\star t=1$ gives $(t\xi)^\star(t\eta)=\xi^\star\eta$,
the formula involves the labels only through $\xi^\star\eta$, and
$a\mapsto(-1)^a$ turns addition into multiplication of signs.
\end{proof}

For $\BBc{1,1,3,1}$,
$\B{x}{x}=\B{1}{x^{-1}x}=\B{1}{1}=-1$,
using the $(1,1)$-channel diagonal entry of
Table~\ref{table:measurement1131}.

\subsection{Non-Degeneracy of the Braiding Pairing}\label{sec:kernel}

The pairing is computed directly from the defining polynomials $(f,g)$: solve
the two period equations~\eqref{eq:parallel-transverse-ribbons}, form $\Delta$,
and take coefficient traces.  For standard-monomial bases
$\{\mu_i\}_{i=1}^Q$ of $V_m$ and $\{\nu_j\}_{j=1}^Q$ of $V_e$ the binary Gram
matrix is
\begin{equation}
G_{ij}
=\phi(\mu_i,\nu_j)
=\operatorname{Tr}_{z,w}\!\left(\mu_i^\star\nu_j\Delta\right),
\label{eq:direct-gram-matrix}
\end{equation}
and \(\B{\xi}{\eta}=(-1)^{u^T Gv}\) for $\xi=\sum_i u_i\mu_i$ and
$\eta=\sum_jv_j\nu_j$.  That the pairing is perfect
(Lemma~\ref{lem:braiding-nondegenerate}) is the workhorse invoked throughout
Secs.~\ref{sec:scaffold}--\ref{sec:transmission}.

\paragraph{Example: $\mathrm{BB}(1,1,3,1)$.}  For
\(f=1+x+x^{-1}y\) and \(g=1+y+x^3y^{-1}\) the quotient relations
\(y=x+x^{2}\) and \(1+x^{2}+x^{3}=0\) of \(V_m\) give the
standard-monomial basis \((1,x,x^{2})\) with \(Q=3\) and \(x^{7}=y^{7}=1\),
so $l_0=m_0=7$.  Moreover $xy=x^{2}+x^{3}=1$, so the affine
monomial of Eq.~\eqref{eq:inversion} is trivial on labels,
\begin{equation}\label{eq:kappa-1131}
    [\kappa]=[x^{-1}y^{-1}]=1
\end{equation}
in both dual quotient spaces: the affine label formulas reduce to pure-antipode
ones for this example, although microscopic block anchors remain affine.

\begin{lemma}[Non-degeneracy of the braiding pairing]
\label{lem:braiding-nondegenerate}
The pairing $\phi:V_m\times V_e\to\mathbb F_2$ is perfect.  Namely, for
$r\in V_m$,
\(\B{r}{X}=1\quad\text{for every }X\in V_e\)
implies $r=0$, and the analogous assertion holds after exchanging
$V_m$ and $V_e$; equivalently, the Gram matrix of
Eq.~\eqref{eq:direct-gram-matrix} is invertible over $\mathbb F_2$.
Consequently anyons are separated by their statistics: if $A,A'\in V_m$
satisfy $\B{A}{X}=\B{A'}{X}$ for every $X\in V_e$, then $A=A'$, and
likewise with the two tones exchanged.
\end{lemma}

\begin{proof}
Set \(I_e=(f^\star,g^\star)\), so that \(V_e=R/I_e\).
The Laurent ring $R=\mathbb F_2[x^{\pm1},y^{\pm1}]$ is a localization of
$\mathbb F_2[x,y]$, hence a Noetherian regular domain of dimension two, and in
particular a unique factorization domain.  Since $V_e$ is nonzero and
finite-dimensional over $\mathbb F_2$, every prime containing $I_e$ is maximal,
and maximal ideals of $R$ have height two; so $\operatorname{ht}I_e=2$.  In
particular $f^\star$ is nonzero, hence a non-zero-divisor in the domain $R$;
and were $g^\star$ a zero-divisor modulo $(f^\star)$, unique factorization
would give $f^\star$ and $g^\star$ a common irreducible factor $p$, whence
$I_e\subseteq(p)$ with $\operatorname{ht}(p)=1$, contradicting
$\operatorname{ht}I_e=2$.  So $f^\star,g^\star$ is a regular sequence; so is
$1+z,1+w$, because $z$ and $w$ translate in independent directions.

For nested regular sequences a determinant controls the comparison of the
two quotients.  If $S$ is a commutative ring, both $a_1,a_2$ and $t_1,t_2$ are
regular sequences and $t=Ca$, then, with $U=S/(t_1,t_2)$ and
$D=S/(a_1,a_2)$, multiplication by $\det C$ is an isomorphism
\begin{equation}\label{eq:ci-determinant-iso}
D\xrightarrow{\ \sim\ }
\operatorname{Ann}_U(a_1,a_2).
\end{equation}
This is the determinant--colon formula of
Wiebe~\cite{wiebeHomologische1969}; see~\cite[Cor.~2.3.10(a), with the
published correction]{brunsCohenMacaulay1998}, and
\cite[Prop.~1.2]{simonStrookerWiebe2007} for the fact that no Noetherian or
local hypothesis is required.

The two period equations~\eqref{eq:parallel-transverse-ribbons} are the single
matrix identity
\begin{equation}\label{eq:ribbon-change-of-basis}
    \begin{pmatrix}1+z\\1+w\end{pmatrix}
    =C\begin{pmatrix}f^\star\\g^\star\end{pmatrix},
    \qquad
    C=\begin{pmatrix}h&v\\c&d\end{pmatrix},
    \qquad
    \det C=hd+vc=\Delta ,
\end{equation}
so $\Delta$ is the determinant expressing the period pair through the defining
pair, and Eq.~\eqref{eq:ci-determinant-iso}, applied with $t=(1+z,1+w)$ and
$a=(f^\star,g^\star)$, makes multiplication by $\Delta$ an isomorphism
\begin{equation}\label{eq:delta-iso}
    V_e=R/(f^\star,g^\star)
    \xrightarrow{\ \sim\ }
    \operatorname{Ann}_{R/(1+z,1+w)}(f^\star,g^\star).
\end{equation}
The quotient $R/(1+z,1+w)$ is a finite \emph{Frobenius} algebra: its
coefficient trace is $\operatorname{Tr}_{z,w}$ and pairs the quotient perfectly
with itself.  Let $\eta\in V_e$ satisfy $\phi(\xi,\eta)=0$ for every
$\xi\in V_m$.  As $\widehat\xi^{\,\star}$ ranges over all of $R$,
$\operatorname{Tr}_{z,w}(F\,\widehat\eta\Delta)=0$ for every $F\in R$, so
$\widehat\eta\Delta=0$ in $R/(1+z,1+w)$ and $\eta=0$ by
Eq.~\eqref{eq:delta-iso}.  The antipode is an involutive
$\mathbb F_2$-algebra automorphism carrying $(f,g)$ to $(f^\star,g^\star)$, so
$\dim V_m=\dim V_e=Q$; an injection $V_e\to V_m^{*}$ between spaces of equal
finite dimension is bijective, the pairing is perfect, and $G$ has full rank.

The Frobenius property is used again below in the following form: the
annihilator of an ideal $I$ is its trace-orthogonal complement
(\(\operatorname{Tr}(Ia\,\cdot)=0\) exactly when $Ia=0$, the form being
non-degenerate), and so has the dimension of the quotient by $I$.

For the separation statement, bilinearity (Lemma~\ref{lemma:bilinear}) gives
$\B{A}{X}\B{A'}{X}=\B{A+A'}{X}$, so the hypothesis says that
$\B{A+A'}{X}=1$ for every $X\in V_e$; non-degeneracy then forces $A+A'=0$,
that is $A=A'$ over $\mathbb F_2$.
\end{proof}

\section{The Defect-Line Scaffold}\label{sec:scaffold}

Localized Abelian anyons at fixed total charge have one-dimensional fusion
spaces, so a protected local degeneracy needs new objects.  We engineer a
\emph{defect line}: a wall along which the code is modified so that an anyon
crossing it becomes its affine electric--magnetic dual.

\begin{figure*}[!t]
    \centering
    \begin{minipage}[t]{0.31\textwidth}
        \centering
        \subfloat[]{\includegraphics[height=0.18\textheight]{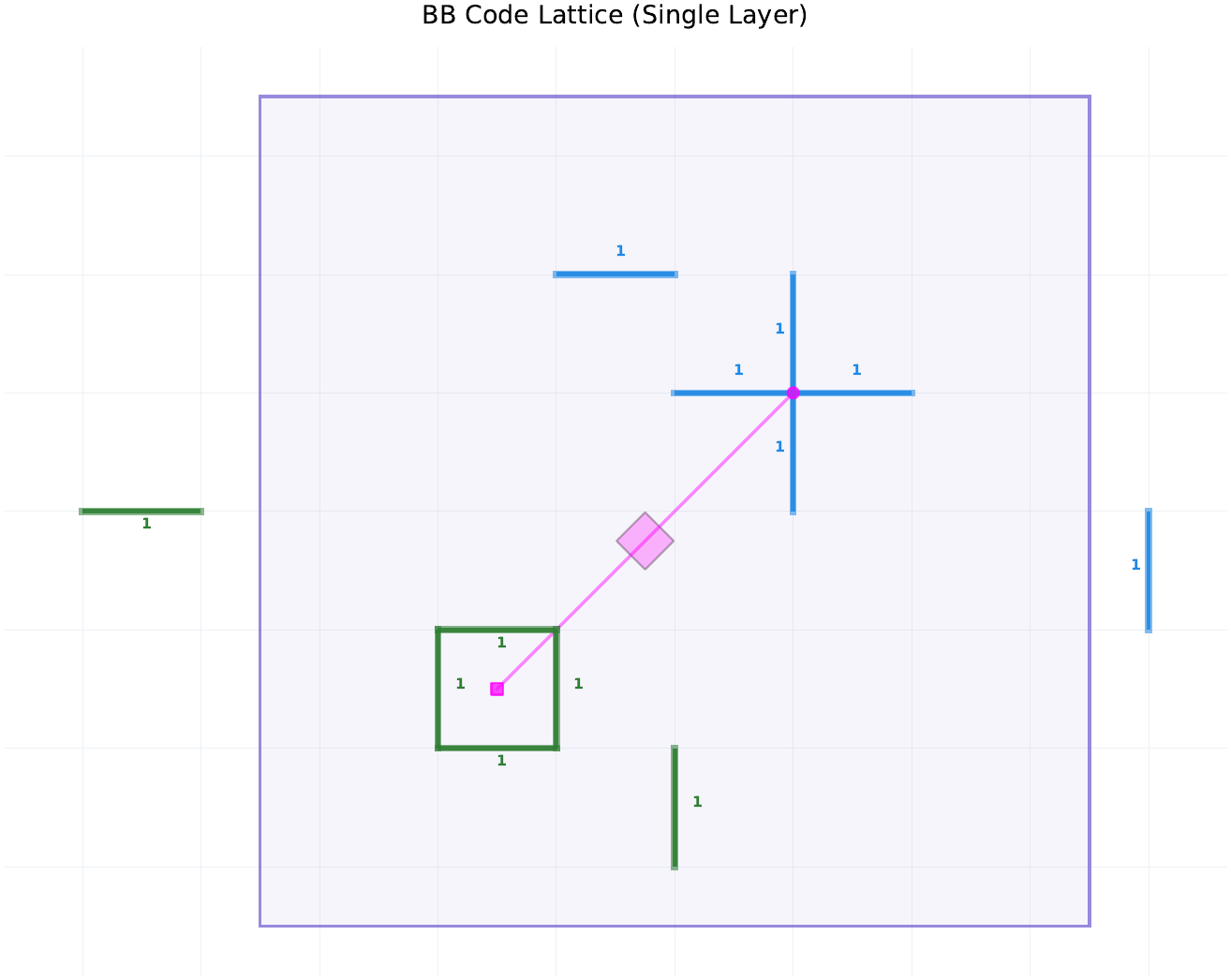}}
        \par\vspace{3pt}
        \small Composite matching inside one block.
    \end{minipage}
    \hfill
    \begin{minipage}[t]{0.31\textwidth}
        \centering
        \subfloat[]{\includegraphics[height=0.18\textheight]{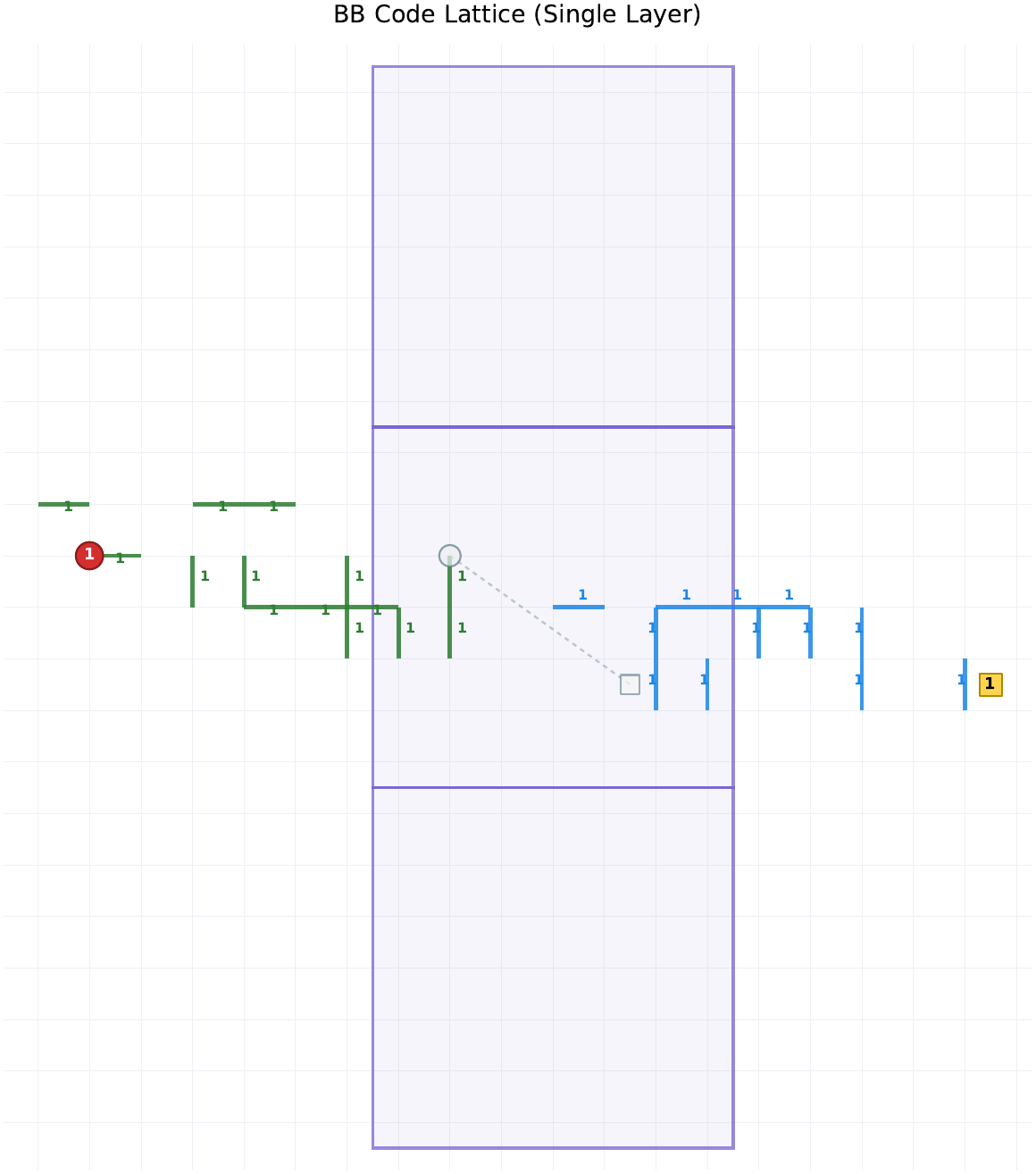}}
        \par\vspace{3pt}
        \small Target affine-dual transmission.
    \end{minipage}
    \hfill
    \begin{minipage}[t]{0.31\textwidth}
        \centering
        \subfloat[]{\includegraphics[height=0.18\textheight]{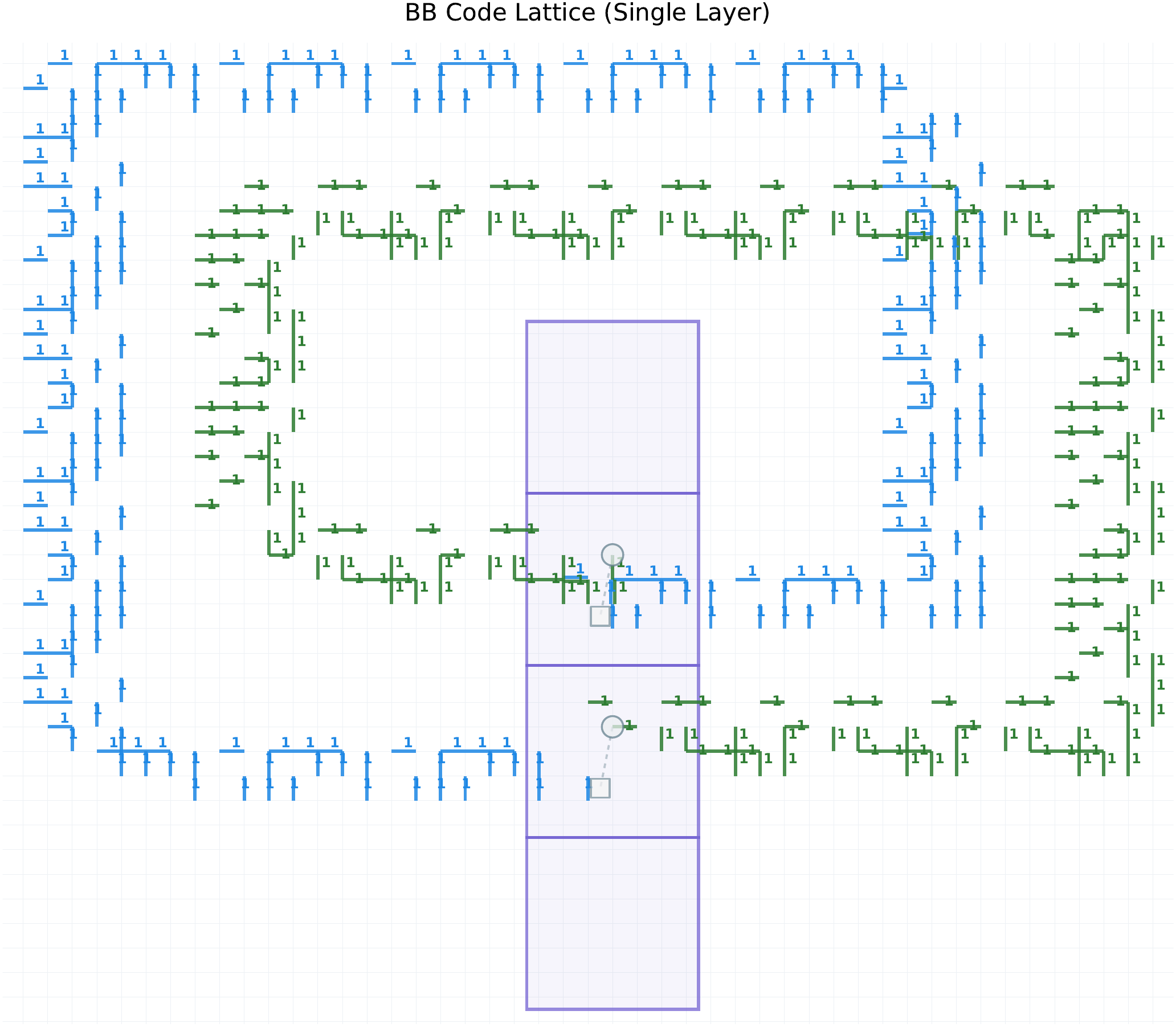}}
        \par\vspace{3pt}
        \small A loop around an endpoint.
    \end{minipage}
    \caption{Overview of the block wall and its target action for
    $\BBc{1,1,3,1}$.  The microscopic offsets use the corner-origin affine
    reflection in Eq.~\eqref{eq:inversion}; for this $7\times7$ example its
    center is the lattice site $(3,3)$.  Colours as in
    Fig.~\ref{fig:closed-loop-BB1131}; violet squares are defect blocks.
    (a) The composite layer matches the $X$ check at $u$
    with the $Z$ check at $\iota(u)$ (magenta link; the diamond marks the
    reflection center).  (b) The target label action is the
    affine duality $e_r\mapsto m_{r^\iota}$ and
    $m_s\mapsto e_{s^\iota}$.  Its microscopic realization requires crossing
    Pauli strings with these remote endpoint syndromes and no residual wall
    syndrome: red circles mark electric (vertex) and yellow squares magnetic
    (plaquette) syndromes, the number being the syndrome bit, and hollow markers
    joined by a grey dashed line an $\iota$-matched pair the composite checks
    absorb.  (c) A loop winding around a termination of the wall; it closes
    only as a duality-invariant composite (Sec.~\ref{sec:absorbable}).  Microscopic interblock and patch
    representatives are shown below in Figs.~\ref{fig:Tmicro} and
    \ref{fig:completion-triptych}.}
    \label{Fig:DefectConstruction}
\end{figure*}

Section~\ref{sec:blockalgebra} fixes the block algebra and the target label
action.  The construction then proceeds in three operator layers at increasing
distance from the \emph{pristine} code --- the BB code as given, before any
check is removed or added (Figs.~\ref{Fig:DefectConstruction}(a), \ref{fig:Tmicro}
and~\ref{fig:completion-triptych}).
Composite checks $C$ (Sec.~\ref{sec:composites}) impose the affine
electric--magnetic matching inside each block; each is a product of two
pristine checks and never leaves the pristine stabilizer group.  Merging a
check pair costs one listed measurement per site --- the wall can no longer tell
which member of the pair was violated --- and what a block therefore cannot
resolve accumulates as its \emph{ledger} of removed-check violations.  Interblock
operators $T$ (Sec.~\ref{sec:junctions}) identify the relative wall ledgers of
adjacent blocks; each is a pair of period ribbons, has left that group, and
sits with its endpoints on removed checks that the composites absorb pairwise.
The construction's one free choice, which period ribbon builds the interblock operators, is
fixed there.  Patch operators $M$ (Sec.~\ref{sec:completion}) supply the
local filling relations for the remaining contractible wall processes.

All three layers commute with the composites for one reason: whatever
syndrome an operator leaves on the removed checks comes in $\iota$-matched
pairs, which a composite reads as $1+1=0$.  A composite itself leaves no
pristine syndrome, being a product of two pristine checks; an interblock operator's four
endpoints form two such pairs; and the patch operator has electric and
magnetic removed-check syndromes related by $\iota$, which its dressing by
pristine checks leaves unchanged.

The operator results of this section concern the straight-wall
\emph{interior} and are proved under their stated lattice hypotheses; natural termination is stated here
for later use.  The crossing law is established separately, in
Sec.~\ref{sec:transmission}, and the termination of the wall and its twist
endpoints, where the non-Abelian physics lives, need separate operator-level
input and are treated in Sec.~\ref{sec:twists}.

\paragraph*{Relation to prior defect constructions.}
General operator-algebra algorithms for gapped boundaries and defects in
$(2{+}1)$D Pauli stabilizer codes~\cite{liangOperator2024}, and the
quantum-double boundaries and condensations behind
them~\cite{beigiQuantum2011}, treat the BB examples at the level of boundaries
and long-period anyon strings, without defect Hamiltonians or twist-endpoint
data.  We give a closed-form realization of one wall --- the affine
electric--magnetic duality wall --- for \emph{every} BB code with
$0<Q<\infty$, together with its twist endpoints.  For the toric code with a
single-edge longitudinal ribbon, its straight wall reduces to the defect line of
Ref.~\cite{kitaevModels2012} once each locked pair of edges is read as one qubit (Sec.~\ref{sec:toric-limit}).

\subsection{Block Algebra and the Affine Duality}\label{sec:blockalgebra}

We build the defect line from a block of size $l_0\times m_0$, where
$l_0,m_0$ are the anyon periods defined in Sec.~\ref{sec:anyontypes}: the
block is the smallest axis-aligned rectangular cell, in the fixed
coordinates, on which every anyon label repeats,
so it is the natural unit from which to tile a wall. Let
\begin{equation}\label{eq:block-algebra}
    \mathsf A
    =
    \frac{\mathbb F_2[x,y]}
    {(x^{l_0}+1,\;y^{m_0}+1)},
    \qquad
    \dim_{\mathbb F_2}\mathsf A=l_0m_0,
\end{equation}
and define \(\mathcal I_m=(f,g)\mathsf A\) and
\(\mathcal I_e=(f^\star,g^\star)\mathsf A\).
The antipode is well defined on $\mathsf A$ because
$x^{-1}=x^{l_0-1}$ and $y^{-1}=y^{m_0-1}$ in the quotient.  It exchanges
$\mathcal I_m$ and $\mathcal I_e$.

Let $J_{\rm per}=(x^{l_0}+1,y^{m_0}+1)\subset R$.  By
Lemma~\ref{lem:common-periods},
\(J_{\rm per}\subseteq\langle f,g\rangle\cap\langle f^\star,g^\star\rangle\).
Because $\mathsf A\cong R/J_{\rm per}$, the quotient map induces
\(\mathsf A/\mathcal I_e\cong R/\langle f^\star,g^\star\rangle=V_e\),
or equivalently the exact sequence
\begin{equation}\label{eq:block-SES}
    0\longrightarrow \mathcal I_e
    \longrightarrow \mathsf A
    \xrightarrow{\;\pi_e\;}V_e
    \longrightarrow0
\end{equation}
and similarly with $e$ and $m$ exchanged.  Therefore
\begin{equation}\label{eq:ideal-dimension}
    \dim_{\mathbb F_2}\mathcal I_e
    =\dim_{\mathbb F_2}\mathcal I_m
    =l_0m_0-Q.
\end{equation}
Physically, Eq.~\eqref{eq:block-SES} splits the $l_0m_0$ bits of one block
into $Q$ topologically meaningful directions, the anyon label in $V_e$, and
$l_0m_0-Q$ label-trivial ones, the ideal $\mathcal I_e$.

\paragraph*{The affine block involution.}
Choose the lower-left corner of the first block as the global
Laurent-coordinate origin, and obtain every later block origin by translation
through one whole wall period.  Relative to these origins write
$u=(i,j)\in\Lambda_0=\mathbb Z_{l_0}\times\mathbb Z_{m_0}$ using the standard
representatives $0\leq i<l_0$ and $0\leq j<m_0$, and define
\begin{equation}\label{eq:inversion}
    \iota(i,j)=(l_0-1-i,m_0-1-j),
    \qquad
    \kappa=x^{l_0-1}y^{m_0-1}.
\end{equation}
Geometrically, $\iota$ is the $180^\circ$ rotation about the center of the
block, a permutation and an involution for every parity of $l_0,m_0$.  The
standard Laurent representatives obey the exact identity
\begin{equation}\label{eq:affine-anchor}
    \widehat{\iota(u)}=\kappa\widehat u^\star
    \quad\text{in }R,
    \qquad
    \kappa\kappa^\star=1.
\end{equation}
Thus no modular wrap or parity-dependent representative enters the
microscopic commutator calculation.  We write
\begin{equation}\label{eq:iota-label}
    r^\iota:=\kappa r^\star
\end{equation}
for the induced additive involution on anyon labels.  It exchanges $V_e$ and
$V_m$ and satisfies $(r^\iota)^\iota=r$.
For a block polynomial $c=\sum_uc_u u$, we use the same symbol
coefficientwise, so that
\begin{equation}\label{eq:affine-anchor-polynomial}
    \widehat{\iota(c)}:=\sum_uc_u\widehat{\iota(u)}
    =\kappa\widehat c^\star
    \quad\text{in }R.
\end{equation}
Composed with $\iota$, the braiding pairing becomes a pairing of $V_m$ with
itself,
\begin{equation}\label{eq:code-channel-pairing}
 \lambda(a,b)=\B{a}{b^\iota}=\B{1}{\kappa a^\star b^\star},
 \qquad a,b\in V_m ,
\end{equation}
the second form by Eq.~\eqref{eq:phi-reduction}.

\begin{lemma}[The wall pairing]\label{lem:pca-perfect}
The map $\lambda:V_m\times V_m\to\{\pm1\}$ is a symmetric non-degenerate
bicharacter; in particular
\begin{equation}
    \phi(b,a^\iota)=\phi(a,b^\iota)
    \qquad\text{for all }a,b\in V_m .
\label{eq:phi-affine-exchange}
\end{equation}
\end{lemma}
\begin{proof}
The last expression in Eq.~\eqref{eq:code-channel-pairing} is symmetric in $a$
and $b$, the quotient ring being commutative, and bilinear by
Lemma~\ref{lemma:bilinear}.  Since $b\mapsto b^\iota$ is a bijection
$V_m\to V_e$, non-degeneracy is Lemma~\ref{lem:braiding-nondegenerate}.
\end{proof}

\paragraph*{The target label action.}
The corresponding label map is
\begin{equation}\label{eq:defect-duality}
    D_{\rm lab}(e_r)=m_{r^\iota},
    \qquad
    D_{\rm lab}(m_s)=e_{s^\iota}.
\end{equation}

\begin{lemma}[Affine electric--magnetic duality]\label{lem:label-duality}
On a general composite anyon the map of Eq.~\eqref{eq:defect-duality} acts by
\begin{equation}\label{eq:defect-duality-composite}
 D_{\rm lab}(e_rm_s)=e_{s^\iota}m_{r^\iota},
 \qquad r\in V_e,\ s\in V_m,
\end{equation}
and it is an involution preserving Abelian fusion and mutual braiding.  On
those composites it also preserves topological spin.
\end{lemma}
\begin{proof}
Let \(a=e_r m_s\), with \(r\in V_e\) and \(s\in V_m\).
Equations~\eqref{eq:iota-label} and~\eqref{eq:defect-duality} give
\[
D_{\rm lab}(a)=e_{s^\iota}m_{r^\iota},
\qquad D_{\rm lab}^2(a)=a.
\]
Thus \(D_{\rm lab}\) is an involution and, since \(\iota\) is additive, preserves fusion.
Eq.~\eqref{eq:phi-affine-exchange}, applied with \(a=v\) and \(b=u^\iota\), gives
\[
\mathcal B(u^\iota,v^\iota)=\mathcal B(v,u)
\quad(u\in V_e,\ v\in V_m). \tag{*}
\]
For \(b=e_{r'}m_{s'}\), the standard CSS braiding and spin formulas, with
\(\mathfrak M\) the full monodromy of Eq.~\eqref{eq:full-anyon-monodromy}, now give
\[
\begin{aligned}
\mathfrak M(D_{\rm lab}a,D_{\rm lab}b)
 &=\B{r^\iota}{s'^\iota}\B{r'^\iota}{s^\iota}
  =\B{s'}{r}\B{s}{r'}=\mathfrak M(a,b),\\
\theta(D_{\rm lab}a)
 &=\B{r^\iota}{s^\iota}=\B{s}{r}=\theta(a),
\end{aligned}
\]
where pure electric and pure magnetic anyons have spin \(+1\).
\end{proof}

The map $D_{\rm lab}$ is the \emph{target} of the whole construction
[Fig.~\ref{Fig:DefectConstruction}(b)]: the wall should transmit $e_r$ to
$m_{r^\iota}$ and $m_s$ to $e_{s^\iota}$, in both crossing directions, with no
residual syndrome on measured wall rows, no absorbed charge and no additional
channel.  That the completed scaffold realizes exactly this map,
microscopically and invertibly, is Theorem~\ref{thm:transmission}.

\paragraph*{Other frames.}
For a monomial $\mu$ put
\begin{equation}\label{eq:defect-duality-mu}
 D_\mu(e_rm_s)=e_{\mu s^\star}m_{\mu r^\star},
 \qquad\text{so that}\qquad D_{\rm lab}=D_\kappa .
\end{equation}
The dependence of the proof of Lemma~\ref{lem:label-duality} on the affine
monomial uses only $\kappa\kappa^\star=1$; replacing $\kappa$ by any monomial
$\mu$, the same braiding-reduction and CSS spin calculations show that $D_\mu$
is an involution preserving fusion, braiding and spin.  A lattice translation by a monomial $o$ acts on labels by
multiplication, $e_rm_s\mapsto e_{or}m_{os}$; we write this action as $o$.
Each $D_\mu$ is the label action of a lattice symmetry.  Let $\tau_\mu$ apply
a Hadamard to every qubit, exchange the two qubit blocks and move cell $w$ to
$\mu w^\star$; this is the Hadamard-type fold-transversal operation of
Ref.~\cite{eberhardtLogicalTIT2025} for the ZX-duality formed by the last two
steps (the standard one composed with a translation).  It acts by
$(u_1,v_1\mid u_2,v_2)\mapsto(\mu v_2^\star,\mu u_2^\star\mid\mu v_1^\star,\mu u_1^\star)$
on the Pauli module of Sec.~\ref{sec:chain}, so it exchanges the $X$ check at
$q$ with the $Z$ check at $\mu q^\star$, preserves the stabilizer group, and
maps syndromes by $(s_e,s_m)\mapsto(\mu s_m^\star,\mu s_e^\star)$, which on
labels is $D_\mu$.  The composite $C_{j,u}$ of Eq.~\eqref{eq:composite} is
$S^X_{j,u}$ times its $\tau_{z^{2(j-1)}\kappa}$-image, a half-turn about its
block centre; as the block translation $z$ along the wall
(Sec.~\ref{sec:junctions}) is a period, all of these induce $D_\kappa$.  The
first identity below is $\tau_\mu$ conjugated by a translation, which moves the
centre of the half-turn by $o$.

\begin{lemma}[Translations and dualities]\label{lem:frame-label}
For monomials $o,\mu,\nu$, as maps on anyon labels,
\begin{equation}\label{eq:frame-label}
 o\,D_\mu\,o^{-1}=D_{o^2\mu},
 \qquad
 D_\mu D_\nu=\mu\nu^{-1},
\end{equation}
and $D_\mu=D_\nu$ if and only if $\mu\equiv\nu\pmod{\langle f,g\rangle}$.
\end{lemma}
\begin{proof}
Since $(o^{-1})^\star=o$, $D_\mu(e_{o^{-1}r}m_{o^{-1}s})=e_{o\mu s^\star}m_{o\mu r^\star}$,
and multiplying by $o$ gives the first identity.  Similarly
$D_\mu(e_{\nu s^\star}m_{\nu r^\star})=e_{\mu\nu^\star r}m_{\mu\nu^\star s}$ with
$\nu^\star=\nu^{-1}$.  Hence $D_\mu=D_\nu$ exactly when $\mu\nu^{-1}$ acts trivially
on $V_m$ and on $V_e$.  As $V_m$ is generated by $1$ and $\nu$ is a unit, the
first condition is $\mu\equiv\nu\pmod{\langle f,g\rangle}$.  The second is
$\mu\equiv\nu\pmod{\langle f^\star,g^\star\rangle}$, which the antipode turns
into $\mu^{-1}\equiv\nu^{-1}\pmod{\langle f,g\rangle}$, the same condition after
multiplication by the unit $\mu\nu$.
\end{proof}

\subsection{Composite Operators and the Removed Checks}\label{sec:composites}

\paragraph*{Structure on the lattice.}
Along the wall the code no longer measures the $X$-check at offset $u$ and
the $Z$-check at $\iota(u)$ separately; it measures only their product.
Index the wall by blocks $j=1,\ldots,N$ and offsets $u\in\Lambda_0$, block
$j$ carrying the longitudinal monomial $z^{j-1}$, and
write $S^X_{j,u}$ and $S^Z_{j,u}$ for the two pristine checks replaced at
each wall site.  The composite checks are
\begin{equation}\label{eq:composite}
    C_{j,u}=S^X_{j,u}S^Z_{j,\iota(u)}.
\end{equation}
Each $C_{j,u}$ is a single Hermitian Pauli operator supported on the union of
two pristine checks of the same block, so its locality is bounded by the block
size plus the pristine check range; the matching is illustrated in
Fig.~\ref{Fig:DefectConstruction}(a).

\paragraph*{Physical mechanism and cost.}
An $X$-check violation at $u$ is thereby identified with a $Z$-check
violation at $\iota(u)$: an electric excitation entering the wall becomes
indistinguishable from the image magnetic excitation at the center-reflected
offset, which is the microscopic origin of the affine duality
\eqref{eq:defect-duality}. The price is one listed measurement per wall
site: each of the $l_0m_0N$ sites of an $N$-block wall now contributes one
check where the pristine code contributed two.

\paragraph*{Why the layer is exact.}
The construction never leaves the pristine stabilizer group.

\begin{lemma}[Composite commutativity]\label{lem:composite}
All $C_{j,u}$ commute with one another and with every retained pristine
check.
\end{lemma}
\begin{proof}
Each factor lies in the commuting pristine stabilizer group and the symplectic
pairing is bilinear, so every such pairing is a sum of vanishing terms.
\end{proof}

\subsection{Interblock Operators and the Single Ledger}\label{sec:junctions}

The wall is a stack of identical defect blocks, and adjacent blocks must
agree: a ledger accumulated in block $j$ must transfer to block $j+1$ without
cost.  That is what the interblock operators $T$ enforce, pinning the relative
wall charge of adjacent blocks to zero and collapsing the extensive
bookkeeping space $\mathsf A^N$ to one intensive copy of $\mathsf A$.

\begin{figure*}[t]
    \centering
    \includegraphics[width=0.82\textwidth,trim=0 25 325 60,clip]
    {\detokenize{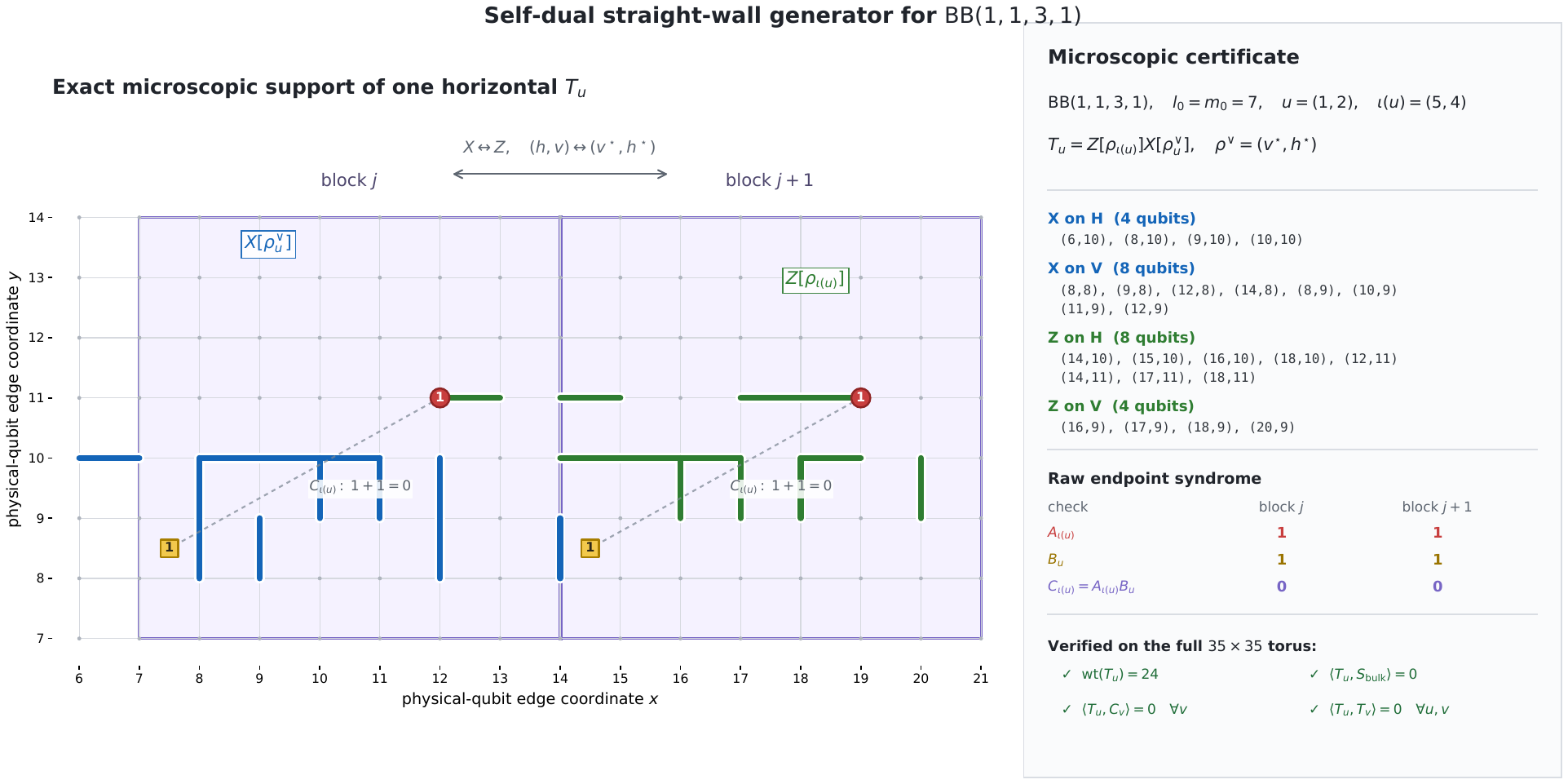}}
    \caption{Exact microscopic support of one horizontal self-dual interblock
    operator for $\BBc{1,1,3,1}$.  The displayed $7\times7$ blocks are
    shown in corner-origin coordinates: the marked offset $u=(1,2)$ is paired
    with $\iota(u)=(5,4)$ by the affine center reflection in
    Eq.~\eqref{eq:inversion}.  Blue and green segments show its $X$ and $Z$
    support.  \emph{Colour convention for every ribbon figure below}: blue is
    the $X$-tone ribbon, carrying magnetic labels, green the $Z$-tone ribbon,
    carrying electric ones; grey is the defect line, and grey dashed segments
    connect the $\iota$-matched endpoint pairs the composite checks absorb.  The two banks carry the period-ribbon data
    $\mathbf r_{\rm per}=(h,v)$ and its CSS dual
    $\mathbf r_{\rm per}^\vee=(v^\star,h^\star)$, anchored by the
    block-periodic convention; the artwork writes $\rho_u^\vee$ and
    $\rho_{\iota(u)}$ for these anchored copies.  Syndrome markers are as in
    Fig.~\ref{Fig:DefectConstruction}(b); vertex syndromes sit at lattice sites
    and plaquette syndromes at plaquette centers, half a spacing off the
    edge coordinates on the axes.  Their paired endpoint syndromes agree, so
    every composite $C$ sees zero syndrome; the self-dual form also makes all
    translated interblock operators mutually commuting by
    Lemma~\ref{lem:selfdual}.}
    \label{fig:Tmicro}
\end{figure*}

\paragraph*{Structure on the lattice.}
Each $T$ is a pair of mutually dual open strings straddling the boundary between blocks $j$ and $j+1$
(Fig.~\ref{fig:Tmicro}). Its $Z$-type half is a period ribbon anchored at
the offset $\iota(u)$ of block $j$; it deposits an electric endpoint
syndrome on the two removed $X$-checks at $(j,\iota(u))$ and
$(j+1,\iota(u))$. Its star/swapped $X$-type half is the CSS-dual ribbon
anchored one block forward at the offset $u$; it deposits the
$\iota$-matched magnetic syndrome on the removed $Z$-checks at $(j,u)$ and
$(j+1,u)$. The composites $C$ see the sum of these two identical bits and
therefore detect no excitation.

Concretely, fix the wall orientation and block origin, and let $z$ and $w$
denote translation by one whole block parallel and transverse to the wall:
\begin{equation}
 (z,w)=
 \begin{cases}
 (x^{l_0},y^{m_0}),&\text{horizontal wall},\\
 (y^{m_0},x^{l_0}),&\text{vertical wall}.
 \end{cases}
\end{equation}
Write $w=t^{p_\perp}$, where $(t,p_\perp)=(y,m_0)$ for a horizontal
wall and $(t,p_\perp)=(x,l_0)$ for a vertical wall; the transverse
fundamental interval occupied by the removed wall block is then
$0\leq t\leq p_\perp-1$.  The period ribbons of
Eq.~\eqref{eq:parallel-transverse-ribbons} are named accordingly:
$\mathbf r_{\rm per}=(h,v)$ is the longitudinal ribbon, $(c,d)$ the transverse
one, and $\Delta=cv+dh$ their comparison polynomial.

\subsubsection{Transverse Normalization}

The interblock rows are built from $(h,v)$, and that ribbon is fixed only up to
a syzygy: two solutions of the longitudinal period equation differ by
$(h,v)\mapsto(h+g^\star t,v+f^\star t)$.  In the pristine code such a change
multiplies each row by a product of checks and is invisible; on the wall those
of them inside the removed block have been merged away, so different solutions
can give different walls, with different comparison polynomials.  We impose
a transverse normalization and fix one finite lift; the normalization is
always available and is the condition on which Theorem~\ref{thm:transmission}
rests.

Write $\overline\Delta$ for the image of $\Delta$ in
$\mathsf B=R/(1+z)$.  It does not depend on the auxiliary transverse
ribbon: two solutions of $f^\star c+g^\star d=1+w$ differ by
$(c,d)\mapsto(c+g^\star s,d+f^\star s)$, which changes $\Delta$ by
$s(1+z)$ and hence leaves $\overline\Delta$ unchanged.  The normalization
below is therefore a statement about $(h,v)$ alone.  In $\mathsf B$,
divide the folded comparison polynomial by $1+w$:
\begin{equation}\label{eq:transverse-normalization-division}
 \overline\Delta=r+(1+w)q,
 \qquad
 \operatorname{supp}_t r\subseteq\{0,1,\ldots,p_\perp-1\}.
\end{equation}
This division always has finite support, and $q$ is unique once the displayed
remainder interval is fixed.  Explicitly, if $k=r_0+np_\perp$ with
$0\leq r_0<p_\perp$, then $t^k+t^{r_0}$ is a finite Laurent multiple of
$1+w$; this covers negative transverse exponents and also the degenerate
case $p_\perp=1$.  When $Q>0$, the remainder is nonzero: $r=0$ would make
$\Delta$ vanish after both period folds, contradicting
Lemma~\ref{lem:braiding-nondegenerate}.

Choose a finite lift $\widetilde q\in R$ of $q$ and replace
\begin{equation}\label{eq:transverse-normalized-ribbon}
 (h,v)\longmapsto
 (h',v')=(h+g^\star\widetilde q,\,v+f^\star\widetilde q).
\end{equation}
The longitudinal period equation is preserved because
\(f^\star h'+g^\star v'=f^\star h+g^\star v
+(f^\star g^\star+g^\star f^\star)\widetilde q=1+z\), while
\(\overline{cv'+dh'}=\overline\Delta+(1+w)q=r\).
We henceforth reuse $(h,v)$ and $\Delta$ for this transverse-normalized ribbon
and comparison polynomial, and fix the resulting wall realization once and for
all; every patch row and every support or buffer constant below is the one
for that ribbon.

The remainder $r$ and the folded transverse-normalized ribbon class are
canonical once the orientation, origin, and transverse fundamental interval
are fixed, but the microscopic wall is not.  Two finite lifts of $q$ differ
by $(1+z)s$ and in general give different interblock and patch rows.
One finite lift is fixed; equivalence of the walls obtained from different
lifts is not established here.  Theorem~\ref{thm:transmission} holds for every choice of lift
regardless.

\subsubsection{The Interblock Operators}

The chosen longitudinal ribbon's CSS dual is
\begin{equation}
    \mathbf r_{\rm per}^\vee=(v^\star,h^\star).
\end{equation}
This is the ordinary CSS ribbon dual, not the block involution $\iota$: the
affine center reflection enters separately, through the anchoring monomials in
Eq.~\eqref{eq:T-definition}.  Its magnetic syndrome is
\begin{equation}
    gv^\star+fh^\star
    =(f^\star h+g^\star v)^\star=1+z^{-1},
\end{equation}
so anchoring it one block forward gives
\begin{equation}
    \operatorname{syn}_m(z\mathbf r_{\rm per}^\vee)
    =z(1+z^{-1})=1+z
    =\operatorname{syn}_e(\mathbf r_{\rm per}).
\end{equation}
For $u\in\Lambda_0$, let $\widehat u$ and
$\widehat{\iota(u)}$ denote the chosen Laurent-monomial representatives in
the fundamental block.  The normalized interblock operator between blocks $j$
and $j+1$ is
\begin{equation}\label{eq:T-definition}
    T_{j,u}
    =\eta_{j,u}\,
     Z[z^{j-1}\widehat{\iota(u)}\mathbf r_{\rm per}]\,
     X[z^j\widehat u\mathbf r_{\rm per}^\vee],
\end{equation}
where $\eta_{j,u}\in\{\pm1,\pm i\}$ is chosen so that $T_{j,u}$ is
Hermitian.  This phase convention does not affect any binary commutator or
rank calculation below.

\subsubsection{Why the Layer Is Exact}

For a Pauli vector $p=(p_Z\mid p_X)$ in the ordering
\eqref{eq:pauli-component-order}, define the polynomial symplectic pairing
\begin{equation}\label{eq:omega-poly}
    \omega_R(p,q)
    =p_Zq_X^{\star T}+p_Xq_Z^{\star T}\in R.
\end{equation}
The coefficient of $x^ay^b$ is the binary commutator of $p$ with the
translate $x^ay^bq$.

\begin{lemma}[Mutual commutation of the self-dual family]\label{lem:selfdual}
For Laurent monomials $a,b$, write
\begin{equation}
    \mathbf t_{a,b}
    =(b\mathbf r_{\rm per}\mid a\mathbf r_{\rm per}^\vee).
\end{equation}
Then
$\omega_R(\mathbf t_{a,b},\mathbf t_{a',b'})=0$ for all monomials
$a,b,a',b'$.  In particular, all normalized interblock operators mutually
commute.
\end{lemma}
\begin{proof}
\begin{align*}
\omega_R(\mathbf t_{a,b},\mathbf t_{a',b'})
&=b(a')^{-1}(hv+vh)\\
&\quad+a(b')^{-1}(v^\star h^\star+h^\star v^\star)=0
\end{align*}
over $\mathbb F_2$, the two cross terms cancelling;
Eq.~\eqref{eq:T-definition} is the case
$b=z^{j-1}\widehat{\iota(u)}$, $a=z^j\widehat u$.
\end{proof}

\begin{lemma}[Commutation with $C$ and the bulk]\label{lem:TC}
Every normalized interblock operator in Eq.~\eqref{eq:T-definition} commutes
with every retained bulk check and every composite $C_{j,u}$.
\end{lemma}
\begin{proof}
The ribbon equation gives the complete electric syndrome of the $Z$ half of
$T_{j,u}$: it consists only of the two removed checks
$S^X_{j,\iota(u)}$ and $S^X_{j+1,\iota(u)}$.  The dual equation gives the
complete magnetic syndrome of the $X$ half: it consists only of
$S^Z_{j,u}$ and $S^Z_{j+1,u}$.  Hence $T_{j,u}$ has no syndrome on a
retained bulk check.  At either endpoint, Eq.~\eqref{eq:composite} pairs the
indicated $S^X$ check with the indicated $S^Z$ check.  The two contributions
to its commutator with $T_{j,u}$ therefore add to $1+1=0$ over
$\mathbb F_2$; every other composite receives zero contributions.
\end{proof}

\subsubsection{Ledger Counting}

The counting role of the interblock family is captured by a path-incidence
sequence.  Let $w_j\in\mathsf A$ denote the wall ledger of block $j$ and
define
\begin{equation}
\begin{split}
    d_{\rm w}:\mathsf A^{N-1}&\longrightarrow\mathsf A^N,\\
    d_{\rm w}(q_1,\ldots,q_{N-1})
    &=(q_1,q_1+q_2,\ldots,q_{N-2}+q_{N-1},q_{N-1}),
\end{split}
\end{equation}
and
\begin{equation}
    \Sigma:\mathsf A^N\longrightarrow\mathsf A,
    \qquad
    \Sigma(w)=\sum_{j=1}^Nw_j.
\end{equation}

\begin{lemma}[Telescoping exact sequence]\label{lem:telescope}
\textup{(i)} For $N\ge2$ the sequence
\begin{equation}\label{eq:telescope}
    0\longrightarrow\mathsf A^{N-1}
    \xrightarrow{\;d_{\rm w}\;}\mathsf A^N
    \xrightarrow{\;\Sigma\;}\mathsf A
    \longrightarrow0
\end{equation}
is exact; in particular
$\mathsf A^N/\operatorname{im}d_{\rm w}\cong\mathsf A$.

\textup{(ii)} The same holds in the compact sector over $\mathbb Z$: on
finitely supported sequences $(q_j)_{j\in\mathbb Z}$ in $\mathsf A$, with
$(d_{\rm w}q)_j=q_{j-1}+q_j$ and $\Sigma(w)=\sum_jw_j$, one has
$\ker\Sigma=\operatorname{im}d_{\rm w}$.
\end{lemma}
\begin{proof}
Exactness is the standard telescoping argument: $\Sigma d_{\rm w}=0$ since every
$q_j$ appears twice, $d_{\rm w}$ is injective by induction from $q_1=0$, and the
cumulative sums $q_j=w_1+\cdots+w_j$ solve $d_{\rm w}q=w$ for $w\in\ker\Sigma$.  Over
$\mathbb Z$ the same cumulative sums $q_j=\sum_{k\le j}w_k$ are finitely supported
precisely when $\Sigma(w)=0$, which is \textup{(ii)}.
\end{proof}

Up to row and column ordering, the commutator matrix of all interblock rows
with the removed $S^X$ checks is $D_N\otimes\Pi_\iota$, where $D_N$ is the
binary path-incidence matrix representing $d_{\rm w}$ in
Lemma~\ref{lem:telescope} and $\Pi_\iota$ is the permutation induced by
$u\mapsto\iota(u)$.  Since $h$ and $v$ are fixed Laurent polynomials, the
support of an interblock row does not grow with the number of wall blocks.

\subsection{Patch Operators and the Recovered Rank}\label{sec:completion}

\begin{figure*}[!t]
    \centering
    \subfloat[]{%
        \includegraphics[trim=5 6 5 5,clip,height=84bp]{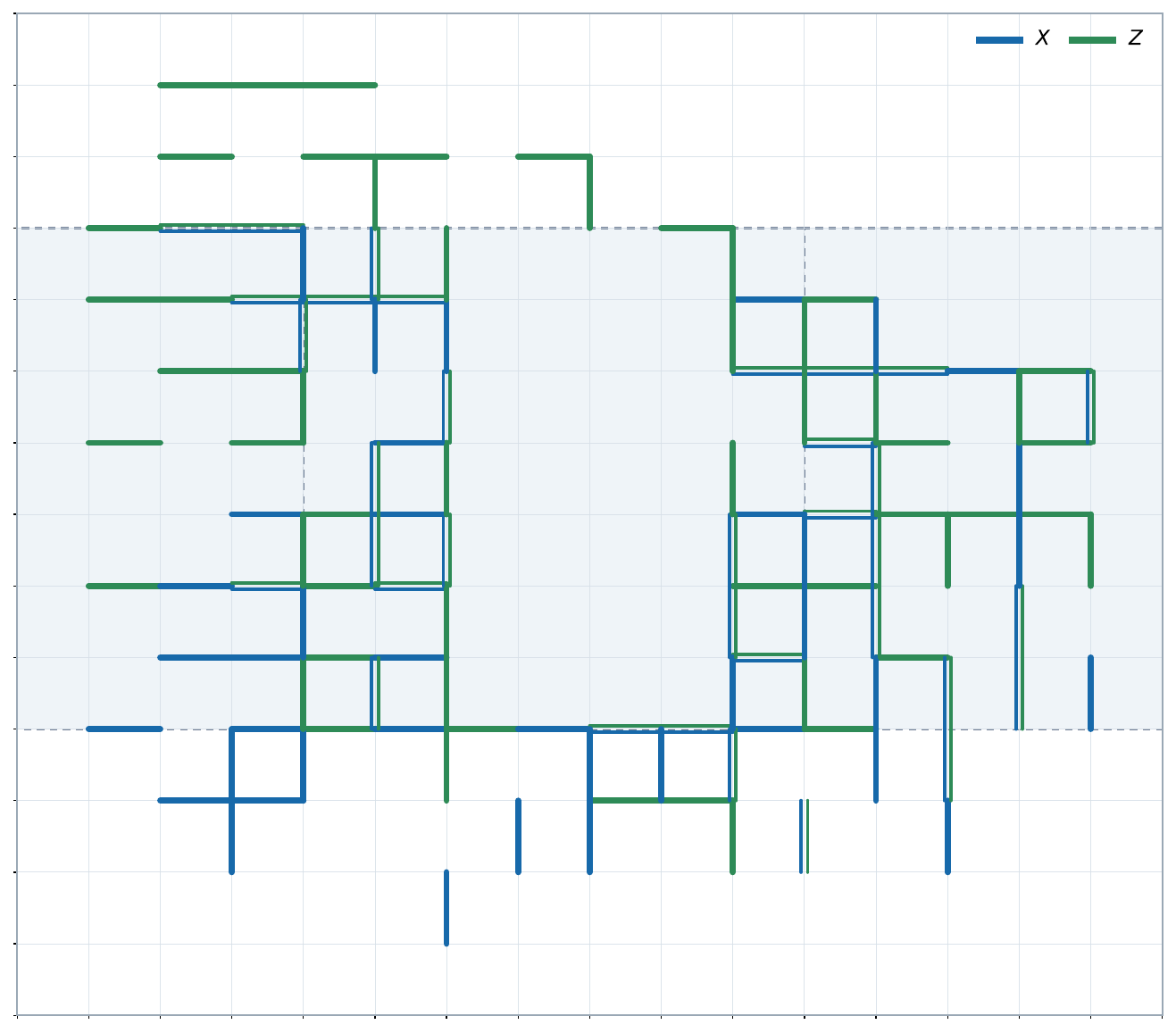}}
    \hfill
    \subfloat[]{%
        \includegraphics[trim=5 5 5 5,clip,height=84bp]{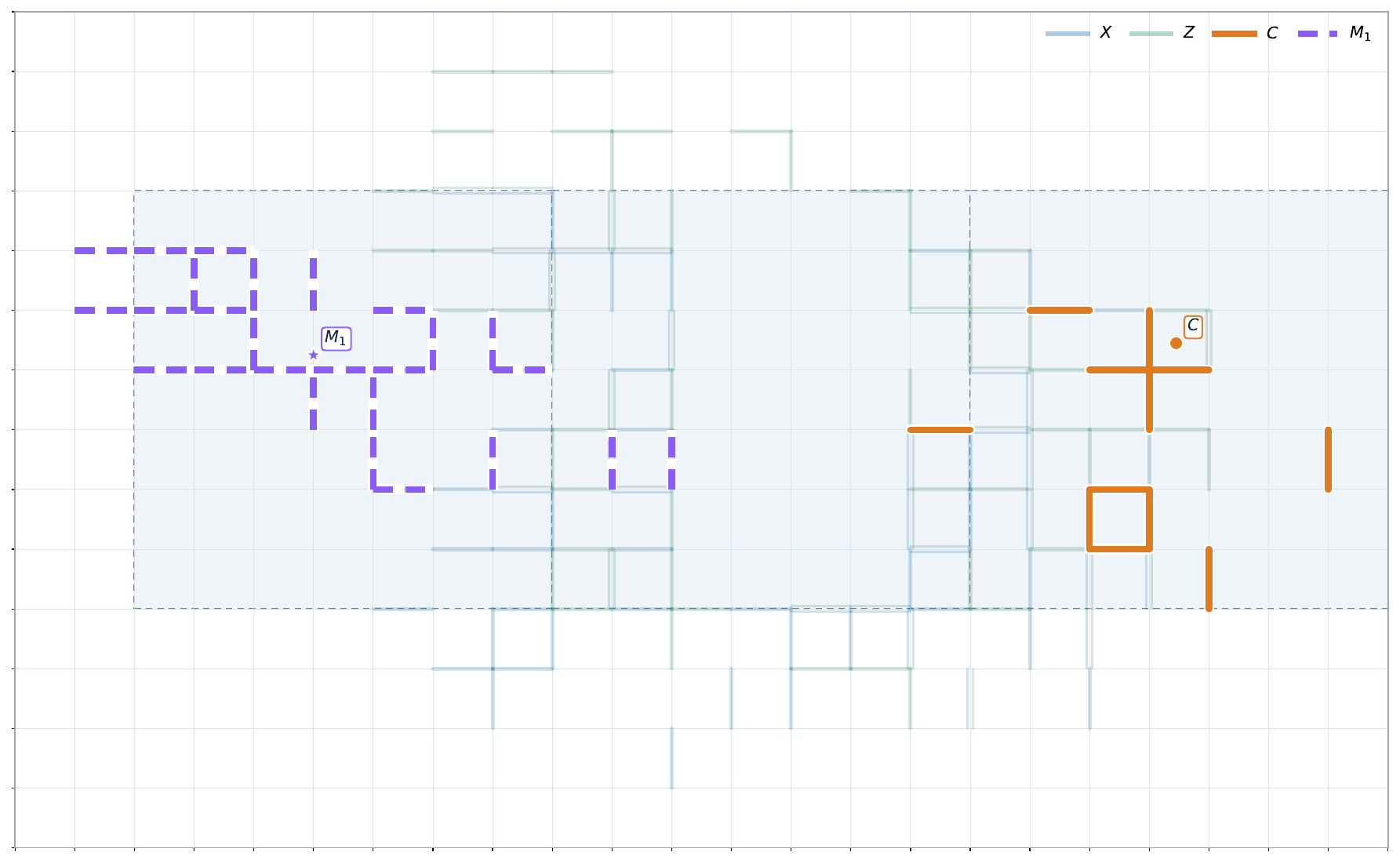}}
    \hfill
    \subfloat[]{%
        \includegraphics[trim=5 13 5 5,clip,height=84bp]{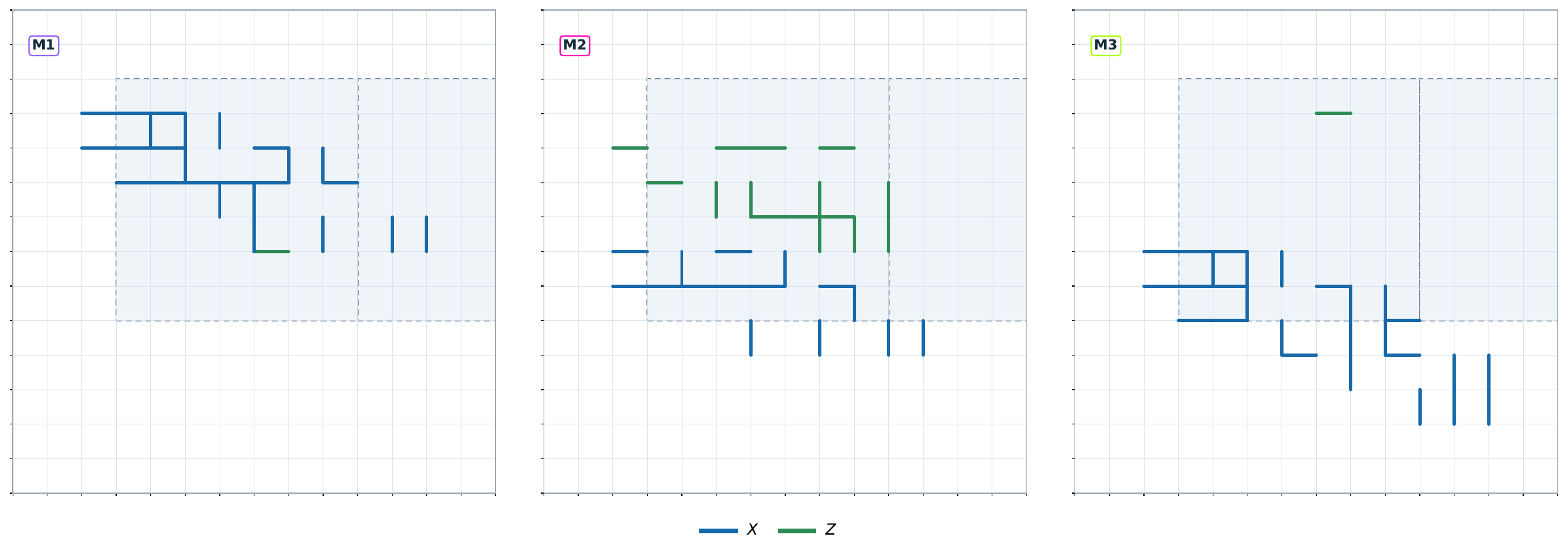}}
    \caption{A microscopic patch certificate for a straight
    $\BBc{1,1,3,1}$ wall; the shaded strip with dotted boundaries is the
    wall.  (a) A closed twice-crossing Pauli process with an
    operator-free interior.  (b) Two representative terms in its exact
    stabilizer decomposition are highlighted: one retained composite $C$ and
    one patch representative $M_1$; all other support is faded.  (c) The
    three patch representatives $M_1,M_2,M_3$ appearing in this
    example.  The exact decomposition uses $45$ scaffold rows and these
    three patch rows.  The three displayed representatives are the
    terms needed for this process, not the full $46$-dimensional patch
    family.}
    \label{fig:completion-triptych}
\end{figure*}

Consider the target closed process in which an anyon crosses the wall,
becomes its dual, turns around without enclosing a twist endpoint, crosses
back, and annihilates.  Such a twice-crossing process carries vacuum
topological charge.  A locally exact wall is required to fill it by local
stabilizer relations, just as a contractible Wilson loop in the toric code
is filled by plaquette or star checks.  The patch operators supply
these filling relations for the straight-wall interior.  A microscopic
patch certificate for $\BBc{1,1,3,1}$ is shown in
Fig.~\ref{fig:completion-triptych}.

Let $\mathcal R_{\rm ret}$ denote the listed retained pristine rows and set
\begin{equation}
    \mathcal R_0
    =\mathcal R_{\rm ret}\cup\{C_{j,u}\}\cup\{T_{j,u}\},
    \qquad
    \mathcal S_0=\langle\mathcal R_0\rangle .
\end{equation}
Thus $\mathcal R_0$ is the elementary local row family and $\mathcal S_0$ is
the stabilizer group that it generates.  Lemma~\ref{lem:telescope} reduces
the $N$ block ledgers to one copy of $\mathsf A$; the exact sequence
\eqref{eq:block-SES} then separates its $Q$ topological directions from the
$l_0m_0-Q$ label-trivial directions in $\mathcal I_e$.

A ledger map $\ell$ measures what the $C$--$T$ scaffold leaves
unconstrained, with values in the ideal $\mathcal I_e$, and zero-ledger exactness
(Theorem~\ref{thm:zero-ledger}) shows that the ledger is the \emph{only}
local invariant of the scaffold.  A uniform lift
(Theorem~\ref{thm:completion-lift}) then constructs local
representatives $M(h)$, one per basis element of $\mathcal I_e$, which are
adjoined to the Hamiltonian; the payoff is straight-wall local exactness, and
with it the affine-dual transmission proved in Sec.~\ref{sec:transmission}.

Choose a basis $h_1,\ldots,h_{l_0m_0-Q}$ of $\mathcal I_e$ and reference
Paulis $M_a=M(h_a)$.  The local Hamiltonian then contains the translated row
family
\begin{equation}
    \mathcal R_M
    =\{M_{j,a}=z^jM_a:
       \operatorname{supp}M_{j,a}\text{ and its fixed translation package}
       \text{ lie in the straight-wall interior}\},
\end{equation}
and no further translate is installed; on a finite wall this is exactly the
natural termination of Definition~\ref{def:natural-termination}.
The completed straight-wall Hamiltonian is
\begin{equation}\label{eq:Hwall}
    H_{\mathrm{wall}}
    = -\sum_{j,u} C_{j,u}
      \;-\sum_{j,u} T_{j,u}
      \;-\!\!\sum_{M_{j,a}\in\mathcal R_M}\!\!M_{j,a}
      \;-\!\!\sum_{(i,j)\notin L}\!\bigl(S_X^{(i,j)}+S_Z^{(i,j)}\bigr).
\end{equation}

\paragraph*{Structure of the patch rows on the lattice.}
Each reference patch operator $M(h)$, $h\in\mathcal I_e$, of
Theorem~\ref{thm:completion-lift} is a compact two-tone Pauli patch pinned to
one wall block.  Throughout, \emph{tone} is the binary CSS grading: a Pauli is
$X$-tone or $Z$-tone according to its half, and correspondingly a charge is
magnetic or electric, $X$-tone matching $m$ and $Z$-tone matching $e$.  Both
labellings are used, whichever the context makes natural, and \emph{two-tone},
\emph{same-tone} and \emph{opposite-tone} refer to that grading.  Its $Z$ half
deposits the electric pattern $\widehat h$ on the removed $X$-checks of that
block and its $X$ half, the affine mirror image, deposits
$\widehat{\iota(h)}$ on the removed $Z$-checks at the center-reflected
offsets: exactly the pairs the composite layer identifies.  A dressing by
removed pristine checks cancels the interblock commutators.

\paragraph*{Ledger and exact rank accounting.}
For a Pauli operator $p$, let
$\epsilon_{\rm rem}(p)\in\mathsf A^N$ record its commutators with the removed
$S^X_{j,u}$ checks, and define
\begin{equation}\label{eq:ledger-map}
    \ell(p)=\Sigma\epsilon_{\rm rem}(p)\in\mathsf A.
\end{equation}
Retained bulk rows and composites have zero removed-check syndrome, while an
interblock syndrome lies in $\operatorname{im}d_{\rm w}$.  Hence
$\ell(s)=0$ for every $s\in\mathcal S_0$ by
Lemma~\ref{lem:telescope}, and $\ell$ is well defined on centralizer classes
modulo $\mathcal S_0$.

\begin{lemma}\label{lem:charge-ledger}
If a compact Pauli $p$ centralizes the infinite straight-wall scaffold, or a
non-wrapping finite-wall scaffold away from its endpoints, then
\begin{equation}\label{eq:contractible-ledger}
    \pi_e(\ell(p))=0,
    \qquad
    \ell(p)\in\mathcal I_e.
\end{equation}
\end{lemma}
\begin{proof}
All retained electric syndromes vanish, so the full pristine electric
syndrome of $p$ is $\sum_j z^{j-1}W^X_j$.  In $V_e$, the period relation $1+z=0$ identifies this class
with $\pi_e(\sum_jW^X_j)=\pi_e(\ell(p))$.  On the other hand, the pristine
syndrome map has the form $f^\star u+g^\star v$, whose class in $V_e$ is
zero.  This proves Eq.~\eqref{eq:contractible-ledger}.
\end{proof}

This is only a necessary label-level condition: it neither constructs a
local representative of every $h\in\mathcal I_e$ nor excludes a local class
with zero ledger.

For a listed check matrix $H$, let $k(H)$ be the number of encoded qubits.
The rows of $\mathcal R_{\rm ret}\cup\{C_{j,u}\}$ and of $\mathcal R_0$
commute pairwise by Lemmas~\ref{lem:composite}, \ref{lem:TC}
and~\ref{lem:selfdual}, which every later scaffold argument uses silently.

\paragraph*{Zero-ledger exactness on the infinite wall.}
To construct local representatives and exclude local classes with zero
ledger, we work first on the infinite plane, extending
$\mathcal R_0$ and $\mathcal S_0$ block-periodically to an infinite straight
wall with $j\in\mathbb Z$.  Let $\mathcal P_c$ be the compactly supported
Paulis and define
\begin{equation}\label{eq:compact-local-homology}
 \mathcal H^{\rm loc}_0=
 \frac{\mathcal Z_c}{\mathcal S_{0,c}},\qquad
 \mathcal Z_c=\{p\in\mathcal P_c:[p,r]=0\ \forall r\in\mathcal R_0\},
 \qquad
 \mathcal S_{0,c}=\left\{\prod_{r\in F}r:F\subset\mathcal R_0
 \text{ finite}\right\},
\end{equation}
where all rows have finite support.  This avoids truncating the commutator
complex at an artificial strip boundary.

For the infinite wall $\ell(p)=\sum_{j\in\mathbb Z}W^X_j$, a finite sum for
$p\in\mathcal P_c$, and Lemma~\ref{lem:telescope}\textup{(ii)} gives
$\ker\Sigma=\operatorname{im}d_{\rm w}$ on finitely supported sequences.

\begin{lemma}[Compact pristine filling]\label{lem:compact-filling}
Assume $0<Q<\infty$.  Every compact Pauli $p\in\ker\cobdry$ has a unique
compact pristine filling $a\in G$ with $p=\bdry a$.  Moreover,
\begin{equation}\label{eq:filling-support}
 \operatorname{supp}a\subseteq
 \operatorname{conv}(\operatorname{supp}p).
\end{equation}
\end{lemma}
\begin{proof}
Finite-dimensionality of $R/\langle f,g\rangle$ makes $f$ and $g$ coprime in
the unique-factorization domain $R$, so the two syzygy equations in
$\cobdry p=0$ give
\begin{equation}
 (u_1,v_1)=(g^\star a_Z,f^\star a_Z),
 \qquad
 (u_2,v_2)=(fa_X,ga_X),
\end{equation}
so $p=\bdry(a_Z,a_X)$, with $a_Z$ and $a_X$ the coefficients of the $Z$- and
$X$-type checks, uniquely since $R$ is a domain.  Finally,
$\operatorname{Newt}(rs)=\operatorname{Newt}(r)+\operatorname{Newt}(s)$
for nonzero Laurent polynomials.  The constant monomial lies in each of
$\operatorname{Newt}(f)$, $\operatorname{Newt}(g)$,
$\operatorname{Newt}(f^\star)$, and $\operatorname{Newt}(g^\star)$, since $f$ and $g$
have constant term~$1$.
Applied componentwise to $\bdry a$, Newton-polytope multiplicativity
therefore gives
\[
 \operatorname{Newt}(a_Z)\subseteq
 \operatorname{Newt}(g^\star a_Z),\qquad
 \operatorname{Newt}(a_X)\subseteq\operatorname{Newt}(fa_X),
\]
with zero components omitted.  Hence
$\operatorname{supp}a\subseteq
\operatorname{conv}(\operatorname{supp}\bdry a)$, proving
Eq.~\eqref{eq:filling-support}.
\end{proof}

\subsubsection{Zero-Ledger Exactness}

Measure distance in complete period blocks, with the block $\ell^\infty$-metric on the
lifted plane,
\[
 \|(i,j)\|_{\rm blk}
 =\max\!\left\{\frac{|i|}{l_0},\frac{|j|}{m_0}\right\}.
\]
For a finite Laurent vector, $\operatorname{supp}$ below means its spatial
support after forgetting the Pauli component.  Let
\[
 r_{\rm chk}
 =
 \left\lceil
 \max_{\alpha\in
 \operatorname{supp}(f)\cup\operatorname{supp}(g)
 \cup\operatorname{supp}(f^\star)\cup
 \operatorname{supp}(g^\star)}
 \|\alpha\|_{\rm blk}
 \right\rceil .
\]
Thus every pristine row is supported in the $r_{\rm chk}$-neighborhood of
its anchor.  For either orientation and $u\in\Lambda_0$, let $E_u$ be the
four removed-check anchors that are the endpoint syndrome of $T_{0,u}$, and
put
\begin{equation}
\begin{split}
 r_T&=
 \max_{\text{orientations},\,u\in\Lambda_0}
 \left\lceil
 \max_{q\in\operatorname{supp}T_{0,u}}
 d_{\rm blk}\!\left(q,\operatorname{conv}E_u\right)
 \right\rceil,\\
 r_{\rm inc}&=r_{\rm chk}+r_T+1.
\end{split}
\label{eq:endpoint-constant-1}
\end{equation}
These are explicit finite-support maxima determined by
$f,g,l_0,m_0$ and the chosen longitudinal ribbons.

\begin{lemma}[Open-path decomposition]\label{lem:open-path-decomposition}
Let $\Omega$ be a convex block region of the lifted plane, and let $p$,
supported in $\Omega$, commute with every pristine check other than the removed
checks of one straight wall, and with that wall's composites and its interblock
rows between consecutive blocks along the whole wall.  Suppose its removed-check
patterns $W^X_j,W^Z_j\in\mathsf A$ obey $\sum_jW^X_j=0$.  Then:

\textup{(i)} commutation with the composites gives
\begin{equation}\label{eq:removed-syndrome-matching}
 W^Z_j=\Pi_\iota W^X_j,
\end{equation}
and the interblock product $J$ with coefficient sequence $\Pi_\iota q$,
$q_j=\sum_{k\le j}W^X_k$, cancels both patterns, uses only interblock rows between
consecutive blocks of the wall, and is supported in $N_{r_{\rm inc}}(\Omega)$;

\textup{(ii)} $p'=pJ$ commutes with every pristine check, and its compact
filling $a=(a_Z,a_X)$ of Lemma~\ref{lem:compact-filling} is supported in
$N_{r_{\rm inc}}(\Omega)$;

\textup{(iii)} with $A_X(j,u)$ and $A_Z(j,u)$ the coefficients of $a$ on the two
removed checks, $m_j(u)=A_X(j,\iota(u))+A_Z(j,u)$ is independent of $j$ along
the wall; if it vanishes on one block of the wall, $p$ is a product of retained
rows, composites and interblock rows, all supported in
$N_{r_{\rm inc}+r_{\rm chk}}(\Omega)$.
\end{lemma}
\begin{proof}
An interblock product with coefficient sequence $\alpha$ has removed-check
syndromes $W_T^X=d_{\rm w}\Pi_\iota\alpha$ and $W_T^Z=d_{\rm w}\alpha$.  The
cumulative sums obey $(d_{\rm w}q)_j=q_{j-1}+q_j=W^X_j$ and, because of the zero
sum, vanish outside the extreme nonzero syndrome blocks (Lemma~\ref{lem:telescope});
on a wall of blocks $1,\ldots,N$ they sit on the internal block boundaries
$1,\ldots,N-1$.  So $\alpha=\Pi_\iota q$ cancels both patterns by
Eq.~\eqref{eq:removed-syndrome-matching}.  A removed check carrying nonzero
syndrome lies within $r_{\rm chk}$ of $\operatorname{supp}p$, one further block
reaches the adjacent interblock endpoints, and each interblock factor lies within
$r_T$ of the convex hull of its anchors; hence
$\operatorname{supp}J\subseteq N_{r_{\rm inc}}(\Omega)$, which is \textup{(i)}.
Then $p'$ commutes with every pristine check, and
Lemma~\ref{lem:compact-filling} places its filling in
$\operatorname{conv}(\operatorname{supp}p')\subseteq N_{r_{\rm inc}}(\Omega)$, a
block neighbourhood of a convex set being convex; this is \textup{(ii)}.
Interblock rows commute with one another, so $p'$ still commutes with every
interblock row of the wall, and the four endpoints of $T_{j,u}$ give
\[
 A_X(j,\iota(u))+A_X(j+1,\iota(u))+A_Z(j,u)+A_Z(j+1,u)=0,
\]
that is, $m_j(u)=m_{j+1}(u)$.  If $m$ vanishes on one block it vanishes on every
block of the wall, and the only removed checks are this wall's, so every
removed-row contribution to $\bdry a$ occurs as
$S^X_{j,\iota(u)}S^Z_{j,u}=C_{j,\iota(u)}$ and every other row of the filling
is a retained check.  Every row lies within $r_{\rm chk}$ of
$\operatorname{supp}a$; restoring $J$ proves \textup{(iii)}.
\end{proof}

\begin{theorem}[Zero-ledger exactness]\label{thm:zero-ledger}
On the infinite straight wall, the ledger map of Eq.~\eqref{eq:ledger-map} is
injective on the compact local quotient
$\mathcal H^{\rm loc}_0=\mathcal Z_c/\mathcal S_{0,c}$ of
Eq.~\eqref{eq:compact-local-homology}:
\begin{equation}\label{eq:compact-ledger-injective}
 \ker\!\left(\ell:\mathcal H^{\rm loc}_0\longrightarrow\mathcal I_e\right)=0.
\end{equation}
Equivalently, a compactly supported Pauli that centralizes every scaffold row
--- retained checks, composites and interblock rows --- and has zero ledger is a
finite product of scaffold rows.
\end{theorem}
\begin{proof}
On the infinite wall every interblock row is installed.  Let $p\in\mathcal Z_c$
obey $\ell(p)=0$; Lemma~\ref{lem:open-path-decomposition} applies with
$\Omega=\operatorname{conv}(\operatorname{supp}p)$.  The filling is compact, so
$m_j(u)=0$ on blocks far from $\operatorname{supp}p$, and part \textup{(iii)}
gives $p\in\mathcal S_{0,c}$.
\end{proof}

\paragraph*{Buffered finite-wall windows.}
For a finite wall, let
$\Omega_{\rm out}=N_R(\Omega_{\rm in})$ be a radius-$R$ neighborhood of a
convex inner window and define
\begin{align}
 \mathcal Z_{\rm in}
 &=\{p:\operatorname{supp}p\subseteq\Omega_{\rm in},
      \ [p,r]=0\text{ for every }r\in\mathcal R_0\},\\
 \mathcal S_{\rm in,out}
 &=\left\{\prod_{r\in F}r:
      F\subseteq\mathcal R_0\text{ finite},
      \operatorname{supp}r\subseteq\Omega_{\rm out}\text{ for all }r\in F,
      \ \operatorname{supp}\!\left(\prod_{r\in F}r\right)
      \subseteq\Omega_{\rm in}\right\},\\
 \mathcal K_{\rm in,out}
 &=\mathcal Z_{\rm in}/\mathcal S_{\rm in,out}.
\end{align}

\begin{corollary}\label{cor:buffered-injectivity}
Put
\begin{equation}\label{eq:effective-R0}
 R_0=2r_{\rm chk}+r_T+1 .
\end{equation}
Then whenever $R\ge R_0$ and
$\Omega_{\rm out}=N_R(\Omega_{\rm in})$ lies in the straight-wall interior
without wrapping around the torus,
\begin{equation}\label{eq:buffered-ledger-injective}
 \ker\!\left(\ell:\mathcal K_{\rm in,out}\longrightarrow\mathcal I_e\right)=0.
\end{equation}
\end{corollary}
\begin{proof}
Lift the non-wrapping window to the plane.  A zero-ledger element of
$\mathcal Z_{\rm in}$ meets the hypotheses of
Lemma~\ref{lem:open-path-decomposition} with $\Omega=\Omega_{\rm in}$; its
filling lies in $N_{r_{\rm inc}}(\Omega_{\rm in})$, inside the straight-wall
interior, so it vanishes on the end blocks of the wall and $m_j$ vanishes there.  Part
\textup{(iii)} places the whole scaffold decomposition in
$N_{r_{\rm inc}+r_{\rm chk}}(\Omega_{\rm in})=N_{R_0}(\Omega_{\rm in})$.
\end{proof}

\subsubsection{The Uniform Patch Lift}

\begin{lemma}[The wall trace, and the summed interblock obstruction]
\label{lem:wall-trace}
For a compact Laurent polynomial $F$ define the \emph{wall trace}
\begin{equation}\label{eq:wall-trace}
    \operatorname{Tr}_z(F)=\sum_{n\in\mathbb Z}\operatorname{tr}(z^nF).
\end{equation}

\textup{(i)} The sum is finite and satisfies
$\operatorname{Tr}_z(z^kF)=\operatorname{Tr}_z(F)$ and
$\operatorname{Tr}_z(F^\star)=\operatorname{Tr}_z(F)$; in particular
$\operatorname{Tr}_z(F)+\operatorname{Tr}_z(F^\star)=0$ over $\mathbb F_2$.

\textup{(ii)} Let $P=(a,b\mid\kappa b^\star,\kappa a^\star)$ be any compact Pauli
in the affine form of Eq.~\eqref{eq:inversion}, write
$\mathbf r_{\rm per}=(h,v)$ and put $K_P=av+bh$.  Then for every interblock label
$u$ the interblock commutators of $P$ sum to zero:
\begin{equation}\label{eq:summed-junction-obstruction}
 \sum_{j\in\mathbb Z}\langle P,T_{j,u}\rangle
 =\operatorname{Tr}_z(\widehat u^\star K_P)
  +\operatorname{Tr}_z(\widehat uK_P^\star)
 =0 .
\end{equation}
\end{lemma}
\begin{proof}
Only finitely many $z^nF$ have a constant term, and $\operatorname{Tr}_z(z^kF)$
reindexes the sum; since $\operatorname{tr}(z^nF^\star)=\operatorname{tr}(z^{-n}F)$
the antipode reverses the summation order, giving \textup{(i)}.  For
\textup{(ii)}, expanding the symplectic pairing with
Eq.~\eqref{eq:T-definition} and summing over interblock translates gives
$\operatorname{Tr}_z(\widehat u^\star K_P)
 +\operatorname{Tr}_z[\widehat{\iota(u)}^\star\kappa K_P^\star]$, and
Eq.~\eqref{eq:affine-anchor} with $\kappa^\star\kappa=1$ turns the second term
into $\operatorname{Tr}_z(F^\star)$ for $F=\widehat u^\star K_P$; apply
\textup{(i)}.
\end{proof}

\begin{theorem}[Uniform affine patch lift]\label{thm:completion-lift}
With $\mathcal Z_c$ the compactly supported Paulis centralizing every
\emph{scaffold} row, $\mathcal I_e=(f^\star,g^\star)\mathsf A$, and $\ell$
the ledger map of
Eq.~\eqref{eq:ledger-map}, there is a linear map
\begin{equation}\label{eq:completion-lift}
 \Gamma:\mathcal I_e\longrightarrow\mathcal Z_c,
 \qquad h\longmapsto M(h),
 \qquad \ell\Gamma=\operatorname{id}_{\mathcal I_e},
\end{equation}
and a radius $r_{\rm lift}$, depending only on $f,g,l_0,m_0$ and the chosen period
ribbons and \emph{not} on the wall length, such that
$\operatorname{supp}M(h)$ lies within $r_{\rm lift}$ of the wall carrier for every
$h\in\mathcal I_e$.  Every block translate $z^jM(h)$ is anchored by the same
global Laurent origin and the same affine convention
Eq.~\eqref{eq:inversion} as $M(h)$ itself.
\end{theorem}
\begin{proof}
The construction is explicit, and its three choices are fixed once, since the
endpoint constants and the witness of Proposition~\ref{rem:shallow-necessity}
are computed from this family.

\emph{Basis.}  Order the standard monomials $x^iy^j$ of
$\mathsf A=R/(1+z,1+w)$ lexicographically in $(i,j)$ and row-reduce
\((f^\star,g^\star):\mathsf A^2\to\mathsf A\), the $f^\star$-columns first:
$\xi_\nu$ is the reduced echelon basis of $\mathcal I_e$, and
$(\bar a_\nu,\bar b_\nu)$ expresses $\xi_\nu$ in the pivot columns.  For
$h=\sum_\nu h_\nu\xi_\nu$ put $M(h)=\prod_\nu M_\nu^{h_\nu}$, with $M_\nu$ built
below.

\emph{Lift.}  Since $J_{\rm per}\subseteq\langle f^\star,g^\star\rangle$, finite
Laurent division, first by $1+z$ and then by $1+w$ with remainders in the
fundamental interval, gives
\(\widehat\xi_\nu+f^\star\widehat{\bar a_\nu}+g^\star\widehat{\bar b_\nu}=U_\nu(1+z)+V_\nu(1+w)\).
With the period ribbons \(f^\star h+g^\star v=1+z\) and \(f^\star c+g^\star d=1+w\),
put \(a_\nu=\widehat{\bar a_\nu}+U_\nu h+V_\nu c\) and
\(b_\nu=\widehat{\bar b_\nu}+U_\nu v+V_\nu d\), so that
$\widehat\xi_\nu=f^\star a_\nu+g^\star b_\nu$, and define, in the component
ordering~\eqref{eq:pauli-component-order}, the matched provisional operator
\begin{equation}\label{eq:affine-provisional-completion}
    M^{(0)}_\nu=(a_\nu,b_\nu\mid\kappa b_\nu^\star,\kappa a_\nu^\star).
\end{equation}
Its electric syndrome is $\widehat\xi_\nu$ and its magnetic syndrome is
$g\kappa b_\nu^\star+f\kappa a_\nu^\star=\kappa\widehat\xi_\nu^\star=\widehat{\iota(\xi_\nu)}$,
so it has no retained-check syndrome, commutes with every composite and has
ledger $\xi_\nu$.

\emph{Dressing.}  $M^{(0)}_\nu$ has the affine form of
Lemma~\ref{lem:wall-trace}\textup{(ii)}, so for every $u$ its interblock
commutators \(t_{\nu,u}(j)=\langle M^{(0)}_\nu,T_{j,u}\rangle\) sum to zero and
\(q_{\nu,u}(j)=\sum_{k<j}t_{\nu,u}(k)\) has finite support.  The removed-check
dressing
\begin{equation}
 M_\nu
 =M_\nu^{(0)}
 \prod_{j,u}
 \left(S^X_{j,\iota(u)}\right)^{q_{\nu,u}(j)}
\label{eq:endpoint-constant-3}
\end{equation}
cancels them; removed pristine checks commute with retained rows and
composites and carry zero ledger, so $M_\nu\in\mathcal Z_c$ with
$\ell(M_\nu)=\xi_\nu$.  Each step is a finite operation on fixed data, which
gives the uniform radius $r_{\rm lift}$.
\end{proof}

\begin{corollary}[Straight-wall local exactness]
\label{cor:straight-wall-completion}
Let $\Omega_{\rm in}$ be a convex inner window and
$\Omega_{\rm out}=N_R(\Omega_{\rm in})$ with $R\ge R_0$ of
Eq.~\eqref{eq:effective-R0}, lifting without wrapping into the straight-wall
interior, and assume $\Omega_{\rm in}$ contains a translate of each
reference patch support together with its bounded translation
certificate.  Then
\begin{equation}\label{eq:local-ledger-isomorphism}
 \ell:\mathcal K_{\rm in,out}\xrightarrow{\;\sim\;}\mathcal I_e,
 \qquad
 zM(h)=M(h)\pmod{\mathcal S_0}.
\end{equation}
The $l_0m_0-Q$ patch classes associated with a basis of $\mathcal I_e$
are independent, their translates admit a mutually commuting Hermitian lift,
and every compact Pauli centralizing the completed infinite-wall rows is,
up to scalar phase, a finite product of those rows.  On a naturally terminated finite wall the same conclusions hold for Paulis
supported in such an interior window, provided the rows used are installed
there in the sense of Definition~\ref{def:natural-termination}; the finite-window form
actually used later states its own conditions, those of a crossing window
(Definition~\ref{def:qualifying-crossing-window}).

\emph{Bounded support.}  Moreover the translation relation in
Eq.~\eqref{eq:local-ledger-isomorphism} admits a \emph{bounded} scaffold
certificate: for each $h\in\mathcal I_e$ there is a finite set
$\mathscr C(h)$ of scaffold rows with
\begin{equation}\label{eq:bounded-translation-certificate}
 zM(h)M(h)=\prod_{r\in\mathscr C(h)}r,
\end{equation}
every member of which is supported in a uniform neighborhood of
$\operatorname{supp}M(h)\cup\operatorname{supp}zM(h)$, with a bound
depending only on the code, the chosen period ribbons and the fixed
patch family, and not on the wall length.
\end{corollary}
\begin{proof}
Surjectivity follows from Eq.~\eqref{eq:completion-lift}, injectivity from
Corollary~\ref{cor:buffered-injectivity}, and
$\ell(zM(h)M(h))=0$ gives the translation relation by
Theorem~\ref{thm:zero-ledger}.  The proof of
Corollary~\ref{cor:buffered-injectivity} bounds the factors around
$\operatorname{conv}(\operatorname{supp}M(h)\cup\operatorname{supp}zM(h))$;
a finite maximum over the fixed patch family gives
Eq.~\eqref{eq:bounded-translation-certificate}, uniformly in wall length.  For $h,h'\in\mathcal I_e$, set
$b_n=\langle M(h),z^nM(h')\rangle$.  The relation
$z^{n+1}M(h')z^nM(h')\in\mathcal S_0$ and scaffold centrality give
$b_{n+1}=b_n$.  For sufficiently large $|n|$ the supports are disjoint, so
$b_n=0$ for every $n$; all patch classes and their translates commute.
Choose phases coherently on an independent binary basis.  Finally, a compact
completed-wall centralizer $p$ has $\ell(p)\in\mathcal I_e$ by
Lemma~\ref{lem:charge-ledger}.  Write $\ell(p)=\sum_a\lambda_ah_a$ in the
basis and set $\widetilde M=\prod_a\bigl(z^{j_a}M(h_a)\bigr)^{\lambda_a}$, each
factor a translate available in the same admissible neighborhood.  Since
$zM(h)=M(h)\pmod{\mathcal S_0}$ and $\ell$ vanishes on $\mathcal S_0$,
$\ell(\widetilde M)=\ell(p)$, so $p\widetilde M$ has zero ledger and is, up to
scalar phase, a scaffold product by Theorem~\ref{thm:zero-ledger}.
\end{proof}

We fix here, once for the whole paper, which rows a finite wall carries.

\subsubsection{Natural Termination}

\begin{definition}[Natural termination]\label{def:natural-termination}
Fix once a basis \(h_1,\ldots,h_{l_0m_0-Q}\) of \(\mathcal I_e\) and, for each
\(a\), one finite bounded scaffold certificate \(\mathscr C_a\) for the
translation relation \(zM(h_a)M(h_a)\), supplied by the bounded-support clause
of Corollary~\ref{cor:straight-wall-completion},
Eq.~\eqref{eq:bounded-translation-certificate}, so that
\(zM(h_a)M(h_a)=\prod_{r\in\mathscr C_a}r\) in the binary Pauli space.  Write
\(\mathscr P_a=\{M(h_a),zM(h_a)\}\cup\mathscr C_a\) for the typed translation
package, and \(A_a\subset\mathbb Z\) for the finite set of longitudinal block
offsets met by its supports and by its required scaffold-row carriers,
relative to carrier exponent \(0\), an interblock row being counted in both blocks
it joins.  A finite wall whose blocks carry exponents
\(0,\ldots,N-1\) relative to its first block \(o\) is \emph{naturally
terminated} when its $C$ rows occupy that finite open path of complete period
blocks, its $T$ rows occupy the internal block boundaries of the path, and its
installed patch rows are precisely the translates whose package fits in
the wall,
\begin{equation}
 M_{j,a}=oz^jM(h_a),
 \qquad
 j\in J_a=\bigl\{j\in\mathbb Z:\ j+A_a\subseteq\{0,\ldots,N-1\}\bigr\}.
 \label{eq:exact-retention}
\end{equation}
No additional endpoint row is postulated, and no other translate is installed.
For wall \(s\) of a configuration we write \(o_s\), \(N_s\) and \(J_{s,a}\).
\end{definition}

Weights are bounded but depend on the realization: for one admissible choice
for $\BBc{1,1,3,1}$ the retained pristine checks have weight $6$, the
composites at most $12$, the interblock rows at most $24$, a bound attained, and the patch
rows at most $20$.

\subsection{The Toric-Code Limit}\label{sec:toric-limit}

\begin{figure}[t]
    \centering
    \includegraphics[width=\columnwidth]{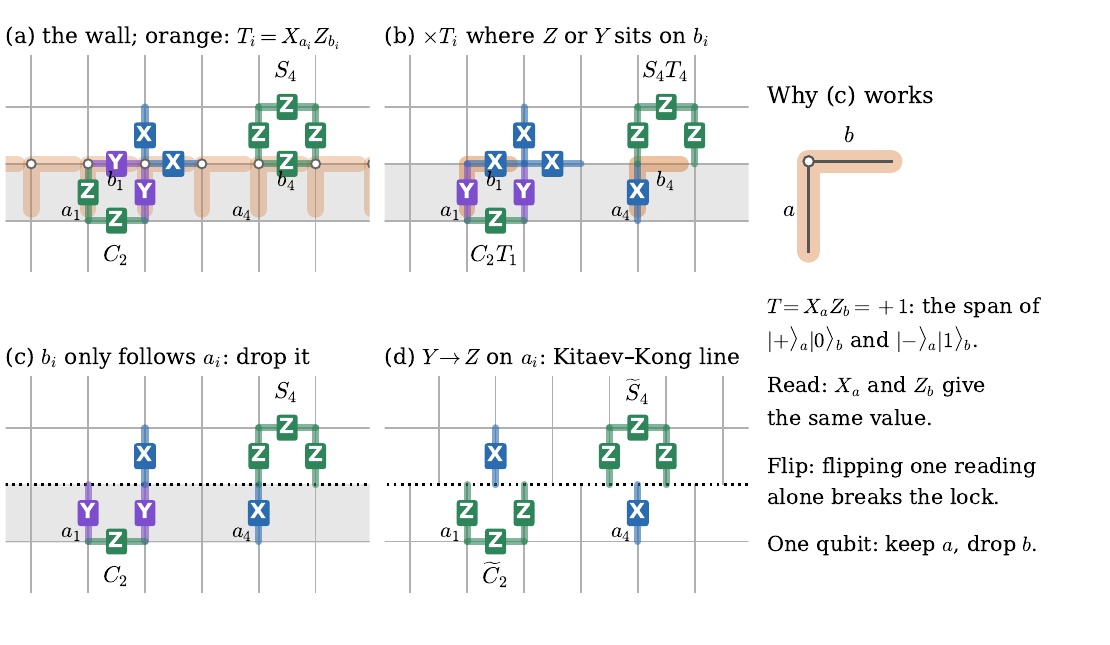}
    \caption{Reduction of the toric-code wall to the defect line of
    Ref.~\cite{kitaevModels2012} (Sec.~\ref{sec:toric-limit}), followed on the
    composite $C_2$ and the plaquette $S_4$ above the wall.  Blue, green and
    violet mark $X$, $Z$ and $Y$; grey cells are wall blocks, and hollow
    vertices carry no separate star.  In (c) and (d) the dotted line carries no
    qubit; in (d) the upper half is offset by half a spacing, as in Fig.~1 of
    Ref.~\cite{kitaevModels2012} rotated by $90^\circ$.}
    \label{fig:toric-limit}
\end{figure}

In the toric code, $f=1+x$ and $g=1+y$, so $Q=1$ and $l_0=m_0=1$: a block is a
single cell, $\iota$ is the identity, there are no patch operators, and the
duality exchanges $e$ and $m$.  Draw the cell $x^iy^j$ as a unit square with its
qubits on its top and right edges; $S^Z$ is then the plaquette of the square and
$S^X$ the star at its upper-right corner, and the composite $C_i$ is a star times
the plaquette below and to its left.  The single-edge solution $(h,v)=(x,0)$ of
the longitudinal period equation makes each interblock row two-body,
\begin{equation}\label{eq:toric-T}
 T_i=X_{a_i}Z_{b_i},
\end{equation}
on the two edges $a_i$ and $b_i$ that meet at a vertex of the wall's upper
boundary [Fig.~\ref{fig:toric-limit}(a)].

Figure~\ref{fig:toric-limit} turns this wall into the defect line of Kitaev and
Kong~\cite{kitaevModels2012} in three steps.  Every check commutes with $T_i$,
and $T_i=+1$ locks the pair: $X_{a_i}$ and $Z_{b_i}$ read the same value, so a
check that flips one reading flips the other.  Once every check with a $Z$ component on
$b_i$ is multiplied by $T_i$, the edge $b_i$ enters only through such joint flips
and is dropped.  Renaming $Y$ as $Z$ on each $a_i$ then gives, up to sign and a
rotation, the line checks $C_{2,5,3|7}$ and $D_{3|7,8,9}$ of
Ref.~\cite[Eq.~(5)]{kitaevModels2012}, on a lattice whose line carries no qubits,
and for a wall wrapping a sufficiently large torus the code space of the wall is
that of the line on the same torus.

The toric code does not need the transverse
normalization~\eqref{eq:transverse-normalized-ribbon} that the general
construction fixes: the two-body wall lies outside
Theorem~\ref{thm:transmission}, and its transmission $e\leftrightarrow m$ is
that of Ref.~\cite{kitaevModels2012}.  On a finite wall of at least two blocks with
natural termination (Definition~\ref{def:natural-termination}) the two terminal
composites become five-body checks at the two ends of the reduced line.

\section{Non-Abelian Twists}\label{sec:twists}

A wall that exchanges $e$ and $m$ must stop somewhere, and where it stops
the exchange has nowhere to go.  That terminal point is the twist $\sigma$, and
this section proves, for the lattice operator algebra, that it is genuinely
non-Abelian in three steps.  It absorbs an
entire group of anyons without changing: the invariant channels
$\epsilon_a=m_ae_{a^\iota}$ of Lemma~\ref{lem:endpoint-caps}.  It carries no
logical degree of freedom: every Pauli in an endpoint window
that commutes with the Hamiltonian is, up to phase, a product of completed rows (Theorem~\ref{thm:endpoint-exactness}).
And for $n_{\rm pair}$ twist pairs on a completed torus, in the configurations
specified below,
\[
 k(H_{\rm fin})-k(H_{\rm bulk})=(n_{\rm pair}-1)Q,
\]
with the pair-loop and transfer algebra of Sec.~\ref{sec:pca} --- the
loops and transfer strings that read and shift a pair's charge --- acting
irreducibly on the resulting $2^{(n_{\rm pair}-1)Q}$-dimensional space, with
no leftover sector (Theorems~\ref{thm:exhaustion} and~\ref{thm:pca-register}).

Two qualifications belong at the outset.  The twist is an \emph{extrinsic}
defect: it is the endpoint of a wall rather than a deconfined finite-energy
excitation, it does not appear in the label set $V_e\oplus V_m$ of
Sec.~\ref{sec:anyontypes}, and the bulk order of the completed model is still
the Abelian order of the pristine code.  Accordingly ``non-Abelian'' refers
throughout to the fusion and endpoint algebra of the defects, in the sense
made precise by Theorems~\ref{thm:exhaustion} and~\ref{thm:pca-weyl}, and not to a change
of topological phase~\cite{barkeshliTwist2013, barkeshliSymmetry2019}.

The affine duality reaches the twists through the wall
\emph{construction} --- the interblock anchoring
$\widehat{\iota(u)}=\kappa\widehat u^\star$ of Eq.~\eqref{eq:T-definition},
which makes $\epsilon_a=m_ae_{a^\iota}$ absorbable --- and not through the
transmitted-label relation.  For the standard configurations
Theorem~\ref{thm:exhaustion} assumes, which exist for every $n_{\rm pair}$
(Sec.~\ref{sec:probes}), let
$\mathcal F_n$ be the wall-induced factor of that theorem.  Throughout, and in the Letter, the
\emph{operational quantum dimension} of a twist means the per-twist growth
rate
\[
 d_\sigma^{\rm op}
 =\lim_{n\to\infty}\bigl(\dim\mathcal F_n\bigr)^{1/(2n)}
 =\lim_{n\to\infty}2^{(n-1)Q/(2n)}=2^{Q/2}.
\]
This operational statement uses Theorem~\ref{thm:exhaustion} and its
geometric hypotheses;
Corollary~\ref{cor:fusion-rules} identifies it with the categorical quantum dimension.

The counting in this section is binary: it asks only whether two operators
commute, never by what phase, and needs no phase convention.
Section~\ref{sec:fusiontwists} reconstructs charge sectors from the pair-loop and
transfer algebra and identifies them with fusion channels.

\paragraph*{Setting for this section.}
The global results of Sec.~\ref{sec:twists} hold in a standard configuration
(Definition~\ref{def:standard-layout}) and state only the hypotheses that go
beyond it; the local results state their own windows, and
Proposition~\ref{prop:frames} states its own geometry.

\subsection{Absorbable Channels and Exact Caps}\label{sec:absorbable}

The endpoint of the duality wall is the twist $\sigma$.  It carries no
Abelian label of its own and is characterized instead by what it can
swallow.  An anyon carried once around the endpoint crosses the wall an odd
number of times and comes back acted on by the duality $D_{\rm lab}$
[Fig.~\ref{Fig:DefectConstruction}(c)]; it can
therefore return to itself only if it is $D_{\rm lab}$-invariant.  The
invariant combinations are exactly
\[
 \epsilon_a=m_a e_{a^\iota},\qquad a\in V_m,
\]
so there are $|V_m|=2^Q$ of them.  Equivalently, they are exactly the pairs
$(A_1,A_2)\in V_m\oplus V_e$ with $\B{A_1}{X^\iota}=\B{X}{A_2}$ for all
$X\in V_m$, for which the magnetic and the electric loop of one test label read
the same sign (Eq.~\eqref{eq:phi-affine-exchange} and
Lemma~\ref{lem:braiding-nondegenerate}).  Here $2^Q$ counts absorbable channels;
that the same number reappears as $d_\sigma^2$ is proved in
Sec.~\ref{sec:fusiontwists}, not a restatement.
Charges are read by closed loops, which evaluate syndrome and not crossings:
by Lemma~\ref{lem:closed-loop-interior} a Pauli $p$ changes the eigenvalue of a
loop with filling $A$ by $(-1)^{\tr(A^\dagger S\,\cobdry p)}$, which is defined
although the ribbons are thick and crossing numbers are not.  For the
period-rectangle fillings of Sec.~\ref{sec:probes}, with a syndrome cluster deep
inside the filling and all others outside it, this sign is a character of the
enclosed charge (Lemma~\ref{lem:plateau-evaluation}).  A \emph{cap} carries $\epsilon_a$ from far away and deposits it
on the endpoint; the construction below is \emph{exact}, creating no excitation
apart from its two ends.

\subsubsection{Exact Caps at a Natural Endpoint}

\begin{lemma}[Exact linear endpoint caps]
\label{lem:endpoint-caps}
Use the fixed orientation, the affine block involution \(\iota\), the
longitudinal period ribbon \(\mathbf r_{\rm per}=(h,v)\) with
\(f^\star h+g^\star v=1+z\) and the patch family of
Theorem~\ref{thm:completion-lift}, and fix once, for the whole paper, a linear
label section \(s_m:V_m\to\mathsf A\).  Write
\(\mathbf r_{\rm per}^{\vee}=(v^\star,h^\star)\),
\(A=\widehat{s_m(a)}\), \(A^\iota=\widehat{\iota(s_m(a))}=\kappa A^\star\) and
\(S_R(z)=1+z+\cdots+z^{R-1}\) for \(R\geq1\).  Use the natural finite
termination and index its blocks by \(1,\ldots,N\).  Provided the indicated exterior block is
pristine and the resulting fixed endpoint collar does not meet another
endpoint, define
\begin{align}
 E_{L,a}
 &=
 Z\!\left[z^{-R}A^\iota S_R(z)(h,v)\right]\,
 X\!\left[z^{1-R}A S_R(z)(v^\star,h^\star)\right],
 \label{eq:left-exact-cap}\\
 E_{R,a}
 &=
 Z\!\left[z^{N-1}A^\iota S_R(z)(h,v)\right]\,
 X\!\left[z^{N}A S_R(z)(v^\star,h^\star)\right].
 \label{eq:right-exact-cap}
\end{align}
Then the exact pristine syndromes are
\begin{align}
 \cobdry E_{L,a}
 &=
 \left(A^\iota(z^{-R}+1),\,A(z^{-R}+1)\right),
 \label{eq:left-cap-syndrome}\\
 \cobdry E_{R,a}
 &=
 \left(z^{N-1}A^\iota(1+z^R),\,
       z^{N-1}A(1+z^R)\right),
 \label{eq:right-cap-syndrome}
\end{align}
so for $a\neq0$ each cap has exactly one terminal cluster \((e_{a^\iota},m_a)\), exactly
one corresponding remote bulk cluster, and no other pristine syndrome.

At the terminal block these two exact check patterns are paired
coefficientwise by the composites, so every composite has zero commutator
with the cap; the caps also commute with every internal interblock row and every
installed patch translate, with no dressing by removed checks.  The
construction is \(\mathbb F_2\)-linear in \(a\), up to the scalar phase
convention for Pauli representatives; \(E_{s,a}\) denotes the cap at endpoint
\(s\).
\end{lemma}

\begin{proof}
Put \(c=s_m(a)=\sum_{u}a_u u\), and write \(\doteq\) for equality modulo
Pauli phase.  From Eq.~\eqref{eq:T-definition} and
\(\widehat{\iota(c)}=A^\iota\),
\[
 \prod_{j=p}^{q}\prod_{u\in\Lambda_0}T_{j,u}^{a_u}
 \doteq
 Z\!\left[A^\iota\Bigl(\sum_{j=p}^{q}z^{j-1}\Bigr)(h,v)\right]
 X\!\left[A\Bigl(\sum_{j=p}^{q}z^j\Bigr)
                 (v^\star,h^\star)\right].
\]
Consequently
\[
 E_{L,a}\doteq
 \prod_{j=1-R}^{0}\prod_uT_{j,u}^{a_u},
 \qquad
 E_{R,a}\doteq
 \prod_{j=N}^{N+R-1}\prod_uT_{j,u}^{a_u}.                 \tag{1}
\]

Now \(g v^\star+f h^\star=(f^\star h+g^\star v)^\star=1+z^{-1}\), while
\((1+z)S_R(z)=1+z^R, \; (1+z^{-1})S_R(z)=z^{-1}+z^{R-1}\).
Applying \(\cobdry\) to the two displayed cap formulas therefore gives
exactly Eqs.~\eqref{eq:left-cap-syndrome} and
\eqref{eq:right-cap-syndrome}.  Hence there are no other pristine
syndromes.  At either terminal block the electric pattern is
\(A^\iota=\widehat{\iota(c)}\) and the magnetic pattern is \(A=\widehat c\);
since \(\iota^2=1\), their coefficients cancel pairwise on
\(C_{j,u}=S^X_{j,u}S^Z_{j,\iota(u)}\).  The remote block is pristine, so
these terms are precisely the stated terminal and bulk clusters.

By (1) and Lemma~\ref{lem:selfdual}, each cap commutes with every internal
interblock row.  Moreover,
Theorem~\ref{thm:completion-lift} places every patch representative,
and hence every translate of it, in the centralizer of the infinite
scaffold; by (1), it therefore commutes with both caps.  Finally, all displayed constructions are
\(\mathbb F_2\)-linear in \(c=s_m(a)\), modulo Pauli phase.
\end{proof}

\subsection{Layout, Rectangles and Evaluation}\label{sec:probes}

This subsection fixes the standard configuration of walls and probes, the
period rectangles and their fillings, and the evaluation lemmas used below;
the pair-charge families are assembled in Sec.~\ref{sec:pca}.

\subsubsection{The Standard Configuration}

\paragraph*{The effective constants.}
Each constant listed below is a block radius or support count obtained by
enumerating finite supports fixed by the code, the chosen period ribbons, the
patch family and the certificates $\mathscr C_a$ --- not by $(f,g)$
alone --- so each is computable in finitely many steps and independent of the
wall length; those two properties are all that is used below.  Put
\[
 r_*=\max\{1,r_{\rm row},r_{\rm rib},r_A,r_{\rm cl},r_{\rm deep}\},
 \quad
 m=8r_*+8,
 \quad
 R=3m,
 \quad
 D_{\rm sep}=2R+4r_*+2,
 \quad
 N_{\rm led}=2r_{\rm pkg}+3 ,
\]
with \(r_{\rm pkg}\) the package radius of Eq.~\eqref{eq:endpoint-package-radius}, so that
\(\max A_a-\min A_a\leq2r_{\rm pkg}\) for every offset set of
Definition~\ref{def:natural-termination}; and let
\(R_{\rm led}=\max\{R_0,r_{\rm lift},r_{\rm rel},r_{\rm row},r_C,r_T,
r_{\rm chk}\}\), which bounds every row and bounded relation used in the
ledger collars.

\begin{quote}\small
$r_{\rm row}$, every composite, interblock and patch row, from its carrier;
$r_{\rm rib}$, the support of every basis-labelled period-ribbon copy, from its
block edge; $s_{\rm rib}$, the qubit count of a one-period labelled piece;
$r_A$, the support of $B\Delta^\star$ for every basis label and the plateau depths of Lemma~\ref{lem:plateau-evaluation}; $r_{\rm cl}$, an
exact terminal or remote syndrome cluster, from its block anchor;
$r_{\rm deep}$, the least collar thickness placing a cluster outside every
non-pristine completed-wall row; $r_C$, a composite row; $r_{\rm lift}$, the
patch lifts; $r_{\rm rel}$, the scaffold decomposition of $zM(h_a)M(h_a)$;
$R_{\rm log}$, the transverse width of a labelled period-ribbon seed about its
core, unchanged by concatenation; $r_{\rm chk}$, $r_T$ and
$R_0=2r_{\rm chk}+r_T+1$, the check, interblock and combined radii; and
$r_{\rm cap}$, the terminal collar beyond which the finite Hamiltonian agrees
with the infinite straight-wall family.
\end{quote}

\begin{definition}[Crossing data at depth $N$]\label{def:crossing-constants}
The depth-independent effective constants are fixed by the code and one wall
realization, before any crossing is chosen.  Fix once bases of $V_e$ and
$V_m$ with their compact lifts.  A \emph{crossing datum} is a depth
$N\geq1$ large enough that the
remote clusters at that depth clear the $r_{\rm rel}$-collar of the patch
relations $zM_a=M_a\pmod{\mathcal S_0}$, together with the radii it
then determines: $r_\times$, bounding the dressed depth-$N$ crossings of
Eq.~\eqref{eq:direct-crossing} for the fixed $V_e,V_m$ bases, both directions
and CSS tones, the factors of their prescribed dressings, their exact
syndrome clusters and the rows meeting them, and
\[
 R_{\rm end}=r_{\rm cap}+r_{\rm rel}+r_\times+r_{\rm row}+1,
 \qquad
 D_\times=2(r_\times+r_{\rm row})+1,
\]
the first specifying a minimum distance from an anchor to either endcap,
the second a sufficient anchor spacing for the radius-\((r_\times+r_{\rm row})\)
neighbourhoods centred at distinct anchors to be disjoint.
\end{definition}

\paragraph*{The master range and the room.}
The effective constants, those of the endpoint block
(Sec.~\ref{sec:endpointexact}) and of the finite-window block
(Eq.~\eqref{eq:finite-window-constants}), and the crossing radius $r_\times$ of
Definition~\ref{def:crossing-constants} are collected into a single scale:
\begin{equation}\label{eq:master-range}
 \rho_{\max}
 =\max\bigl\{1,\;
   r_{\rm chk},\,r_T,\,d_T,\,r_C,\,r_{\rm lift},\,r_{\rm rel},\,r_{\rm row},\,
   r_{\rm rib},\,s_{\rm rib},\,r_A,\,r_{\rm cl},\,r_{\rm deep},\,
   r_{\rm pkg},\,R_{\rm log},\,
   r_{\rm cap},\,r_\times\bigr\}.
\end{equation}
Fix the depth-independent constants from the microscopic realization, and
compute \(r_*\), \(m\), \(R\) and \(D_{\rm sep}\).  Then fix a crossing datum
(Definition~\ref{def:crossing-constants}) whose depth \(N_\times\) exceeds
\(C_\times=\max\{m+r_A+r_{\rm cl}+r_{\rm deep},\,r_{\rm row},\,r_{\rm cap}\}+r_{\rm rib}\);
it determines \(r_\times\), \(R_{\rm end}\) and \(D_\times\).  Form \(\rho_{\max}\)
after these choices.  The
collar radius obeys \(r_{\rm deep}\leq r_{\rm row}+r_{\rm cl}+1\), since every
non-pristine row lies within \(r_{\rm row}\) of the wall carrier and every
cluster within \(r_{\rm cl}\) of its anchor.

\begin{definition}[Standard configuration]
\label{def:standard-layout}
With the crossing datum just fixed, a \emph{standard configuration at room
$\rho$} on $n\geq1$ walls consists of the following, on a commensurate torus
\(\mathbb T=\mathbb Z_{L_1}\times\mathbb Z_{L_2}\).
\begin{enumerate}
\item[\textup{(S1)}]
\emph{Walls in one frame.}  The walls are $n$ disjoint, naturally terminated
walls of complete period blocks, all of one orientation, in one global Laurent
origin, one period-block lattice and one affine monomial \(\kappa\), so that
the composite matching on every wall imposes the same affine diagonal
\(p=\kappa q^\star\) on pristine torus relations; wall $s$ is the single-wall
construction translated by the monomial \(o_s\) of its first block, with the
same patch convention.  The period ribbons
\[
 f^\star h+g^\star v=1+z,
 \qquad
 f^\star c+g^\star d=1+w,
 \qquad
 \Delta=cv+dh
\]
are those of the single-wall construction, \(z\) being the block translation
along the walls and \(w\) the transverse one.

\item[\textup{(S2)}]
\emph{Probes.}  Around wall $t$ lies the period rectangle \(D_t=N_m(C_t)\),
and, on one endpoint side common to all walls, the exact cap of
Lemma~\ref{lem:endpoint-caps} at depth \(R\).

\item[\textup{(S3)}]
\emph{Transfer routes} ($n\geq2$).  Wall $n$ is the \emph{reference}.  Each
$s<n$ has designated anchors $\tau_s$ on wall $s$ and $\tau_{n,s}$ on wall $n$,
the latter distinct for distinct $s$; write $\tau_u$ for either.  Each
$\tau_u$ has a crossing direction $w_u\in\{w,w^{-1}\}$ and remote anchors
\(\rho^{e}_{u}=\tau_u w_u^{N_\times}\) and
\(\rho^{m}_{u}=\tau_u w_u^{-N_\times}\).  For each $s<n$ and each tone
$\chi\in\{e,m\}$, a label-independent axis-aligned block path joins
\(\rho^{\chi}_s\) to \(\rho^{\chi}_{n,s}\), its $r_{\rm rib}$-tube outside
every \(N_{r_A+r_{\rm cl}+r_{\rm deep}}(D_t)\) and every non-pristine row.  The
two paths and the two transverse segments from \(\rho^m_u\) to \(\rho^e_u\)
form the closed route $\gamma_s$, which therefore crosses wall $s$ and the
reference wall once each and no other wall.

\item[\textup{(S4)}]
\emph{One region.}  A block rectangle $U$ with a non-wrapping lift contains the
$\rho_{\max}$-neighbourhood of every wall carrier, rectangle, cap and route,
and a horizontal and a vertical coordinate core cycle $\gamma_x$, $\gamma_y$,
meeting once, lie in \(\mathbb T\setminus U\).

\item[\textup{(S5)}]
\emph{Room.}  Every wall has at least $\rho$ blocks, and the block distance
between distinct wall carriers, from a lift of a wall carrier to its other
periodic images, between distinct designated anchors, from a designated anchor
to an endcap, and from $U$ to \(\gamma_x\cup\gamma_y\) is at least $\rho$.
\end{enumerate}
A \emph{standard configuration} is one at room \(\rho_0\), a constant fixed
once, depending only on the code and the choices above, not on the torus or
on the number, lengths or positions of the walls.  It is taken large enough for
the lower bounds \(\rho_0\geq X\) used in the proofs below, finitely many, each
\(X\) an expression in those constants alone.
\end{definition}

\begin{figure}[htbp]
    \centering
    \includegraphics{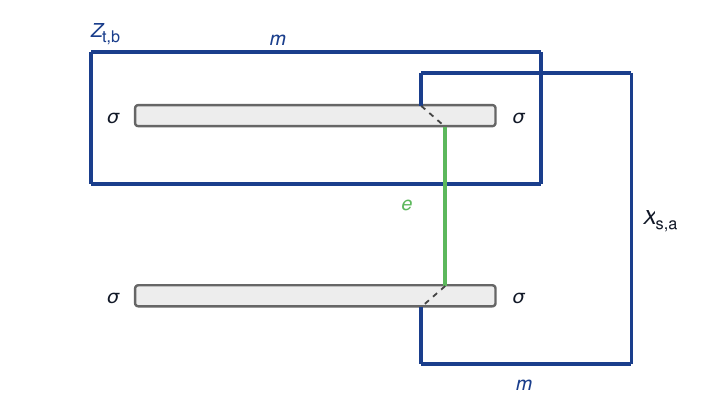}
    \caption{The two families of Definition~\ref{def:pca-families}; colours as in
    Fig.~\ref{fig:Tmicro}.  Grey bars are two of the walls, their rounded ends
    the twists $\sigma$.  The charge loop $Z_{t,b}$ encircles one whole wall
    and crosses none, so it is a closed pristine loop with an explicit filling.
    The transfer loop $X_{s,a}$ must separate one twist of each wall from the
    other two, so it crosses wall $s$ and the reference wall once each; at each
    crossing the ribbon is broken and the grey dashed segment marks the
    $\iota$-matched endpoint pair the composite checks absorb, the label passing from
    $m_a$ to $e_{a^\iota}$ and back.  Those two crossings carry its only
    ledgers (Lemma~\ref{lem:pca-ledger}); their placement is what
    Lemma~\ref{lem:room}\textup{(L6)} records, and Theorem~\ref{thm:pca-weyl} evaluates the
    resulting pairing.}
    \label{fig:AdjacentPairWeyl}
\end{figure}

Every wall pairs sites by the same involution \(\iota\), so a single label
group acts on every wall at once and a charge carried from one wall to another
keeps its name.  Room says that, far from the boundary, nothing local can tell
that a finite wall is finite; far from the machinery, charge travels on
pristine paths clear of the rows that build the walls; and far from each
other, distinct twist pairs are addressable one at a time.
Figure~\ref{fig:AdjacentPairWeyl} sketches a probe rectangle and a transfer
route for two walls.

\paragraph*{Construction.}
Standard configurations exist for every \(n\geq1\) and every \(\rho\geq\rho_0\).
Place $n$ walls of at least $\rho$ blocks in order on one $z$-line, with gaps
of at least $\rho$ blocks and the reference wall last; for \(n\geq2\) put one
designated anchor on each other wall and \(n-1\) on the reference wall,
pairwise at least $\rho$ apart and at least $\rho$ from the endcaps, all with
\(w_u=w\).  Take \(D_t=N_m(C_t)\) and the depth-\(R\) caps on the side of
increasing $z$, which run straight into the gaps; join the remote anchors
\(\tau_uw^{\pm N_\times}\), which lie on two exterior block lines, along those
lines; enclose everything in a block rectangle \(U\); and enlarge the torus
until \(U\) injects with core cycles at distance \(\rho\).  Since
\(N_\times>C_\times\), every tube and remote cluster of the routes clears every
\(N_{r_A+r_{\rm cl}+r_{\rm deep}}(D_t)\) and every non-pristine row, the
clusters by the bank-distance bound of the proof of
Theorem~\ref{thm:transmission}.  In the plane of $U$, the route $\gamma_s$
encloses the facing ends of wall $s$ and of the reference wall, the walls
between them, and no other endpoint.  For $n=2$, numbering the twists
$1,2,3,4$ along the line, $\gamma_1$ separates $\{2,3\}$ from $\{1,4\}$: it is
the transfer route \emph{realizing the $(23)(14)$ cut}.

\begin{lemma}[What room provides]
\label{lem:room}
In a standard configuration the following hold.
\begin{enumerate}
\item[\textup{(L1)}]
\emph{Collars.}  Each wall $s$ has \(N_s\geq N_{\rm led}\) blocks and a
non-wrapping \(R_{\rm led}\)-collar \(\Omega_s\) of its carrier; the collars
are pairwise disjoint; each \(J_{s,a}\) of Eq.~\eqref{eq:exact-retention} is
nonempty; and \(\Omega_s\) contains the support of every operator in
\(o_sz^j\mathscr P_a\) for every \(j\in J_{s,a}\), so each certificate row for
\(M_{j+1,a}M_{j,a}\) is a legal finite-wall row supported in \(\Omega_s\).

\item[\textup{(L2)}]
\emph{Probe clearance.}  For \(u\neq t\),
\(\operatorname{dist}_{\rm blk}(D_t,C_u)>\max\{r_{\rm rib}+r_{\rm row},\,
r_A+r_{\rm cl}+r_{\rm deep}\}\); each cap leaves its own \(D_t\) once, and its
remote cluster lies outside every \(N_{r_A+r_{\rm cl}+r_{\rm deep}}(D_u)\).

\item[\textup{(L3)}]
\emph{Complement clearance.}
\begin{equation}
 N_{R_{\rm log}}(\gamma_x\cup\gamma_y)\subseteq\mathbb T\setminus U.
 \label{eq:complement-clearance}
\end{equation}

\item[\textup{(L4)}]
\emph{Crossing windows.}  At every designated anchor \(\tau_s\),
\(\Omega_{s,\rm in}=N_{L_\times^{\rm room}}(\{\tau_s\})\) and
\(\Omega_{s,\rm out}=N_{\rho_{\rm win}}(\Omega_{s,\rm in})\) form a crossing
window (Definition~\ref{def:qualifying-crossing-window}).

\item[\textup{(L5)}]
\emph{Isolation.}  Each \(\Omega_{s,\rm out}\) is disjoint from every other
wall's non-pristine and terminal collars, and no check retained at one anchor
is removed at another.

\item[\textup{(L6)}]
\emph{Ledger placement.}  For each $t<n$, the wall-$t$ ledger of $\gamma_t$
 is anchored in the $(r_A{+}1)$-plateau of $D_t$; and every other ledger --- the
 reference-wall ledger of $\gamma_t$, and both ledgers of every $\gamma_s$ with
 $s\ne t$ --- has \emph{electric} support disjoint from
 $\operatorname{supp}A_{t,b}$ for every $b$.  At the reference wall the
 ledger of every $\gamma_s$ at $\tau_{n,s}$ is anchored in the
 $(r_A{+}1)$-plateau of $D_n$, and the wall-$s$ ledgers' electric supports miss
 $\operatorname{supp}A_{n,b}$.  The magnetic halves are
 unconstrained: the filling has only its second component nonzero, so the
 pairing never reads them.

\item[\textup{(L7)}]
\emph{Endpoint windows.}  At every endpoint $s$, the block containing $s$ is an
endpoint window (Definition~\ref{def:hull-controlled-window}).
\end{enumerate}
\end{lemma}

\begin{proof}
Each bound \(\rho_0\geq X\) below is one for which \(\rho_0\) was taken
large, and \textup{(S5)} transfers it to a wall length or a separation.  The
core cycles \(\gamma_x\) and \(\gamma_y\) lie at distance at least
\(\rho_0\) from $U$, which contains a wall block, so both torus periods exceed
\(2\rho_0\) by \textup{(S4)} and \textup{(S5)}; hence the block
\(\ell^\infty\)-ball of radius below \(\rho_0\) about a block of a lift projects
injectively to the torus.

\textup{(L1)}: \(N_s\geq\rho_0\geq N_{\rm led}\); carriers at least
\(\rho_0\geq2R_{\rm led}+1\) apart, and from their own images, have disjoint
non-wrapping collars; \(J_{s,a}\) is a nonempty interval because
\(\max A_a-\min A_a\leq2r_{\rm pkg}<N_{\rm led}\); and \(R_{\rm led}\) bounds every row and bounded
relation of the packages.  \textup{(L2)}: \(D_t=N_m(C_t)\) and carriers at least
\(\rho_0\geq D_{\rm sep}\) apart give the first clause.  A cap is a straight ribbon from
the endpoint to depth \(R\), so it leaves \(D_t\) once at depth \(m\); its remote
cluster lies \(R-m=2m>r_A+2r_{\rm cl}+r_{\rm deep}\) beyond \(D_t\), and at least
\(D_{\rm sep}-R-m=2m+4r_*+2\) from every foreign \(D_u\).  \textup{(L3)}:
\(\rho_0\geq R_{\rm log}+1\).  \textup{(L4)}: the wall has at least
\(\rho_0\geq N_{\rm win}\) blocks; each anchor is at least
\(\rho_0\geq L_\times^{\rm room}+\rho_{\rm win}+N_{\rm win}\) from the endcaps
and from every foreign carrier, so \(\Omega_{s,\rm out}\), a ball of radius
\(L_\times^{\rm room}+\rho_{\rm win}<\rho_0\) about \(\tau_s\), lies in the straight-wall interior
with both endcaps at least \(N_{\rm win}\) blocks away, and it lifts without
wrapping.  \textup{(L5)}: outer windows have radius
\(L_\times^{\rm room}+\rho_{\rm win}\) about their anchors, which are at least
\(\rho_0\geq2(L_\times^{\rm room}+\rho_{\rm win}+r_{\rm chk})+1\) from each other
and \(\rho_0\geq L_\times^{\rm room}+\rho_{\rm win}+\max\{R_{\rm led},r_{\rm cap}\}+1\)
from foreign carriers, while the non-pristine collars have radius at most
\(R_{\rm led}\) and the terminal collars \(r_{\rm cap}\) about the carrier, and
\(\rho_0\geq\max\{R_{\rm end},D_\times\}\).
\textup{(L6)}: every anchor lies on its carrier, hence in the
\((r_A+1)\)-plateau of \(D_t=N_m(C_t)\), since \(m>r_A+1\); this depth is the
label-uniform bound \(d_q\leq r_A+1\) of
Lemma~\ref{lem:plateau-evaluation}\textup{(iii)}.  Every other ledger lies
within \(r_\times\) of an anchor at least \(\rho_0\geq r_\times+m+r_A+1\) from
\(C_t\), hence off \(\operatorname{supp}A_{t,b}\subseteq N_{r_A}(D_t)=N_{m+r_A}(C_t)\).  \textup{(L7)}: the wall
has at least \(\rho_0\geq N_{\rm end}\) blocks, so the endpoint block is at
least \(\rho_{\rm end}\) from the opposite endpoint, and its
\(\rho_{\rm end}\)-neighbourhood lifts without wrapping and, by
\(\rho_0\geq\rho_{\rm end}+r_{\rm chk}+1\), contains no removed check of another
wall.
\end{proof}

\begin{remark}[What the conditions exclude]
On a \emph{fixed} torus these conditions bound how many walls of a given length
fit.  Without the foreign-carrier clearance of \textup{(L2)}, foreign syndrome
clusters could enter the commutator of \(Z_{t,b}\), breaking
Lemma~\ref{lem:pca-pairing}, or \(Z_{t,b}\) could fail to centralize a foreign
wall row; without the common \(\kappa\) of \textup{(S1)},
Definition~\ref{def:pca-families} supplies transfers only for labels with
\(\iota_s(a)=\iota_n(a)\).
\end{remark}

\subsubsection{Pair Loops and Corridor Edge Pieces}

\paragraph*{Construction.}
The construction uses the layout, period ribbons and label section of the Setting.

For \(k\in\mathbb Z\), define the finite Laurent polynomial
\[
 S_k(u)=\frac{1+u^k}{1+u},
\]
so \(S_k=1+\cdots+u^{k-1}\) for \(k>0\), \(S_0=0\), and
\(S_k=u^k+\cdots+u^{-1}\) for \(k<0\).

Let the lower-left anchor and side lengths of \(D_t\) be
\(\rho_t,z^{L_t},w^{W_t}\). Define the logically named \(Z\)-operator,
which is physically pure \(X\)-tone (pure magnetic), by
\begin{align}
 Z_{t,b}
 =X\!\Big[B\rho_t\big(&zS_{L_t}(z)(1+w^{W_t})(v^\star,h^\star)
 \notag\\[-2mm]
 &+wS_{W_t}(w)(1+z^{L_t})(d^\star,c^\star)\big)\Big].
 \label{eq:period-rectangle-loop}
\end{align}
Equivalently, this is the product of the four labelled magnetic
period-ribbon sides of \(\partial D_t\). Its pristine filling is explicitly
\begin{equation}
 A_{t,b}
 =\left(0,
 B\rho_tzwS_{L_t}(z)S_{W_t}(w)\Delta^\star\right)\in G,
 \qquad
 Z_{t,b}=\bdry A_{t,b}.
 \label{eq:period-rectangle-filling}
\end{equation}

Let \(a\in V_m\), put
\(A=\widehat{s_m(a)}\) and \(A^\iota=\kappa A^\star\). For an oriented block edge
\(\rho\to\rho u^k\), \(u\in\{z,w\}\), put \((p_z,q_z)=(h,v)\) and
\((p_w,q_w)=(c,d)\), and define
\begin{align}
 P_e(\rho,u,k;A^\iota)
 &=Z\!\left[\rho A^\iota S_k(u)(p_u,q_u)\right],
 \\
 P_m(\rho,u,k;A)
 &=X\!\left[\rho A\,uS_k(u)(q_u^\star,p_u^\star)\right].
 \label{eq:corridor-edge-pieces}
\end{align}

\begin{lemma}[Period-rectangle pair loops and corridor edge pieces]\label{lem:period-rectangle-corridors}
With the objects just constructed, the following hold.

\textup{(i) The pair loops.}  For \(b\ne0\), \(Z_{t,b}\) is a compact contractible closed pristine
period-ribbon loop on the carrier \(\partial D_t\), enclosing wall \(t\) and
no other wall. It commutes with every row of the completed finite-wall
Hamiltonian. Its support lies in
\(N_{r_{\rm rib}}(\partial D_t)\), and the support of its filling lies in
\(N_{r_A}(D_t)\), with the effective weight bound
\(\operatorname{wt}Z_{t,b} \le 2s_{\rm rib}(L_t+W_t)\).
Hence the terminal cluster of wall \(t\) is
\(r_{\rm deep}\)-deep in the filling carrier \(D_t\), whereas every other wall
cluster and every remote cap cluster is outside the
\(r_{\rm deep}\)-collar of \(\operatorname{supp}A_{t,b}\). The formulas are linear
in \(b\). For \(b=0\), one has \(Z_{t,0}=I\) (the zero binary Pauli) and
\(A_{t,0}=0\); the same geometric carrier is retained only as a routing
convention.

\textup{(ii) Edge pieces and paths.}  Each labelled edge piece
\(P_e(\rho,u,k;A^\iota)\) and \(P_m(\rho,u,k;A)\) of
Eq.~\eqref{eq:corridor-edge-pieces} has exact syndrome only at the two endpoints
of its edge.  Hence the product of edge pieces along an axis-aligned block path
has exact syndrome only at the two ends of the path: every internal vertex
occurs in two consecutive pieces and cancels over \(\mathbb F_2\).  This holds
tonewise --- for either tone separately --- and neither the choice of path nor
its length enters.
\end{lemma}

\begin{proof}
The identities \((1+u)S_k(u)=1+u^k\) and \(gu(q_u^\star)+fu(p_u^\star)=1+u\)
show that each labelled edge piece of Eq.~\eqref{eq:corridor-edge-pieces} has
syndrome only at the two endpoints of its edge, which gives \textup{(ii)}, and
that the four factors of Eq.~\eqref{eq:period-rectangle-loop} have cancelling
corner syndromes.  For the filling, the antipodes of the two period equations,
\(fh^\star+gv^\star=1+z^{-1}\) and \(fc^\star+gd^\star=1+w^{-1}\), give
\begin{equation}
 z(1+w)(v^\star,h^\star)
 +w(1+z)(d^\star,c^\star)
 =zw\Delta^\star(f,g),
 \label{eq:rectangle-adjugate-identity}
\end{equation}
since the first component is
\(zw[(fc^\star+gd^\star)v^\star+(fh^\star+gv^\star)d^\star]=zwf\Delta^\star\),
the cross terms cancelling in characteristic two, and the second is
\(zwg\Delta^\star\).  Multiplying by \(B\rho_tS_{L_t}(z)S_{W_t}(w)\) and using
\(\bdry(0,P)=(0,0\mid fP,gP)\) gives Eq.~\eqref{eq:period-rectangle-filling}:
a finite boundary in the non-wrapping lift, hence contractible, and the unique
compact filling by Lemma~\ref{lem:compact-filling}.  The support tube, the
weight bound (\(s_{\rm rib}\) qubits per period piece on the four sides) and
the filling collar are read off the two formulas.  The \(m\)-deep placement puts
the terminal cluster of wall \(t\) in the eroded rectangle and, since
\(m>r_{\rm row}+r_{\rm rib}\), separates the loop tube from the interblock and
patch rows of wall \(t\); the foreign-carrier clearance of \textup{(L2)}
separates it from every foreign wall-row tube and places every foreign terminal
cluster outside \(N_{r_A+r_{\rm deep}}(D_t)\); and the remote-cluster exclusions
are those of \textup{(L2)}.  So all inside/outside assertions follow from these
support inclusions.  As \(Z_{t,b}=\bdry A_{t,b}\), it also commutes with every
pristine check, retained row and composite.  Every formula is additive in the
label section.
\end{proof}

\begin{definition}[Block index and canonical representatives]
\label{def:block-index}
Every monomial of $R$ factors uniquely as
\[
 x^{al_0}y^{bm_0}\,x^uy^v,
 \qquad
 a,b\in\mathbb Z,
 \quad
 0\leq u<l_0,
 \quad
 0\leq v<m_0 ,
\]
its \emph{canonical within-block representative}; $(a,b)$ is its \emph{block
index}, and the block $\ell^\infty$-radius of a finite Laurent polynomial is the
largest $\max\{|a|,|b|\}$ over its monomials; the zero polynomial has block
radius $0$.  The radii $d_q$ and the plateau depths below use these block indices.
Choose $r_A$ to bound, for every basis label, both the block-index
radius and the ceiling of the geometric block-metric radius of
$B\Delta^\star$, and $d_q-1$ for every $q\in\mathsf A$, with $d_q$ as in
Lemma~\ref{lem:plateau-evaluation}; these are finitely many finite radii. Geometric support distances use the block metric
defined before Corollary~\ref{cor:buffered-injectivity}.
\end{definition}

\begin{definition}[Filling plateau]
\label{def:filling-plateau}
Let $D_t$ be a period rectangle with lower-left block anchor $\rho_t$ and side
lengths $z^{L_t},w^{W_t}$, as in the construction of
Lemma~\ref{lem:period-rectangle-corridors}.  For an integer $d\geq0$, the
\emph{$d$-plateau} of $D_t$ is the set of block positions
\[
 \tau=\rho_tz^{\,i+1}w^{\,j+1},
 \qquad
 d\leq i\leq L_t-1-d,
 \quad
 d\leq j\leq W_t-1-d .
\]
A cluster anchored in the $d$-plateau lies \emph{$d$-deep inside} the filling
of $D_t$.
\end{definition}

\begin{lemma}[Plateau evaluation]
\label{lem:plateau-evaluation}
Fix $b\in V_m$, write $B=\widehat{s_m(b)}$, and let $A_{t,b}$ and
$Z_{t,b}=\bdry A_{t,b}$ be the period-rectangle filling and loop of
Eq.~\eqref{eq:period-rectangle-filling}.  Let $q\in\mathsf A$ and write $\widehat q=\sum_{0\leq u<l_0,\,0\leq v<m_0}
q_{uv}x^uy^v\in R$ for its standard representative, which has block radius $0$
by Definition~\ref{def:block-index}.  Let $\tau$ be a block position and let
$p$ be any finite Pauli whose exact electric syndrome is the polynomial
$\widehat q\tau$, supported in a single block when $q\neq0$.  Let $d_q$ be the
block $\ell^\infty$-radius of $\widehat qB^\star\Delta$.  Then:

\textup{(i)} if $\tau$ lies in the $d_q$-plateau of $D_t$,
\begin{equation}
 \left\langle Z_{t,b},p\right\rangle
 =\operatorname{Tr}_{z,w}\!\left(\widehat qB^\star\Delta\right);
\label{eq:plateau-evaluation}
\end{equation}

\textup{(ii)} if $\operatorname{supp}(\widehat q\tau)$ is disjoint from
$\operatorname{supp}A_{t,b}$, then $\left\langle Z_{t,b},p\right\rangle=0$;

\textup{(iii)} $d_q\leq r_A+1$ by the choice of $r_A$ in
Definition~\ref{def:block-index}, uniformly in $q$ and $b$.

\textup{(iv)} \emph{Summand-wise.}  If the exact electric syndrome of $p$ is
$\sum_j\widehat q_j\tau_j$, with each $q_j\in\mathsf A$ and each $\tau_j$ a block
position, and each summand falls under \textup{(i)} or under \textup{(ii)} with
its own $d_{q_j}$, then
$\left\langle Z_{t,b},p\right\rangle=\sum_j\varepsilon_j$, with
$\varepsilon_j=\operatorname{Tr}_{z,w}(\widehat q_jB^\star\Delta)$ in the first
case and $\varepsilon_j=0$ in the second.

The magnetic component of the syndrome of $p$ never enters, whatever its
position.
\end{lemma}

\begin{proof}
Only the second component of $A_{t,b}$ can be nonzero, and the exchange of
Eq.~\eqref{eq:adjoint-swap} pairs it against the electric syndrome, so
Lemma~\ref{lem:closed-loop-interior} gives
$\langle Z_{t,b},p\rangle=\tr(F^\star\widehat q\tau)$ with
$F=B\rho_tzwS_{L_t}(z)S_{W_t}(w)\Delta^\star$, whatever the magnetic syndrome.
This is an identity in the electric syndrome alone: disjoint support gives
\textup{(ii)}, and linearity gives \textup{(iv)}, whose summands need not
themselves be syndromes of finite Paulis.  For \textup{(i)} put
$C=\widehat qB^\star\Delta$ and $\tau=\rho_tz^{\,i+1}w^{\,j+1}$; then
$\tr(F^\star\widehat q\tau)=\sum_{\ell=0}^{L_t-1}\sum_{k=0}^{W_t-1}\tr(Cz^{\,i-\ell}w^{\,j-k})$,
and the index window $[i-L_t+1,i]\times[j-W_t+1,j]$ contains $[-d_q,d_q]^2$
exactly when $\tau$ lies in the $d_q$-plateau, so each block translate
contributing to $\operatorname{Tr}_{z,w}(C)$ is counted once.  Part \textup{(iii)} holds by definition.
\end{proof}

\subsection{The Pair-Charge Algebra}\label{sec:pca}

A standard configuration of $n\geq2$ walls carries two families of
finite Pauli operators: one that reads the charge of a twist pair, and one that
moves charge between a pair and a fixed reference pair.  Together they are a
Pauli register on the fusion space.  Theorem~\ref{thm:pca-register} gives that
register $2(n-1)Q$ independent logical Pauli classes with a non-degenerate form;
Theorem~\ref{thm:pca-weyl} gives its commutation phase, the braiding form
composed with the affine involution.  Sec.~\ref{sec:exhaustion} uses the first,
Sec.~\ref{sec:TY} the second.

The geometric input is the route clause \textup{(S3)} of
Definition~\ref{def:standard-layout} and items \textup{(L4)}--\textup{(L6)} of
Lemma~\ref{lem:room}.

\subsubsection{The Register and Its Phase}

\begin{definition}[Charge loops and transfer loops]\label{def:pca-families}
In a standard configuration with $n\geq2$, write $A=\widehat{s_m(a)}$ and let the crossing
at an anchor $\tau_u$ be $W_u(a)=W^{(N_\times)}_{u,0}(a)D^{\rm fin}_u(a)$, with
bare part
\begin{equation}
 W^{(N_\times)}_{u,0}(a)
 =Z\!\left[\tau_uA^\iota(c_{u,N},d_{u,N})\right]
  X\!\left[\tau_uA(d_{u,N}^\star,c_{u,N}^\star)\right],
\label{eq:direct-crossing}
\end{equation}
where $w_u$ points from the magnetic to the electric remote bank,
$(c_{u,N},d_{u,N})$ is $S_{N_\times}(w)(c,d)$ for $w_u=w$ and
$w^{-N_\times}S_{N_\times}(w)(c,d)$ for $w_u=w^{-1}$, with $c,d$ the common
transverse ribbon of \textup{(S1)}.  For both seeds, with interblock patterns those
of Lemma~\ref{lem:affine-crossing} for $\Delta_u=c_{u,N}v+d_{u,N}h$, the proof of
Theorem~\ref{thm:transmission} gives compact linear dressings and patch
commutation at the datum's common depth; $r_\times$ bounds them in both
directions and tones, so the crossing window of \textup{(L4)} contains them and
Lemma~\ref{lem:finite-window-straight-wall-agreement} carries them to the
finite wall.  The crossing has dressing a
product of removed pristine checks, so that
\begin{equation}
 \cobdry W^{(N_\times)}_{u,0}(a)
 =\left(
 \tau_uA^\iota(1+w_u^{N_\times}),\;
 \tau_uA(1+w_u^{-N_\times})
 \right).
\label{eq:direct-crossing-syndrome}
\end{equation}
For $s<n$ and $a\in V_m$ put
\[
 X_{s,a}=W_s(a)\,R_e(a)\,W_{n,s}(a)\,R_m(a),
\]
the closed loop along $\gamma_s$, in which $W_s(a)$ and $W_{n,s}(a)$ are the
crossings at $\tau_s$ and $\tau_{n,s}$, and $R_e(a)$ and $R_m(a)$ are the
products of the standard edge pieces of Eq.~\eqref{eq:corridor-edge-pieces}
along the two paths of \textup{(S3)}, joining their like-tone remote clusters.
\end{definition}

\begin{theorem}[Pair-charge register]\label{thm:pca-register}
Fix an $\mathbb F_2$-basis $\mu_1,\dots,\mu_Q$ of $V_m$.  In a standard
configuration the $2(n-1)Q$ classes
\[
 [Z_{t,\mu_i}],\qquad [X_{s,\mu_i}],
 \qquad s,t<n,\quad 1\le i\le Q,
\]
are linearly independent in $\mathcal L(H_{\rm fin})$, and the binary symplectic
form on their span is non-degenerate.
\end{theorem}

\begin{theorem}[Pair-charge Weyl relation]\label{thm:pca-weyl}
In a standard configuration, for all $s,t<n$ and all $a,b\in V_m$,
\[
 Z_{t,b}\,X_{s,a}
 =\lambda(a,b)^{\delta_{st}}\,X_{s,a}\,Z_{t,b},
 \qquad
 \lambda(a,b)=\B{a}{b^{\iota}}.
\]
\end{theorem}

Both follow from one pairing computation.  Neither asserts that the classes are
canonical: both hold for one fixed admissible construction, and no invariance
under a change of crossing depth, corridor routing or rectangle is claimed.  The
group law of the transfer family is established in Sec.~\ref{sec:TY}, where it
is used.

\begin{lemma}[Exact ledger]\label{lem:pca-ledger}
Write $\widehat a=\widehat{s_m(a)}$ for the standard representative of the
label section, and $\widehat{a^{\iota}}=\kappa\widehat a^{\,\star}$ for its
affine partner, as in Eq.~\eqref{eq:iota-label}; these are the exact lifts the
crossing construction uses, not arbitrary representatives of the classes.  In a
standard configuration,
\[
 \cobdry X_{s,a}=\Lambda_s(a)+\Lambda_n(a),
 \qquad
 \Lambda_u(a)=\left(\widehat{a^{\iota}}\,\tau_u,\;\widehat a\,\tau_u\right),
\]
with $\tau_s,\tau_n$ the two crossing anchors of $\gamma_s$.  $X_{s,a}$ has no
other exact syndrome cluster.
\end{lemma}

\begin{proof}
By Definition~\ref{def:pca-families} each crossing is the dressed representative whose bare part
Eq.~\eqref{eq:direct-crossing} has exact syndrome, by
Eq.~\eqref{eq:direct-crossing-syndrome},
$\left(\widehat{a^{\iota}}\tau_u(1+w_u^{N_\times}),\,
\widehat a\tau_u(1+w_u^{-N_\times})\right)$: a ledger at the anchor and one
remote cluster displaced transversally, in each tone.  The transmission dressing
supplied there is a product of removed pristine checks and has zero pristine
syndrome, so it does not contribute.

By \textup{(S3)} each corridor is a product of the standard edge pieces of
Eq.~\eqref{eq:corridor-edge-pieces} along an axis-aligned path, so
Lemma~\ref{lem:period-rectangle-corridors}\textup{(ii)} applies to it directly,
once to $R_e(a)$ on the electric anchors and once to $R_m(a)$ on the magnetic
ones: each has no wall syndrome and no exact syndrome cluster beyond its two
ends.  The four displaced terms then cancel against the crossings' remote
clusters and the two anchor terms survive.  Which path is chosen never enters.
\end{proof}

\begin{lemma}\label{lem:pca-logical}
Every $Z_{t,b}$ and every $X_{s,a}$ is a finite Pauli commuting with every row of
the completed finite-wall Hamiltonian.
\end{lemma}

\begin{proof}
For $Z_{t,b}$ with $b\ne0$ this is
Lemma~\ref{lem:period-rectangle-corridors}\textup{(i)}; for $b=0$ it is trivial.  For $X_{s,a}$, Definition~\ref{def:pca-families} fixes the dressed
crossings constructed in the proof of Theorem~\ref{thm:transmission}.
Their supports lie in the crossing windows of \textup{(L4)}, so
Lemma~\ref{lem:finite-window-straight-wall-agreement} carries their
commutation with the straight-wall rows to the installed finite-wall
rows. The isolation \textup{(L5)} excludes foreign wall rows, and
the corridors are pristine and clear of every non-pristine row by
\textup{(S3)}.  By Lemma~\ref{lem:pca-ledger} what remains is
supported on removed checks, which the completed Hamiltonian does not contain.
\end{proof}

\begin{lemma}[Loop--transfer pairing]\label{lem:pca-pairing}
For all $s<n$, all $t\le n$ and all $a,b\in V_m$,
\[
 \left\langle Z_{t,b},X_{s,a}\right\rangle
 =\begin{cases}
   \phi(a,b^{\iota}) & t=s \text{ or } t=n,\\
   0 & \text{otherwise.}
  \end{cases}
\]
\end{lemma}

\begin{proof}
For $b=0$ both sides vanish, so take $b\ne0$.  By
Lemma~\ref{lem:closed-loop-interior} the pairing is
$\tr(A_{t,b}^{\dagger}S\,\cobdry X_{s,a})$, so by Lemma~\ref{lem:pca-ledger} it
reads the two ledgers of $X_{s,a}$ against $A_{t,b}$ and nothing else.  That
syndrome is a sum of two clusters, not one, so the evaluation used is
Lemma~\ref{lem:plateau-evaluation}\textup{(iv)} and not \textup{(i)} itself.  The
filling has only its second component nonzero, so only the electric half
$\widehat{a^{\iota}}\tau_u$ of each ledger contributes.

For $t<n$, every ledger other than the wall-$t$ ledger of $\gamma_t$ has electric support
disjoint from $\operatorname{supp}A_{t,b}$ by \textup{(L6)}, so
Lemma~\ref{lem:plateau-evaluation}\textup{(iv)} makes its contribution zero; when $s\ne t$ both
ledgers of $X_{s,a}$ are of that kind and the pairing vanishes.  For $t=n$ it is
the reference-wall ledger of $\gamma_s$ that \textup{(L6)} anchors in the
plateau of $D_n$, for every $s$, while the electric support of the wall-$s$ ledger misses
$\operatorname{supp}A_{n,b}$; so exactly one ledger contributes in that case
too, which is why the value does not depend on $s$ there.

In the contributing case the ledger is anchored in the $(r_A{+}1)$-plateau of its
rectangle by \textup{(L6)}, and $d_q\le r_A+1$ by
Lemma~\ref{lem:plateau-evaluation}\textup{(iii)}, so
part~\textup{(iv)} evaluates that contribution as
$\operatorname{Tr}_{z,w}(\widehat{a^{\iota}}\widehat b^{\,\star}\Delta)$, which
is $\phi(b,a^{\iota})$ by Eq.~\eqref{eq:direct-braiding-form}, the period
ribbons of \textup{(S1)} satisfying
Eq.~\eqref{eq:parallel-transverse-ribbons}; and
$\phi(b,a^{\iota})=\phi(a,b^{\iota})$ by
Eq.~\eqref{eq:phi-affine-exchange}.
\end{proof}

\begin{proof}[Proof of Theorem~\ref{thm:pca-weyl}]
For binary Paulis $PQ=(-1)^{\langle P,Q\rangle}QP$.  Insert
Lemma~\ref{lem:pca-pairing} and $\mathcal B=(-1)^{\phi}$.  The identification
of Sec.~\ref{sec:TY} reads $\chi=\lambda$.
\end{proof}

\begin{corollary}[Action on pair-charge labels]\label{cor:pca-charge-action}
Let $P_{s,c}$ be the spectral projector of the commuting family
$\{Z_{s,b}\}_{b\in V_m}$ on the character $c$, as
Lemma~\ref{lem:pair-loop-characters} normalizes it, and
$P_{\mathbf c}=\prod_{s<n}P_{s,c_s}$.  Then
\[
 X_{s,a}\,P_{\mathbf c}\,X_{s,a}^{-1}=P_{\mathbf c+a\mathbf e_s},
\]
so $X_{s,a}$ shifts the label of pair $s$ by $a$ and, by
Lemma~\ref{lem:pca-pairing} at $t=n$, the reference label with it, leaving the
total charge fixed.
\end{corollary}

\begin{proof}
The projectors are the spectral projectors of the commuting family
$\{Z_{s,b}\}$, so it suffices to conjugate a character.  By
Theorem~\ref{thm:pca-weyl}, conjugation by $X_{s,a}$ multiplies the $Z_{t,b}$
eigenvalue by $\lambda(a,b)^{\delta_{st}}$, which is the character of the
label shifted by $a$ at $t=s$ and unchanged elsewhere.  The reference label
moves with it because Lemma~\ref{lem:pca-pairing} gives the same factor at
$t=n$ for every $s$.
\end{proof}

\begin{proof}[Proof of Theorem~\ref{thm:pca-register}]
By Lemma~\ref{lem:pca-logical} all $2(n-1)Q$ operators centralize $H_{\rm fin}$,
so their classes and the induced form are defined.  By
Lemma~\ref{lem:pca-pairing} the $Z$--$X$ block of that form is $\delta_{st}$ times
the matrix $\phi(\mu_j,\mu_k^{\iota})$, invertible by
Lemma~\ref{lem:pca-perfect}; and the $Z$--$Z$ block vanishes because a pair loop
is pristine, so has zero exact pristine syndrome, and
Lemma~\ref{lem:closed-loop-interior} reads only that.

Suppose $\sum_{t,i}\beta_{ti}[Z_{t,\mu_i}]+\sum_{s,j}\alpha_{sj}[X_{s,\mu_j}]$
vanishes.  Pair it first with the loops.  The $Z$--$Z$ block being zero, pairing
with $[Z_{t,\mu_k}]$ leaves only the transfers at wall $t$ and gives
$\sum_j\alpha_{tj}\phi(\mu_j,\mu_k^{\iota})=0$ for all $k$, whence
$\alpha_{tj}=0$ for all $t,j$ by Lemma~\ref{lem:pca-perfect}.  What remains is a
combination of loops alone; pairing it with $[X_{t,\mu_k}]$ gives $\beta_{ti}=0$
in the same way.  The classes are therefore independent, and the same two
computations show the radical of the form on their span is zero.  The order
matters: the $X$--$X$ block is not controlled here, so the loops must be the
first probe.
\end{proof}

\subsection{Exhaustion of the Probe Algebra}\label{sec:exhaustion}

The probes form a perfect binary Weyl algebra, so the space they act on has
dimension at least $2^{(n_{\rm pair}-1)Q}$.  That lower bound alone
does not establish a protected, complete non-Abelian defect sector; the substance of what follows is the matching upper bound, that the
walls create no other local degree of freedom.  We first excise: the logical
algebra splits orthogonally into what the underlying torus already had and
what the walls added.  Then we compare the dimension of the added summand with
the span of the probes.

\subsubsection{A Logical Basis Routed Clear of the Walls}

\begin{lemma}
\label{lem:routed-logical-basis}
Work in a standard configuration, and let
\(\mathcal S(H_{\rm bulk})\) be the binary span of the pristine torus
stabilizers.  Write
\(\mathcal L(H_{\rm bulk})
 =\mathcal S(H_{\rm bulk})^\perp/\mathcal S(H_{\rm bulk})\), and put
\(u_x=x^{l_0}\), \(u_y=y^{m_0}\), \(N_x=L_1/l_0\), \(N_y=L_2/m_0\) and
\(\Sigma_\mu=\sum_{a=0}^{N_\mu-1}u_\mu^a\) for \(\mu=x,y\).
Reorder the fixed period-ribbon seeds of
Eq.~\eqref{eq:parallel-transverse-ribbons} by coordinate direction as
\((a_x,b_x)\) and \((a_y,b_y)\), with
\(f^\star a_\mu+g^\star b_\mu=1+u_\mu\), so that
\(a_xb_y+b_xa_y=\Delta\), independently of which wall orientation makes
\(z=u_x\) or \(z=u_y\).

Let \(\{\eta_i\}_{i=1}^Q\) and \(\{q_j\}_{j=1}^Q\) be the bases of
\(V_e\) and \(V_m\) fixed in Definition~\ref{def:crossing-constants}, with
their compact lifts denoted by hats; these are the bases over which
\(R_{\rm log}\) is taken.  Anchor all four families at the
unique meeting point of \(\gamma_x\) and \(\gamma_y\) and define, in the finite torus ring,
\[
 E_\mu(\eta_i)
 =Z\!\left[\widehat\eta_i\Sigma_\mu(a_\mu,b_\mu)\right],
 \qquad
 M_\mu(q_j)
 =X\!\left[\widehat q_j u_\mu\Sigma_\mu(b_\mu^\star,a_\mu^\star)\right],
 \qquad \mu=x,y,
\]
and let \(B\) be their binary span.  Ordering the four families as
\(E_x,M_x,E_y,M_y\), each in the order of its index, the symplectic Gram
matrix of \(B\) is
\begin{equation}
 \Omega_B=
 \begin{pmatrix}
  0&0&0&D\\
  0&0&D^T&0\\
  0&D&0&0\\
  D^T&0&0&0
 \end{pmatrix},
 \qquad
 D_{ij}
 =\phi(q_j,\eta_i)
 =
 \operatorname{Tr}_{z,w}\!\left(
   \widehat\eta_i\,\widehat q_j^{\,\star}\Delta
 \right),
 \label{eq:routed-logical-gram}
\end{equation}
where every zero denotes the \(Q\times Q\) zero matrix.  The matrix \(D\) is
invertible, so
\(\det\Omega_B=(\det D)^4\ne0\).
Consequently the four families are independent in
\(\mathcal L(H_{\rm bulk})\), their classes span that space, and
\(\dim B=4Q=2k(H_{\rm bulk})\), so \(k(H_{\rm bulk})=2Q\).  Moreover, under
Eq.~\eqref{eq:complement-clearance}, all four families have representatives
supported in \(\mathbb T\setminus U\).
\end{lemma}

\begin{proof}
Work in \(\overline R=\mathbb F_2[x^{\pm1},y^{\pm1}]/(1+x^{L_1},1+y^{L_2})\)
with coefficient trace \(\operatorname{tr}_{\overline R}\).  Since
\(u_\mu^{N_\mu}=1\), Eq.~\eqref{eq:parallel-transverse-ribbons} gives
\(f^\star\widehat\eta\Sigma_\mu a_\mu+g^\star\widehat\eta\Sigma_\mu b_\mu
=\widehat\eta(1+u_\mu^{N_\mu})=0\) and likewise
\(g\widehat q\,u_\mu\Sigma_\mu b_\mu^\star+f\widehat q\,u_\mu\Sigma_\mu a_\mu^\star=0\),
so all four families lie in \(\mathcal S(H_{\rm bulk})^\perp\).  Same-tone
pairings vanish, and
\(\langle Z[p],X[r]\rangle=\operatorname{tr}_{\overline R}(p_1r_1^\star+p_2r_2^\star)\);
with \(\Sigma_\nu^\star=\Sigma_\nu=u_\nu^{-1}\Sigma_\nu\) in \(\overline R\) this
gives \(\langle E_\mu(\eta_i),M_\nu(q_j)\rangle=\operatorname{tr}_{\overline R}
(\widehat\eta_i\widehat q_j^{\,\star}\Sigma_\mu\Sigma_\nu(a_\mu b_\nu+b_\mu a_\nu))\),
whose last factor is \(0\) for \(\mu=\nu\) and \(\Delta\) otherwise.  For every
compact \(F\), coefficient extraction gives
\(\operatorname{tr}_{\overline R}(\Sigma_x\Sigma_yF)=\sum_{r,s\in\mathbb Z}
\tr(u_x^ru_y^sF)=\operatorname{Tr}_{z,w}(F)\), in either wall orientation; this is
Eq.~\eqref{eq:routed-logical-gram}, and
Lemma~\ref{lem:braiding-nondegenerate} makes \(D\) invertible, whence
\(\det\Omega_B=(\det D)^4\ne0\).  Commensurability and
Lemma~\ref{lem:common-periods} put the torus period ideal inside \((f,g)\) and
\((f^\star,g^\star)\), so both quotients of \(\overline R\) have dimension
\(Q\); the coefficient trace makes \(\overline R\) Frobenius, so
annihilator--quotient duality (Sec.~\ref{sec:kernel}) gives each one-tone
stabilizer-generator map rank \(L_1L_2-Q\), and the CSS rank formula on
\(2L_1L_2\) qubits gives \(k(H_{\rm bulk})=2Q\).  A stabilizer in \(B\) pairs
trivially with \(B\) and so vanishes by non-degeneracy of \(\Omega_B\); hence
\(B\) injects into \(\mathcal L(H_{\rm bulk})\), and since
\(\dim B=4Q=\dim\mathcal L(H_{\rm bulk})\) its classes form a basis.  Every
fixed labelled seed lies within \(R_{\rm log}\) of its core segment and
concatenation keeps that width, so routing along \(\gamma_x,\gamma_y\) puts the
supports in \(N_{R_{\rm log}}(\gamma_x\cup\gamma_y)\subseteq\mathbb T\setminus U\)
by Eq.~\eqref{eq:complement-clearance}.
\end{proof}

\subsubsection{Excision and the Symplectic Splitting}

\paragraph*{The no-excitation phase normalization.}
Theorem~\ref{thm:exhaustion} below shows that the total pair loop
$Z_{{\rm tot},b}$ is a stabilizer class, so it acts on the code space by a
single character $\sigma$ of $A=V_m$.  \emph{Which} character is not fixed by the
binary algebra: $\mathcal L(H)=\mathcal S(H)^\perp/\mathcal S(H)$ is a binary
symplectic quotient, so a vanishing class determines the Hermitian lift only up
to a sign, $Z_{{\rm tot},b}=\sigma(b)\prod_rR_r^{\alpha_r(b)}$ over Hermitian Hamiltonian terms $R_r$, with
$\sigma(b)\in\{\pm1\}$, and the positive tensor-product lift of
Lemma~\ref{lem:pair-loop-characters} fixes phases only within the loop family.
Say that the Hamiltonian lifts carry the \emph{no-excitation phase
normalization} when, for a basis of $A$,
\begin{equation}\label{eq:no-excitation-normalization}
 Z_{{\rm tot},b}\lvert\Omega\rangle=\lvert\Omega\rangle
\end{equation}
for a common $+1$ eigenvector $\lvert\Omega\rangle$ of the completed
Hamiltonian; equivalently, when each positive lift $Z_{{\rm tot},b}$ lies in the
group generated by the chosen Hermitian Hamiltonian terms with no scalar factor.
State independence then carries Eq.~\eqref{eq:no-excitation-normalization} to
the whole code space, making $\sigma$ trivial.

\begin{theorem}[Multi-wall excision, global rank and exhaustion of the local fusion factor]
\label{thm:exhaustion}
Work in a standard configuration on $n=n_{\rm pair}\geq1$ walls
(Definition~\ref{def:standard-layout}), with the pair rectangles and fillings
of Lemma~\ref{lem:period-rectangle-corridors}.

For \(n\geq2\) wall \(n\) is the reference; for \(n=1\) the transfer family is
empty.  Then
\begin{equation}
 k(H_{\rm fin})-k(H_{\rm bulk})=(n-1)Q.
 \label{eq:splitting-R}
\end{equation}
Moreover, with \(\mathcal L(H)=\mathcal S(H)^\perp/\mathcal S(H)\), there is
an orthogonal symplectic splitting
\begin{equation}
 \mathcal L(H_{\rm fin})
   =\mathcal L_{\rm bulk}\mathbin{\perp}\mathcal L_U,
 \qquad
 \dim\mathcal L_U=2(n-1)Q,
 \label{eq:splitting-S}
\end{equation}
where $\mathcal L_{\rm bulk}$ is generated by routed pristine torus
logical representatives in $\mathbb T\setminus U$, and
\begin{equation}
 \mathcal L_U
 =\operatorname{im}\!\left(
   \mathcal S(H_{\rm fin})^\perp\cap\mathcal P(U)
   \longrightarrow\mathcal L(H_{\rm fin})\right)
 =\operatorname{span}\{[Z_{s,b}],[X_{s,a}]:s<n\}.
 \label{eq:splitting-U}
\end{equation}
Both summands are non-degenerate.  For $n=1$, $\mathcal L_U=0$; for $n\ge2$,
non-degeneracy of $\mathcal L_U$ follows from Theorem~\ref{thm:pca-register},
which gives independence of the $2(n-1)Q$ basis-labelled classes and a
non-degenerate form.  Hence the charge-loop and
transfer-loop algebra is the full matrix algebra on a
$2^{(n-1)Q}$-dimensional factor and acts irreducibly there; there is no
endpoint, wall-local, radical, or reference-pair spectator sector.  (The
register operators $Z_{s,b}$, $X_{s,a}$ are not the code's Pauli
constructors $Z[\,\cdot\,]$, $X[\,\cdot\,]$.)

Finally, let $Z_{{\rm tot},b}=\prod_{s=1}^n Z_{s,b}$.  Then
$[Z_{{\rm tot},b}]=0$ in $\mathcal L(H_{\rm fin})$ for every $b$.  Thus the
total pair charge is fixed, and the reference-pair charge is determined by
the other $n-1$ pair charges.  With the no-excitation phase normalization
of Eq.~\eqref{eq:no-excitation-normalization}, this fixed character is the
vacuum-total-charge sector.
\end{theorem}

\begin{proof}
Equation~\eqref{eq:splitting-R} is the common-frame case of
Proposition~\ref{prop:frames}, whose hypotheses every configuration assumed
here satisfies: the walls are translates of one construction whose first-block
origins lie in the common period-block lattice, and its room contains the
separations assumed there.  Here $J_L=0$, and each $o_s/o_1$ is a monomial
in $x^{l_0}$ and $y^{m_0}$, so $(o_s/o_1)^2$ acts trivially on $V_m$
(Lemma~\ref{lem:common-periods}); Eq.~\eqref{eq:frames-k-dim} then gives
$k(H_{\rm fin})=nQ+Q$, counting qubits for Hermitian signs that keep $-I$ out
of the stabilizer group, which is the binary count $k(H)$.  Since
$k(H_{\rm bulk})=2Q$ (Lemma~\ref{lem:routed-logical-basis}), this is
Eq.~\eqref{eq:splitting-R}.

Apply Lemma~\ref{lem:routed-logical-basis} to the same region $U$, and
let $B$ be the routed representative space supplied there.  The lemma
gives
\(B\subseteq\mathcal S(H_{\rm bulk})^\perp, \; \dim B=4Q=2k(H_{\rm bulk})\),
shows that the form on $B$ is non-degenerate, identifies its image with
all of $\mathcal L(H_{\rm bulk})$, and places its support in
$\mathbb T\setminus U$.  Retained pristine rows commute with $B$, while
every non-pristine row of $H_{\rm fin}$ is supported in $U$ and hence is
disjoint from $B$.  Thus $B\subseteq\mathcal S(H_{\rm fin})^\perp$.  If an
element of $B$ belonged to $\mathcal S(H_{\rm fin})$, it would pair
trivially with all of $B$; non-degeneracy would force it to vanish.
Consequently the quotient map injects $B$ into $\mathcal L(H_{\rm fin})$,
and its image is a non-degenerate summand denoted by
$\mathcal L_{\rm bulk}$.

Let $W$ be the span of the displayed loop and transfer classes for $s<n$.
For $n=1$, $W=0$ and the assertion is immediate.  For $n\ge2$, by
Theorem~\ref{thm:pca-register}, which rests on the perfect pairing of
Lemma~\ref{lem:pca-perfect}, these $2(n-1)Q$ classes are independent and
the symplectic form on $W$ is non-degenerate.  Since $W$ is supported in $U$ and $B$ outside $U$,
$W\perp\mathcal L_{\rm bulk}$.  Equation~\eqref{eq:splitting-R} gives
\(\dim\mathcal L(H_{\rm fin}) =2k(H_{\rm bulk})+2(n-1)Q =\dim\mathcal L_{\rm bulk}+\dim W\).
Therefore
$\mathcal L(H_{\rm fin})=\mathcal L_{\rm bulk}\perp W$.
Every $H_{\rm fin}$ centralizer supported in $U$ is orthogonal to
$\mathcal L_{\rm bulk}$, so its class lies in $W$; the reverse inclusion is
already supplied by the $Z$ and $X$ representatives.  This proves
Eqs.~\eqref{eq:splitting-S}--\eqref{eq:splitting-U} and excludes every local spectator sector.

The Pauli algebra of a non-degenerate binary symplectic space of dimension
$2m$ is the full matrix algebra on $\mathbb C^{2^m}$.  Taking
$m=(n-1)Q$ proves irreducibility and the asserted fusion-factor dimension.

Finally, $Z_{{\rm tot},b}$ commutes with every pair loop.  It also commutes
with each transfer loop because the source and reference commutators
are two equal copies of $\phi(a,b^\iota)$ (Lemma~\ref{lem:pca-pairing}) and
cancel over $\mathbb F_2$.  It is
supported in $U$, so its class lies in $\mathcal L_U=W$; since it commutes
with all generators of the non-degenerate space $W$, its class is zero.
Perfectness of $(a,b)\mapsto\phi(a,b^\iota)$ (Lemma~\ref{lem:pca-perfect})
then makes the reference charge the unique value
whose sum with the other pair charges equals the fixed total character.
Thus the local fusion factor has $|V_m|^{n-1}=2^{(n-1)Q}$ states at fixed
total charge, not an additional factor of $|V_m|$.
\end{proof}

\subsubsection{Pair-Loop Characters}

Throughout this subsection put $A=V_m$, with $\lambda$ the pairing of
Lemma~\ref{lem:pca-perfect}; Fig.~\ref{fig:AdjacentPairWeyl} shows the two
families and the crossings that produce it.

\begin{lemma}[Pair loops form a character family]
\label{lem:pair-loop-characters}
Work in the configuration of Theorem~\ref{thm:exhaustion}.
Each $Z_{u,b}$ of Eq.~\eqref{eq:period-rectangle-loop} is a pure-$X$ binary
Pauli; give it the positive tensor-product lift, so that $Z_{u,b}$ is
Hermitian and $Z_{u,b}Z_{u,b'}$ carries no extra sign.  With that convention,
for every wall $u=1,\ldots,n_{\rm pair}$ and all $b,b'\in A$,
\(Z_{u,0}=I, \; Z_{u,b}Z_{u,b'}=Z_{u,b+b'}\),
and loops on distinct walls commute.  Consequently
$\{Z_{u,b}\}$ is a commuting family, its joint eigenspaces are labelled by
tuples $(c_1,\ldots,c_{n_{\rm pair}})\in A^{n_{\rm pair}}$, and on the
eigenspace of such a tuple
\begin{equation}\label{eq:pair-loop-eigenvalue}
 Z_{u,b}\psi=\lambda(c_u,b)\,\psi
 \qquad\text{for every }u\text{, the reference pair included.}
\end{equation}
We call $c_u$ the charge label of pair $u$.
\end{lemma}
\begin{proof}
Lemma~\ref{lem:period-rectangle-corridors} gives $Z_{u,0}=I$ and formulas
linear in $b$; the binary Pauli of $Z_{u,b}Z_{u,b'}$ is therefore that of
$Z_{u,b+b'}$, and the positive lift fixes the scalar to $1$, so
$b\mapsto Z_{u,b}$ is a representation of the additive group $A$.  For $u\neq u'$, write
$Z_{u,b}=\bdry A_{u,b}$; since $Z_{u',b'}$ is a closed pristine loop, its
exact pristine syndrome vanishes, and
Lemma~\ref{lem:closed-loop-interior} gives
$\langle Z_{u,b},Z_{u',b'}\rangle=\tr(A_{u,b}^{\dagger}S\,\cobdry
Z_{u',b'})=0$.    A commuting family of this form is simultaneously
diagonalizable, and on each joint eigenspace $b\mapsto$ (eigenvalue of
$Z_{u,b}$) is a character of $A$.  By Lemma~\ref{lem:pca-perfect},
$c\mapsto\lambda(c,\cdot)$
identifies $A$ with its full character group; hence each such character is
$\lambda(c_u,\cdot)$ for a unique $c_u\in A$, which is
Eq.~\eqref{eq:pair-loop-eigenvalue}.  Nothing here uses
Theorem~\ref{thm:pca-register}.
\end{proof}

\subsection{Walls in Different Frames and on Incommensurate Tori}\label{sec:frames}

In a BB code the label of an anyon is tied to its position modulo the anyon
periods: translating an excitation by $o$ multiplies its label by $o$.  A
closed path that carries a translation therefore acts on labels, and this has
two consequences for walls.

\paragraph*{The position of a wall fixes its type.}
A wall whose block layout is translated by a monomial $o$ from the frame of
Sec.~\ref{sec:blockalgebra} is the translate of a wall in that frame: its
composites pair sites by the same block rotation $\iota$, now about the
centres of the translated blocks.  By translation covariance of the whole
normalized realization (Theorem~\ref{thm:transmission}, and
Corollary~\ref{cor:finite-transmission} for a finite wall in a crossing
window) it transmits by $o\,D_\kappa\,o^{-1}=D_{o^2\kappa}$
(Lemma~\ref{lem:frame-label}): moving a mirror by $o$ moves its reflection by
$o^2$.  In a common planar lift, crossing wall 1 and then wall 2 acts by
$D_{o_2^2\kappa}D_{o_1^2\kappa}=(o_2/o_1)^2$; the two walls enact the same
duality if and only if $o_1^2\equiv o_2^2\pmod{\langle f,g\rangle}$, and any
additional winding on a frustrated torus must be included separately.  Every
absorbable bulk charge can be transferred unchanged between twist pairs of the
two walls by a process admitting a common planar lift only when the two
dualities agree.  The common frame \textup{(S1)} of a standard configuration ensures this
agreement; the commensurability and geometric hypotheses of
Theorem~\ref{thm:exhaustion} supply the corresponding transfer operators.

\paragraph*{Frustrated tori.}
On a torus that is not commensurate (Sec.~\ref{sec:anyontypes}), magnetic total
charge is conserved only modulo $J_L$ and electric total charge modulo
$J_{e,L}$.  Under the hypotheses below, winding processes realize exactly the
$Q-Q_L$ directions of $J_L$ as changes of one twist pair's channel with every
other channel held fixed; on a commensurate torus no such change exists.

\begin{proposition}[Walls in arbitrary frames]\label{prop:frames}
Fix the code and the planar construction data: one wall orientation, the
transverse-normalized ribbon templates, the label section, and the
patch data and convention of Definition~\ref{def:natural-termination}.  On an
arbitrary $L_1\times L_2$ torus, let the checks be the retained pristine checks
and $n\geq1$ disjoint, naturally terminated translated copies of this wall
construction, with prescribed nonwrapping lifts whose first-block origins
$o_1,\ldots,o_n$ need not lie in one period-block lattice.  Assume
that every wall has at least $\rho_0$ blocks and that the block distance
between distinct walls, and from a lift of a wall to its other periodic
images, is at least $\rho_0$.  Choose Hermitian signs so that the
stabilizer group does not contain $-I$, and let $k$ be the number of qubits of
its simultaneous $+1$ space.  Then every logical class of the pristine code on
the same torus is represented in the completed code, and the walls add exactly
$\dim_{\mathbb F_2}K$ logical qubits to its $2Q_L$:
\begin{equation}\label{eq:frames-k}
 k=2Q_L+\dim_{\mathbb F_2}K,\qquad
 K=\Bigl\{(a_1,\ldots,a_n)\in V_m^{\,n}:\
 \textstyle\sum_s o_sa_s\in J_L,\ \sum_s o_s^{-1}a_s\in J_L\Bigr\},
\end{equation}
where $a_s$ is the change of the channel of the twist pair of wall $s$
under a completed logical operator: its syndrome on that wall, read in the
wall's own frame, has total label $m_{a_s}e_{\kappa a_s^\star}$, and absolute
channels need a reference sector.  The two conditions are conservation of total
magnetic and, after the antipode, total electric syndrome charge on the torus.  Equivalently,
\begin{equation}\label{eq:frames-k-dim}
 k=nQ+\dim_{\mathbb F_2}\frac{V_m}{J_L+\sum_{s>1}\bigl(1+(o_s/o_1)^2\bigr)V_m}.
\end{equation}
Two cases carry the physics.
\begin{enumerate}
\item On a commensurate torus ($J_L=0$), a second wall adds
 $\dim\ker_{V_m}\!\bigl(1+(o_2/o_1)^2\bigr)$ qubits: $Q$ if the two walls
 have the same type, and none when $V_m$ is a field and the types differ.
\item One wall gives $k=Q+Q_L$, that is $Q-Q_L$ qubits beyond the pristine
 $2Q_L$.  When $V_m$ is a field and the torus is frustrated, the bulk encodes
 nothing and a single twist pair encodes $Q$ qubits.
\end{enumerate}
\end{proposition}

For $\BBc{2,2,2,2}$, $V_m\cong\mathbb F_4$, $xy=1$ and $l_0=m_0=3$.  On a
commensurate torus, two sufficiently separated walls whose relative origin is
congruent to $(1,1)$ modulo $(3,3)$ gain two qubits over the one-wall code,
and relative residue $(1,0)$ gains none.  On every $L\times L$ torus with $3\nmid L$,
$Q_L=0$, so a one-wall installation satisfying
Proposition~\ref{prop:frames}, which needs $L$ large enough for
the room $\rho_0$, encodes two qubits.

For a single wall ($n=1$) on a frustrated torus, the $Q-Q_L$ released
qubits have character readers paired perfectly with winding writers of
channel changes in $J_L$; dual bases, with writers multiplied by readers where
needed to cancel their mutual commutators, give $Q-Q_L$ logical Pauli pairs.  To read the channel of the
wall $s=1$, let $E_s$ be its removed
electric anchors, with their prescribed wall lifts.  For $\chi\in V_m^*$ put
$\ell_\chi=\chi\circ\iota^{-1}$ and
$P_\chi=\prod_{v\in E_s}A_v^{\ell_\chi([v/o_s]_e)}$, where $A_v$ is the
pristine electric-detecting check; whenever an embedded rectangle enclosing
exactly these anchors is available, multiplying by retained electric checks
extends $P_\chi$ to its character filling.  By bilinearity in the finite-torus
Pauli space, the torus version of Lemma~\ref{lem:closed-loop-interior},
\(\langle P_\chi,W\rangle=\sum_{v\in E_s}\ell_\chi([v/o_s]_e)\operatorname{syn}_eW(v)=\chi(a)\)
for a completed logical operator $W$ of channel change $a$, and the completed
logical class of $P_\chi$ is nonzero exactly when $\chi|_{J_L}\neq0$.  To
write the channel, electric and magnetic winding transports produce the
matched wall syndrome with labels $m_ae_{\kappa a^\star}$, $a\in J_L$; the two
torus cycles supply the generators of $J_L$, finite cluster corrections give
the prescribed exact syndrome, and a product of removed pristine checks
cancels the remaining interblock and patch commutators.  A nonzero channel
writer cannot admit a syndrome-preserving joint finite planar lift of its
support and all affected pristine checks, with the removed-check syndromes
lifted in the wall's frame: such a planar Pauli has zero total electric and
magnetic label.

\begin{proof}
Let $S_{\rm bulk}$ be the pristine stabilizer span, $C\subset S_{\rm bulk}$
the span of retained and composite rows, and $F$ the completed span.  On each
wall the interblock syndromes span the open-path incidence image
(Lemma~\ref{lem:telescope}), and the installed references add the $l_0m_0-Q$
independent ledger directions of $\mathcal I_e$
(Theorem~\ref{thm:completion-lift}).  The collars of \textup{(L1)}, whose proof uses only the lengths and
separations assumed here, are disjoint and lift without wrapping, so distinct walls have disjoint removed anchors and
each wall's syndromes are computed in a planar lift of its collar.  If a
combination of interblock and reference syndromes on wall $s$ vanishes, the block
sum $\Sigma$ removes the interblock part and leaves $\sum_a\lambda_ah_a=0$, so
every $\lambda_a$ vanishes, and injectivity of $d_{\rm w}$ removes the interblock
coefficients.  Hence the removed-check electric syndromes of the interblock and
reference rows are independent.  By Eq.~\eqref{eq:exact-retention} the
installed translates of $M(h_a)$ on wall $s$ are those with $j$ in the interval
$J_{s,a}$, and for adjacent $j,j+1\in J_{s,a}$ the translated certificate of
Eq.~\eqref{eq:bounded-translation-certificate} writes $M_{j+1,a}M_{j,a}$ as a
product of legal rows of the finite wall \textup{(L1)}.  So every installed
patch row is a reference translate modulo legal rows, so
$F=C+\operatorname{span}\{\text{interblock rows, references}\}$ and
$F\cap S_{\rm bulk}^\perp=C$.

A completed centralizer has pristine syndrome only on removed checks, matched
by the composites, so its syndrome on wall $s$, read in that wall's own frame,
has total label $m_{a_s}e_{\kappa a_s^\star}$.  A pristine torus syndrome is realizable exactly when its
magnetic residue vanishes in $V_m/J_L$ and its electric residue vanishes in
$V_e/J_{e,L}$: the two syndrome images over $R/(1+x^{L_1},1+y^{L_2})$ are
generated by $(f,g)$ and $(f^\star,g^\star)$.  Applying the antipode to the
electric condition and cancelling the common unit $\kappa^\star$ gives
\[
 \textstyle\sum_s o_sa_s\in J_L,\qquad \sum_s o_s^{-1}a_s\in J_L .
\]
For every tuple $(a_s)$ satisfying both conditions there is therefore a
pristine Pauli $W_0$ with exactly the matched syndrome.  Let $A_v$ run over the
removed electric checks; by the independence above the matrix
$\langle B_i,A_v\rangle$, with $B_i$ the interblock and reference rows, has full
row rank, so a product $\prod_vA_v^{\beta_v}$ cancels every commutator of $W_0$
with an installed row without changing its pristine syndrome.  Completed
stabilizer rows change no channel, and the channel changes of completed
logical operators form exactly $K$.  With $b_s=o_sa_s$ the two conditions read
$\sum_sb_s\in J_L$ and $\sum_so_s^{-2}b_s\in J_L$; putting $c=\sum_sb_s$, the
second becomes $o_1^{-2}c+\sum_{s>1}o_s^{-2}\bigl(1+(o_s/o_1)^2\bigr)b_s\in J_L$
with $c$ and the $b_{s>1}$ free.  Hence the image of $V_m^{\,n}$ in
$(V_m/J_L)^2$ has dimension $Q_L+\dim I/J_L=2Q_L-\dim V_m/I$, where
$I=J_L+\sum_{s>1}(1+(o_s/o_1)^2)V_m$, so $\dim K=nQ-2Q_L+\dim V_m/I$,
which turns Eq.~\eqref{eq:frames-k} into Eq.~\eqref{eq:frames-k-dim}.

A zero-channel centralizer has the same pristine syndrome as a product of
interblock and reference rows; subtracting that product gives a pristine
centralizer in the same completed logical class.  Conversely, the removed-check
correction applies to every pristine centralizer and preserves its pristine
logical class.  Since $F\cap S_{\rm bulk}^\perp=C\subset S_{\rm bulk}$, this
defines a surjective pairing-preserving map from the zero-channel logical
subspace $\mathcal A$ onto $S_{\rm bulk}^\perp/S_{\rm bulk}$, of dimension
$4Q_L$, whose kernel is the space $\mathcal T$ of completed logical classes
represented in $S_{\rm bulk}$.  Moreover $\mathcal A^\perp=\mathcal T$, because
$(F^\perp\cap S_{\rm bulk}^\perp)^\perp=F+S_{\rm bulk}$; since the channel map
is onto $K$, $\operatorname{codim}\mathcal A=\dim K$ and hence
$\dim\mathcal T=\dim K$, and products of the wall character operators realize
all restricted characters of $K$, so they span $\mathcal T$.  Therefore
$2k=4Q_L+2\dim K$, which is Eq.~\eqref{eq:frames-k}.
\end{proof}

\subsection{Local Exactness at a Natural Endpoint}\label{sec:endpointexact}

Theorem~\ref{thm:endpoint-exactness} rules out additional local logical
Pauli classes: in an endpoint window, every Pauli
commuting with the Hamiltonian is, up to phase, a product of completed
rows.

\paragraph*{Endpoint constants.}
Theorem~\ref{thm:endpoint-exactness} uses the check radius $r_{\rm chk}$, the
radii $r_T$ and $r_{\rm inc}$ of Eq.~\eqref{eq:endpoint-constant-1}, and the
radius of the patch packages.  Take the pivot basis \(\{\xi_\nu\}\) of
\(\mathcal I_e\) and the patch rows \(M_\nu\) of
Eq.~\eqref{eq:endpoint-constant-3} as the fixed basis and family of
Definition~\ref{def:natural-termination}, with translation packages
\(\mathscr P_\nu=\{M_\nu,zM_\nu\}\cup\mathscr C_\nu\) and offset sets
\(A_\nu\subset\mathbb Z\), and write \(\alpha_\nu=\min A_\nu\) and
\(\beta_\nu=\max A_\nu\).  With $\pi_\nu$ the anchor block of $\mathscr P_\nu$,
the wall block of carrier exponent $0$, let
\begin{equation}\label{eq:endpoint-package-radius}
 r_{\rm pkg}
 =
 \max_{\nu}
 \left\lceil
 \max_{\chi}\;
 d_{\rm blk}\!\left(\chi,\pi_\nu\right)
 \right\rceil ,
\end{equation}
where $\chi$ ranges over every support point of every row of $\mathscr P_\nu$
\emph{and} over every block of every required scaffold-row carrier of that
package, the same data that $A_\nu$ collects longitudinally, but measured in
the two-dimensional block metric; in particular
$\max_{k\in A_\nu}|k|\leq r_{\rm pkg}$.  Put
\begin{equation}
 \rho_{\rm end}=2r_{\rm pkg}+r_{\rm inc}+r_{\rm chk}+2,
 \qquad
 N_{\rm end}=2\rho_{\rm end}+3 .
\label{eq:endpoint-constant-5}
\end{equation}
The finite-window form of the transmission theorem also uses the interblock-row
diameter
\begin{equation}\label{eq:junction-row-diameter}
 d_T
 =
 \max_{\text{orientations},\,u\in\Lambda_0}
 \left\lceil
 \max_{q,q'\in\operatorname{supp}T_{0,u}}
 d_{\rm blk}(q,q')
 \right\rceil ,
\end{equation}
so that an interblock row with one support point in a set lies entirely within
$d_T$ blocks of that set.  When $l_0m_0=Q$ the ideal $\mathcal I_e$ vanishes and
there is no patch basis; we then set $r_{\rm pkg}=0$, and every patch
assertion below is vacuous.

Let the blocks of a naturally terminated wall carry exponents
\(0,\ldots,N-1\), and denote its endpoints at block \(0\) and at block
\(N-1\) by \(s_-\) and \(s_+\).  By Eq.~\eqref{eq:exact-retention} the
installed translates of \(M_\nu\) are \(z^jM_\nu\) with
\(-\alpha_\nu\leq j\leq N-1-\beta_\nu\).  The two extreme ones are the
installed rows nearest the endpoints,
\begin{equation}
 M_{\nu,s_-}=z^{-\alpha_\nu}M_\nu,
 \qquad
 M_{\nu,s_+}=z^{N-1-\beta_\nu}M_\nu,
 \qquad
 K_s=\bigcup_\nu\operatorname{supp}M_{\nu,s}.
\label{eq:endpoint-nearest-packages}
\end{equation}

\begin{lemma}[Patch rows at an endpoint]
\label{lem:endpoint-completion-retention}
On a naturally terminated wall (Definition~\ref{def:natural-termination}) of
\(N\geq N_{\rm end}\) complete period blocks, for either endpoint $s$:

\textup{(i)} every \(M_{\nu,s}\) is an installed row, and \(K_s\) lies within
block distance \(2r_{\rm pkg}\) of the block of $s$;

\textup{(ii)} every installed translate of \(M_\nu\) has ledger \(\xi_\nu\), so
for every \(h=\sum_\nu h_\nu\xi_\nu\in\mathcal I_e\) the operator
\(M_s(h)=\prod_\nu(M_{\nu,s})^{h_\nu}\) is a product of installed rows,
supported in \(K_s\), with \(\ell(M_s(h))=h\).  It need not itself be an
installed row.
\end{lemma}

\begin{proof}
If \(\mathcal I_e=0\) there is nothing to prove.  Otherwise
\(\beta_\nu-\alpha_\nu\leq2r_{\rm pkg}<N_{\rm end}\leq N\), so the index
interval above is nonempty and its two extreme members are installed.  The
package of \(z^{-\alpha_\nu}M_\nu\) lies within \(r_{\rm pkg}\) of its anchor
block, which has exponent \(-\alpha_\nu\) with
\(|\alpha_\nu|\leq r_{\rm pkg}\), hence within \(2r_{\rm pkg}\) of block \(0\);
likewise at \(s_+\), where the anchor has exponent \(N-1-\beta_\nu\) with
\(|\beta_\nu|\leq r_{\rm pkg}\).  This is \textup{(i)}.  A translation along
the wall shifts the removed-check syndrome sequence without changing its sum
in Eq.~\eqref{eq:ledger-map}, so Eq.~\eqref{eq:completion-lift} gives
\(\ell(z^jM_\nu)=\xi_\nu\), and the ledger is additive, which is
\textup{(ii)}.
\end{proof}

Throughout, ``commuting with the completed Hamiltonian'' means centralizing
every completed row separately, as is standard for a Hamiltonian listed as a
sum of commuting Pauli rows.

\begin{definition}[Endpoint window]
\label{def:hull-controlled-window}
Let $0<Q<\infty$, let the wall have at least $N_{\rm end}$ blocks, and fix
either endpoint $s$.  An \emph{endpoint window} at $s$ is a block rectangle
$\Omega_{\rm in}$ whose lift contains $s$ and no other endpoint, at
longitudinal block distance at least $\rho_{\rm end}$ from the opposite
endpoint $s_{\rm opp}$, together with
\(\Omega_{\rm out}=N_{\rho_{\rm end}}(\Omega_{\rm in})\), provided
$\Omega_{\rm out}$ lifts to the plane without wrapping and contains no removed
check of another wall.
\end{definition}

\begin{theorem}[Natural finite-endpoint exactness]\label{thm:endpoint-exactness}
Use the natural termination of
Definition~\ref{def:natural-termination}.
Let $\Omega_{\rm in}\subset\Omega_{\rm out}$ be an endpoint window at an
endpoint $s$ (Definition~\ref{def:hull-controlled-window}).
Then every Pauli supported in $\Omega_{\rm in}$ and commuting with the completed finite-wall Hamiltonian is, up to scalar phase, a product of completed rows, each supported in $\Omega_{\rm out}$.
\end{theorem}

This is a local-topological-order statement of the type posited as TQO-1 and
TQO-2 in the stability
literature~\cite{bravyiTopologicalStability2010, bravyiShort2011}, where it
serves as a hypothesis of the perturbative gap-stability theorem;
Corollary~\ref{cor:distance} below establishes it for this model up to the room of a
standard configuration.  For translation-invariant Pauli
stabilizer Hamiltonians the corresponding condition is equivalent to
exactness of the associated module complex and admits a constructive proof
by polynomial division~\cite[Def.~7 and Lem.~3.1]{haahCommuting2013}; that
argument assumes translation invariance throughout and covers the
homogeneous bulk only.  The content of Theorem~\ref{thm:endpoint-exactness} is that a
buffer of fixed width around the hull $\Xi_s$ survives at a wall endpoint, where translation invariance
fails locally.

\begin{proof}
We treat one endpoint; the argument at the other is the same.

Put \(\Xi_s=\operatorname{conv}\!\left(\widetilde\Omega_{\rm in}\cup K_s\right)\)
in the lifted plane, with $K_s$ as in Eq.~\eqref{eq:endpoint-nearest-packages}.  The
proof uses three properties of the window.
\begin{enumerate}
\item[\textup{(H0)}] \emph{One wall.}  No removed check of another wall lies in
$\widetilde\Omega_{\rm out}$.
\item[\textup{(H1)}] \emph{The hull stays inside.}
\(N_{r_{\rm inc}+r_{\rm chk}}(\Xi_s)\subseteq\widetilde\Omega_{\rm out}\).
\item[\textup{(H2)}] \emph{One internal block survives.}  Some complete
internal wall block $j_0$ before $s_{\rm opp}$ satisfies
\(j_0\cap N_{r_{\rm inc}}(\Xi_s)=\emptyset\).
\end{enumerate}
\textup{(H0)} is part of Definition~\ref{def:hull-controlled-window}.  Every
point of $K_s$ lies within block distance $2r_{\rm pkg}$ of the block of $s$ by
Lemma~\ref{lem:endpoint-completion-retention}\textup{(i)}, and that block lies in the block rectangle
$\widetilde\Omega_{\rm in}$; so a point $\lambda p+(1-\lambda)k$ of $\Xi_s$,
with $p\in\operatorname{conv}\widetilde\Omega_{\rm in}$ and $k$ within
$\ell^\infty$-distance $2r_{\rm pkg}$ of a point $s'$ of that block, lies within
$2r_{\rm pkg}$ of the point $\lambda p+(1-\lambda)s'$ of
$\operatorname{conv}\widetilde\Omega_{\rm in}$.  Hence
\(\Xi_s\subseteq N_{2r_{\rm pkg}+1}(\widetilde\Omega_{\rm in})\), and
\(\rho_{\rm end}=2r_{\rm pkg}+r_{\rm inc}+r_{\rm chk}+2\) gives \textup{(H1)}.
The wall block at longitudinal distance \(2r_{\rm pkg}+r_{\rm inc}+2\) from
$\widetilde\Omega_{\rm in}$ towards $s_{\rm opp}$ lies outside
\(N_{r_{\rm inc}}(\Xi_s)\) and at least \(r_{\rm chk}\geq1\) blocks before
$s_{\rm opp}$, which is \textup{(H2)}.

Let $p$ be supported in $\Omega_{\rm in}$ and centralize every completed
row.  Let $W_j^X,W_j^Z\in\mathsf A$ be its commutator patterns with the
removed pristine checks, and put
\(h=\sum_jW_j^X\).
After lifting the non-wrapping window, the full pristine syndrome of $p$
has trivial anyon label.  As in Lemma~\ref{lem:charge-ledger}, the relation
$z=1$ in $V_e$ gives
\(\pi_e(h)=0,\; h\in\mathcal I_e\).
This uses only the complete pristine syndrome of the compact Pauli; no interblock row beyond either endpoint is used.

Write $h=\sum_\nu h_\nu\xi_\nu$, and let
\(M_s(h)=\prod_\nu\left(M_{\nu,s}\right)^{h_\nu}\)
be the product of the nearest installed patch basis rows; by
Lemma~\ref{lem:endpoint-completion-retention} it is a product of installed rows
supported in $K_s$ with ledger $h$.
Put \(q=pM_s(h)\).  Then $\ell(q)=0$, and
\begin{equation}
 \operatorname{supp}q
 \subseteq\widetilde\Omega_{\rm in}\cup K_s
 \subseteq\Xi_s.
\label{eq:endpoint-constant-9}
\end{equation}

Since $q$ centralizes every installed row, its pristine syndrome lies on
removed checks within $r_{\rm chk}$ of $\Xi_s$, hence in $\widetilde\Omega_{\rm out}$
by \textup{(H1)}, and these belong to this wall by \textup{(H0)}.  With
$\ell(q)=0$, Lemma~\ref{lem:open-path-decomposition} applies to $q$ with
$\Omega=\Xi_s$; only interblock rows between installed blocks are used.
Its filling lies in $N_{r_{\rm inc}}(\Xi_s)$, so hypothesis \textup{(H2)}
leaves a complete internal block $j_0$ before the opposite endpoint on which
\begin{equation}
 A_X(j_0,u)=A_Z(j_0,u)=0\qquad\text{for every }u,
\label{eq:endpoint-constant-13}
\end{equation}
hence $m_{j_0}=0$; no interblock row beyond either endpoint is required.  Part
\textup{(iii)} of the lemma expresses $q$, and with $M_s(h)$ restored $p$, as a
product of completed rows; the patch factors lie in $K_s\subseteq\Xi_s$ by
Eq.~\eqref{eq:endpoint-constant-9} and all other factors in
\(N_{r_{\rm inc}+r_{\rm chk}}(\Xi_s)\subseteq\widetilde\Omega_{\rm out}\) by
\textup{(H1)}.  This proves the stated buffered decomposition.
\end{proof}

\subsection{Non-Abelian Fusion}\label{sec:fusiontwists}

With four twists at fixed total
charge, one pair carries $2^Q$ orthogonal charge sectors, each containing
exactly one state, together exhausting the local space --- the statement an
experiment would meet, since the Wilson loops encircling one pair both label
the state and determine it.  In the standard description of extrinsic
defects~\cite{barkeshliTwist2013, barkeshliSymmetry2019}, the twist is a self-dual
object $\sigma$ of a $\mathbb Z_2$-crossed Tambara--Yamagami theory over $A=V_m$,
and $\mathcal F_n$ is its fusion space, in which the sectors read by the Wilson
loops are the fusion channels of every cut, with one labelling of $A$.  Channels add because $V_m$ is a group, and absorption is
Lemma~\ref{lem:endpoint-caps}; the work is in the single state per sector: the
transfers make all $2^Q$ charge projectors nonzero, and being orthogonal and
summing to the identity their ranks add to $\dim\mathcal F_2=2^Q$.

\paragraph*{Setting for this subsection.}
Throughout, work in a standard configuration
(Definition~\ref{def:standard-layout}), whose endpoint windows are those of
Lemma~\ref{lem:room}\textup{(L7)} and whose region $U$ the splitting
Eq.~\eqref{eq:splitting-U} consumes through the transfer loops.  Put
\(A=V_m, \; |A|=2^Q, \; \epsilon_a=(a,a^\iota)=m_a e_{a^\iota}\),
let $\sigma$ denote the natural endpoint species, and use the no-excitation
phase normalization of Eq.~\eqref{eq:no-excitation-normalization}, so that the
fixed total pair charge of Theorem~\ref{thm:exhaustion} is called vacuum.
Write $\mathcal F_n$ for the local fusion factor of
Theorem~\ref{thm:exhaustion} at $n=n_{\rm pair}$ walls, after the pristine
torus logical factor has been removed by the exact splitting of
Theorem~\ref{thm:exhaustion}.

\begin{theorem}[Rank-one charge sectors]
\label{thm:fusion-reconstruction}
Choose wall $n$ as reference.  For every
$\mathbf c=(c_1,\ldots,c_{n-1})\in A^{n-1}$ the operator
\begin{equation}
 P_{\mathbf c}=\prod_{s=1}^{n-1}P_{s,c_s},
 \qquad
 P_{s,c}=2^{-Q}\sum_{b\in A}\lambda(c,b)Z_{s,b},
\label{eq:fusion-charge-projectors}
\end{equation}
is a rank-one projector on $\mathcal F_n$, and
\(\mathcal F_n=\bigoplus_{\mathbf c\in A^{n-1}}\operatorname{im}P_{\mathbf c}\),
with \(\dim\mathcal F_n=|A|^{n-1}=2^{(n-1)Q}\).
The reference-pair charge is uniquely fixed by the total charge; in the vacuum
sector it is $c_n=\sum_{s<n}c_s$.  Every label occurs: the transfer loop
$X_{s,a}$ carries any sector to the sector shifted by $a$.  No further sector
exists, and no sector has dimension greater than one.  In particular, for $n=2$ the
four-twist space at fixed total charge carries exactly $|A|$ states, one for
each label $\epsilon_a$ of the selected pair.
\end{theorem}

\begin{proof}
Lemma~\ref{lem:pair-loop-characters} gives the canonical Hermitian phases,
the group law
\(Z_{s,0}=I, \; Z_{s,b}Z_{s,b'}=Z_{s,b+b'}\),
commutation of loops on distinct walls, and the identification of $A$ with
its complete character group by $c\mapsto\lambda(c,\mathord\cdot)$.
Character orthogonality gives
\begin{equation}
 P_{s,c}P_{s,c'}=\delta_{c,c'}P_{s,c},
 \qquad
 \sum_{c\in A}P_{s,c}=I.
\label{eq:charge-projector-orthogonality}
\end{equation}

The transfer loops of Definition~\ref{def:pca-families} exist for every $a$,
and Corollary~\ref{cor:pca-charge-action}, whose projectors are those of
Eq.~\eqref{eq:fusion-charge-projectors}, gives
\(X_{s,a}P_{\mathbf c}X_{s,a}^{-1}=P_{\mathbf c+a\mathbf e_s}\).
The transfers therefore act transitively on the projectors in
Eq.~\eqref{eq:fusion-charge-projectors}.  At least one is nonzero because
they sum to the identity, so all $|A|^{n-1}$ are nonzero and have equal
rank.

Theorem~\ref{thm:exhaustion} identifies the full local logical algebra
with the loop--transfer algebra and gives
$\dim\mathcal F_n=2^{(n-1)Q}=|A|^{n-1}$.  The sum of the ranks of the
$|A|^{n-1}$ nonzero orthogonal projectors is this same number, so every rank
is one.  This proves rank-one microscopic sectors and exclusion: any extra
sector would add an orthogonal summand, while a hidden microscopic degree of
freedom invisible to the loop characters would raise the rank of one
displayed summand.  The exhausted non-degenerate algebra permits neither
possibility.

For $n=2$, start in the calibrated vacuum summand and apply $X_{1,a}$.
The resulting nonzero rank-one summand has the exact cap label
$\epsilon_a=(a,a^\iota)$, so every displayed label occurs physically, and
the triviality of the total loop fixes the reference charge to the same
label.  This completes the proof.

\end{proof}

\begin{corollary}[Fusion reconstruction]
\label{cor:fusion-rules}
The endpoint object is categorically simple, every
$\epsilon_a$ is absorbed by it, and these objects obey the fusion rules
\begin{equation}
 \epsilon_a\times\epsilon_b=\epsilon_{a+b},
 \qquad
 \sigma\times\epsilon_a=\sigma,
 \qquad
 \sigma\times\sigma=\bigoplus_{a\in A}\epsilon_a,
\label{eq:fusion-rule}
\end{equation}
with categorical multiplicity one in both rules containing $\sigma$, and
\[
 d_{\epsilon_a}=1,
 \qquad
 d_\sigma=|A|^{1/2}=2^{Q/2}.
\]
\end{corollary}

\begin{proof}
Bulk charges add and the caps of Lemma~\ref{lem:endpoint-caps} are linear in
the label, which is the first rule.  For $a\neq0$ the cap $E_{s,a}$ has as its
only pristine syndrome the matched terminal cluster $\epsilon_a$ and one remote
cluster of the same label; applied after a bulk $\epsilon_a$ is brought to the
remote cluster, it removes that syndrome without changing the completed
endpoint sector, so the endpoint absorbs $\epsilon_a$.

Cutting the four-twist system along the loop that encloses
one twist pair, with its charge held at $\epsilon_a$, gives
\begin{equation}
 P_{1,a}\mathcal F_2
 \simeq
 V_{\sigma\sigma}^{\epsilon_a}\otimes
 \bigl(V_{\sigma\sigma}^{\epsilon_a}\bigr)^*
\label{eq:channel-identification}
\end{equation}
compatibly with the loop-charge projectors: the usual cut-and-glue dictionary
between a ground space and a fusion space.  Theorem~\ref{thm:fusion-reconstruction}
gives $\operatorname{rank}P_{1,a}=1=(N_{\sigma\sigma}^{\epsilon_a})^2$, hence
$N_{\sigma\sigma}^{\epsilon_a}=1$ for every $a$, while the resolution of the
identity leaves no further summand.  In particular
$N_{\sigma\sigma}^{\mathbf 1}=1$.  Since $\sigma$ is self-dual, rigidity gives
\[
 \operatorname{End}(\sigma)
 \cong\operatorname{Hom}(\mathbf 1,\sigma^*\!\times\sigma)
 \cong\operatorname{Hom}(\mathbf 1,\sigma\times\sigma),
\]
whose dimension is one; hence $\sigma$ is categorically simple.  Each
$\epsilon_a$ is invertible, so $\sigma\times\epsilon_a$ is also simple.
The cap, read as a nonzero morphism, therefore gives
$\sigma\times\epsilon_a\cong\sigma$ with a one-dimensional morphism space.
Every $\epsilon_a$ is
Abelian, so $d_{\epsilon_a}=1$ and
Eq.~\eqref{eq:fusion-rule} gives
\(d_\sigma^2=\sum_{a\in A}d_{\epsilon_a}=|A|=2^Q, \; d_\sigma=2^{Q/2}\),
and iterating the multiplicity-one rule reproduces the fixed-total-charge
count $|A|^{n-1}=2^{(n-1)Q}$ already established there.
\end{proof}

\section{Affine-Dual Transmission through the Completed Wall}\label{sec:transmission}

The wall interior is locally exact (Sec.~\ref{sec:completion}).  This section
realizes the target label action of Sec.~\ref{sec:blockalgebra} by explicit
crossing operators and shows that the remote label pairs an admissible
crossing of the transverse-normalized wall can realize are exactly the graph
of $D_{\rm lab}$, which justifies calling the
construction a duality wall; Sec.~\ref{sec:FR} consumes its crossing
construction.

The proof rests on one idea.  Fold a crossing with the affine antipode, so that
its magnetic syndrome joins its electric syndrome on one side.  Summed over the
translates of an interblock row at one offset, the interblock commutators obey a
transverse Gauss law in this folded syndrome
(Lemma~\ref{lem:sigma-syndrome}).  The transverse normalization fixed in the
interblock construction, which confines the transverse support of
$\overline\Delta$ to the removed block's fundamental interval, then leaves only
the folded syndrome beyond the wall, and the sum becomes the braiding pairing
(Lemma~\ref{lem:placement-aware-junction-cut}) evaluated on the mismatch between
the label that went in and the label the duality prescribes.  That pairing is
perfect (Lemma~\ref{lem:braiding-nondegenerate}), so the mismatch vanishes,
which forces one half of the duality; charge accounting
(Lemma~\ref{lem:star-accounting}) forces the other half.
No general transmission theorem for unnormalized realizations is
established here.

\subsection{The Infinite Straight Wall}\label{sec:infinite-wall}

\subsubsection{Crossing Operators and Their Only Obstruction}

\paragraph*{Setting for this section.}
Fix once, for the whole of Sec.~\ref{sec:transmission}: a BB code with
$0<Q<\infty$; the block algebras
\(\mathsf B=R/(1+z),\; \mathsf A=R/(1+z,1+w)\),
with the fixed block fundamental domain and the ideals
$\mathcal I_e=(f^\star,g^\star)\mathsf A$ and
$\mathcal I_m=(f,g)\mathsf A$; the finite longitudinal and transverse
ribbons, satisfying
\(f^\star h+g^\star v=1+z, \; f^\star c+g^\star d=1+w, \; \Delta=cv+dh\);
and the affine anchoring $\widehat{\iota(u)}=\kappa\widehat u^\star$ of
Eq.~\eqref{eq:affine-anchor}.  Results in this section assume this Setting
and state only the hypotheses that go beyond it.

For each class $r\in V_e$, choose once and linearly a standard block lift
$\widetilde r=\sum_ur_u u\in\mathsf A$, and abbreviate
\begin{equation}
 \widehat r=\sum_ur_u\widehat u,
 \qquad
 \overline r=\sum_ur_u\widehat{\iota(u)}
             =\kappa\widehat r^\star.
\end{equation}
Changing the lift by an element of $\mathcal I_e$ does not change the
crossing's remote anyon classes.

\begin{lemma}[Matched crossing obstruction]\label{lem:affine-crossing}
For $r\in V_e$ and its standard lift above, the provisional crossing
operator
\begin{equation}\label{eq:provisional-crossing}
 W_0(r)
 =Z[\widehat r(c,d)]\,
  X[\overline r(d^\star,c^\star)]
\end{equation}
commutes with every composite and with every retained row except the two
designated remote charge-syndrome check clusters.  Its only possible scaffold
obstruction is its interblock-commutator pattern, and its remote clusters
represent $e_r$ and $m_{r^\iota}$ on the two banks.  If
$t_{j,u}(r)=\langle W_0(r),T_{j,u}\rangle$, then
\begin{align}
 t_{j,u}(r)
 &=\tr\!\left(
 z^j\widehat r^\star\widehat u\,\Delta^\star
 +z^{j-1}\overline r^{\,\star}\overline u\,\Delta
 \right),
 \qquad \overline u=\widehat{\iota(u)},
 \label{eq:crossing-junction-pattern}\\
 \sum_{j\in\mathbb Z}t_{j,u}(r)&=0
 \qquad\text{for every }u.
 \label{eq:crossing-obstruction-zero}
\end{align}
\end{lemma}
\begin{proof}
Equation~\eqref{eq:parallel-transverse-ribbons} and its antipode give
\begin{equation}
 \operatorname{syn}_eW_0(r)=\widehat r(1+w),
 \qquad
 \operatorname{syn}_mW_0(r)=\overline r(1+w^{-1}).
\end{equation}
The two clusters adjacent to the wall are $\iota$-matched and hence cancel
in the composites; the other two are the stated remote charges.  Direct
substitution of Eqs.~\eqref{eq:provisional-crossing} and
\eqref{eq:T-definition} into the binary symplectic pairing gives
Eq.~\eqref{eq:crossing-junction-pattern}.  Finally $W_0(r)$ has the affine form of
Lemma~\ref{lem:wall-trace}\textup{(ii)} with $(a,b)=\widehat r(c,d)$, which gives
Eq.~\eqref{eq:crossing-obstruction-zero}.
\end{proof}

Reduce $\Delta=cv+dh$ modulo $1+z$, without quotienting by $1+w$.  For the
transverse-normalized wall fixed in the construction,
Eq.~\eqref{eq:transverse-normalized-ribbon} gives $\overline\Delta=r$, the
remainder of Eq.~\eqref{eq:transverse-normalization-division}, so
\begin{equation}\label{eq:normalized-transverse-support}
 \operatorname{supp}_t\overline\Delta\subseteq\{0,1,\ldots,p_\perp-1\}.
\end{equation}

For a finite crossing Pauli $W$, decompose its exact pristine syndrome into
the wall block and the two strict banks, \(E=E_-+E_0+E_+\) and
\(M=M_-+M_0+M_+\).  The $+$ and $-$ banks are the right and left sides: $E_+,M_+$
are supported at transverse coordinates at least $p_\perp$ and $E_-,M_-$ at
transverse coordinates at most $-1$, and the remote labels of $W$ are
\(r_R=\pi_e(E_+)\), \(s_R=\pi_m(M_+)\), \(r_L=\pi_e(E_-)\), \(s_L=\pi_m(M_-)\).

\subsubsection{Remote Crossings}

\begin{definition}[Remote crossing relation]
\label{def:separated-crossing-relation}
For the fixed transverse-normalized wall, choose after the crossing datum an
admissibility block radius \(r_{\rm adm}\geq r_\times\) of
Definition~\ref{def:crossing-constants}; every statement below holds for any
larger value.  A finite crossing Pauli is \emph{admissible} if its support
lies within \(r_{\rm adm}\) of the crossing anchor and it commutes with every completed-wall row (that is, with every
composite \(C_{j,u}\), every interblock row \(T_{j,u}\), and every installed
patch row) and with every retained check other than its designated remote
syndrome checks.  The designated remote syndrome checks are retained checks
and hence lie strictly outside the removed wall block.  Let
\(\operatorname{Rel}_{\rm wall}\) be the additive remote-label relation
realized by finite admissible crossings with such designated remote syndromes.

An admissible crossing may also have matched pristine syndrome inside the
removed wall block; this is ledger syndrome carried
by \(E_0,M_0\) and handled by the composites, not designated remote syndrome.
Closure under addition follows by multiplying crossing Paulis at a common
anchor.  Leaving a completed-wall row or a non-designated retained check
excited is excluded: such an operator flips a measured stabilizer, so the
charge is a visible excitation, not a hidden channel.  The condition therefore
constrains the final crossing Pauli, not the microscopic wall ledger.
\end{definition}

\subsubsection{The Transverse Gauss Law}

For a finite Pauli $p$ and an offset $u\in\Lambda_0$ set
\begin{equation}\label{eq:sigma-obstruction}
    \sigma_u(p)=\sum_{j\in\mathbb Z}\langle p,T_{j,u}\rangle
    \in\mathbb F_2 ,
\end{equation}
a finite sum because compact $p$ meets only finitely many translates $T_{j,u}$.

\begin{lemma}[The interblock obstruction is a syndrome invariant]
\label{lem:sigma-syndrome}
Let $p$ be a finite Pauli with exact syndrome $(E,M)$ and folded syndrome
$H=E+\kappa M^\star$.  Then
\begin{equation}\label{eq:gauss-sum}
 \sigma_u(p)=\sum_{n\geq1}
 \operatorname{Tr}_z\!\left(w^{-n}\widehat u^{\,\star}\Delta H\right).
\end{equation}
Thus $\sigma_u(p)$ depends only on the exact syndrome $\cobdry p$.  It is linear
in the syndrome, vanishes on every product of pristine checks, and is unchanged
by dressing with removed pristine checks.  Consequently a finite Pauli with a
prescribed syndrome can commute with every interblock row only if $\sigma_u=0$
for every $u\in\Lambda_0$.
\end{lemma}
\begin{proof}
Write $p=(u_1,v_1\mid u_2,v_2)$, so that $E=f^\star u_1+g^\star v_1$ and
$M=gu_2+fv_2$, and fold it:
\begin{equation}\label{eq:fold}
 (U,V)=(u_1+\kappa v_2^\star,\;v_1+\kappa u_2^\star),
 \qquad
 f^\star U+g^\star V=H,
 \qquad
 A=Uv+Vh.
\end{equation}
Expanding the pairing with Eqs.~\eqref{eq:T-definition}
and~\eqref{eq:affine-anchor} and the invariance of $\tr$ under $\star$ gives
\[
 \langle p,T_{j,u}\rangle
 =\tr\!\left(z^{-j}\widehat u^{\,\star}(u_1v+v_1h)\right)
 +\tr\!\left(z^{j-1}\widehat u^{\,\star}\kappa(v_2^\star v+u_2^\star h)\right),
\]
so $\sigma_u(p)=\operatorname{Tr}_z(\widehat u^{\,\star}A)$.  Multiplying out with
Eq.~\eqref{eq:parallel-transverse-ribbons} gives the exact identity
\begin{equation}\label{eq:transverse-gauss}
 (1+w)A=\Delta H+(1+z)(cV+dU).
\end{equation}
Put $\alpha_n=\operatorname{Tr}_z(w^{-n}\widehat u^{\,\star}A)$.  Applying
$\operatorname{Tr}_z(w^{-n}\widehat u^{\,\star}\,\cdot\,)$ to
Eq.~\eqref{eq:transverse-gauss} and using Lemma~\ref{lem:wall-trace}\textup{(i)}
on the last term gives
$\alpha_n+\alpha_{n-1}
 =\operatorname{Tr}_z(w^{-n}\widehat u^{\,\star}\Delta H)$.  Since $A$ is finite,
$\alpha_n=0$ for large $n$, and summing over $n\geq1$ gives
Eq.~\eqref{eq:gauss-sum}.  Its right-hand side is linear in the syndrome, and a
product of pristine checks, in particular a removed one, has zero syndrome.
\end{proof}

\subsubsection{The Folded Syndrome of a Crossing}

Let $W$ be an admissible crossing, and split its folded syndrome by banks:
\begin{equation}\label{eq:H-bank-split}
 H_+=E_++\kappa M_-^\star,
 \qquad
 H_-=E_-+\kappa M_+^\star,
 \qquad
 H_0=E_0+\kappa M_0^\star.
\end{equation}
The affine antipode $u\mapsto\kappa u^\star$ sends a transverse exponent $k$ to
$p_\perp-1-k$, so $H_+$ is supported at transverse coordinates at least $p_\perp$
and $H_-$ at transverse coordinates at most $-1$.  By Eq.~\eqref{eq:composite}
and the affine anchoring, commutation with the composites matches the wall
syndromes coefficientwise: writing \(E_0=\sum_jz^j\widehat e_j\) and
\(M_0=\sum_jz^j\widehat m_j\), it gives
\(\widehat e_j+\kappa\widehat m_j^{\,\star}=0\) for every \(j\), hence
\(\overline{H_0}=0\) in $\mathsf B$.  Since $1+z\in\langle f^\star,g^\star\rangle$,
also $\pi_e(H_0)=0$.  The other two parts carry the mismatches
\begin{equation}\label{eq:mismatches}
 \pi_e(H_+)=r_R+s_L^\iota=:\eta_e,
 \qquad
 \pi_e(H_-)=r_L+s_R^\iota=:\eta_m^\iota,
\end{equation}
where $\eta_m=s_R+r_L^\iota$.  The law $a_R=D_{\rm lab}(a_L)$ is exactly the pair
of equations $\eta_e=0$ and $\eta_m=0$.

\begin{lemma}[Charge-accounting constraint on transmitted labels]
\label{lem:star-accounting}
Let $W$ be an admissible finite crossing Pauli as in
Definition~\ref{def:separated-crossing-relation}, and assume its support,
its syndrome clusters, and every check whose support can meet it lie in a
buffered region on which the scaffold agrees with the lifted infinite
straight-wall scaffold.  Write its remote labels as
$a_L=e_{r_L}m_{s_L}$ and $a_R=e_{r_R}m_{s_R}$.  Then
\begin{equation}\label{eq:star-accounting}
    s_L+s_R=(r_L+r_R)^\iota ,
\end{equation}
that is, $\eta_m=\eta_e^\iota$.
\end{lemma}
\begin{proof}
Both $E$ and $\kappa M^\star$ lie in $\langle f^\star,g^\star\rangle$, so
$\pi_e(H)=0$.  As $\pi_e(H_0)=0$, Eq.~\eqref{eq:mismatches} gives
$\eta_e+\eta_m^\iota=0$.
\end{proof}

\begin{remark}[Why the interblock layer is needed]
A crossing takes a label pair \((a_L,a_R)\) in the \(4Q\)-dimensional space
\((V_e\oplus V_m)^2\).  Charge accounting,
Eq.~\eqref{eq:star-accounting}, imposes \(Q\) independent conditions and leaves
\(3Q\) dimensions, while \(\operatorname{Graph}(D_{\rm lab})\) has \(2Q\).  The
missing \(Q\) conditions are what the interblock rows supply.
\end{remark}

\begin{lemma}[Interblock sum identity]
\label{lem:placement-aware-junction-cut}
Let $W$ be an admissible crossing of the fixed transverse-normalized wall, on the
infinite straight wall or in a non-wrapping lifted window containing its exact
syndrome together with every interblock translate that can meet it.  Then for every
offset $u\in\Lambda_0$,
\begin{equation}\label{eq:placement-aware-junction-formula}
 \sigma_u(W)
 =\operatorname{Tr}_{z,w}\!\left(\widehat\eta_e\,\widehat u^{\,\star}\Delta\right)
 =\phi(u,\eta_e),
 \qquad \eta_e=r_R+s_L^\iota ,
\end{equation}
where $u$ also denotes the class of the monomial $\widehat u$ in $V_m$.
\end{lemma}
\begin{proof}
Insert $H=H_++H_0+H_-$ in Eq.~\eqref{eq:gauss-sum}.  The term in $H_0$ vanishes
because $\overline{H_0}=0$.  By
Eq.~\eqref{eq:normalized-transverse-support}, $\overline{\Delta H_+}$ is supported
at transverse coordinates at least $p_\perp$ and $\overline{\Delta H_-}$ at
transverse coordinates at most $p_\perp-2$, while the monomial $w^n\widehat u$ has
transverse coordinate at most $p_\perp-1$ for $n\leq0$ and at least $p_\perp$ for
$n\geq1$.  Thus $H_-$ contributes nothing to Eq.~\eqref{eq:gauss-sum}, and $H_+$
contributes nothing to the same sum over $n\leq0$, so
\[
 \sigma_u(W)
 =\sum_{n\in\mathbb Z}
  \operatorname{Tr}_z\!\left(w^{-n}\widehat u^{\,\star}\Delta H_+\right)
 =\operatorname{Tr}_{z,w}\!\left(\widehat u^{\,\star}\Delta H_+\right).
\]
By Lemma~\ref{lemma:bilinear} the last trace depends on $H_+$ only through its
class, which is $\eta_e$ by Eq.~\eqref{eq:mismatches}.
\end{proof}

\subsubsection{The Transmission Theorem}

\begin{theorem}[Microscopic affine straight-wall transmission]
\label{thm:transmission}
Let the BB code satisfy \(0<Q<\infty\), fix either wall orientation, and let
the wall be a transverse-normalized microscopic longitudinal-ribbon
realization produced by the interblock construction, with any choice of lift
and the corresponding patch rows.  Then finite admissible crossing
Paulis with designated remote syndromes on retained checks exist for every
electric and magnetic label in both crossing directions, and
\begin{equation}\label{eq:transmission-graph}
 \operatorname{Rel}_{\rm wall}=\operatorname{Graph}(D_{\rm lab}).
\end{equation}
Two finite admissible crossing Paulis with the same exact remote syndrome
differ, up to phase, by a completed-wall stabilizer.
\end{theorem}

\begin{proof}
For the forward inclusion, put
\(c_N=S_N(w)c\) and \(d_N=S_N(w)d\).  Then
\begin{equation}\label{eq:deep-transverse-ribbon}
 f^\star c_N+g^\star d_N=1+w^N.
\end{equation}
For \(r\in V_e\), define
\begin{equation}\label{eq:deep-provisional-crossing}
 W_0^{(N)}(r)
 =Z[\widehat r(c_N,d_N)]
  X[\overline r(d_N^\star,c_N^\star)].
\end{equation}
Its exact syndromes are
\(\operatorname{syn}_eW_0^{(N)}(r)=\widehat r(1+w^N)\) and
\(\operatorname{syn}_mW_0^{(N)}(r)=\overline r(1+w^{-N})\).
The block-zero clusters are affine matched, while the strict-bank clusters
lie at distance at least \((N-1)p_\perp+1\), and hence outside the removed
wall block for every \(N\geq1\).  Moreover, \(W_0^{(N)}(r)\) has the affine form
of Lemma~\ref{lem:wall-trace}\textup{(ii)}, so its interblock pattern, that of
Lemma~\ref{lem:affine-crossing} with \(\Delta_N=c_Nv+d_Nh=S_N(w)\Delta\) in place
of \(\Delta\), sums to zero at every offset.

The compact path recurrence therefore supplies removed-check dressing that
cancels every pointwise interblock commutator.  Choose \(N\) large enough that
the remote clusters also lie beyond the uniform support collar of the
relations \(zM_a=M_a\pmod{\mathcal S_0}\).  For the dressed crossing,
\(b_n=\langle W^{(N)}(r),z^nM_a\rangle\) satisfies \(b_{n+1}=b_n\); it
vanishes for large \(\lvert n\rvert\) by support disjointness and hence for
every \(n\).  Thus \(W^{(N)}(r)\) is admissible and commutes with every
patch row.  Reversing \(w\) and exchanging the CSS halves gives both
crossing directions and both tones.  Since the label spaces are finite
dimensional, the standard lifts, incidence dressings, and final choice of
\(N\) may be fixed on a basis with one uniform support bound and extended
linearly.  Hence
\[
 \operatorname{Graph}(D_{\rm lab})
 \subseteq\operatorname{Rel}_{\rm wall}.
\]

Conversely, let \(W\) be an admissible crossing and write its remote labels as
\(a_L=e_{r_L}m_{s_L}\) and \(a_R=e_{r_R}m_{s_R}\).
Commutation with every interblock row gives $\sigma_u(W)=0$, so
Lemma~\ref{lem:placement-aware-junction-cut} gives $\phi(u,\eta_e)=0$ for every
$u\in\Lambda_0$.  The classes of the monomials $\widehat u$ span $V_m$, and
$\phi$ is perfect (Lemma~\ref{lem:braiding-nondegenerate}), so
\(\eta_e=0\).  Lemma~\ref{lem:star-accounting} then gives \(\eta_m=0\).  These are
\(r_R=s_L^\iota\) and \(s_R=r_L^\iota\), that is,
\(a_R=D_{\rm lab}(a_L)\), proving the reverse inclusion.

Finally, if \(W\) and \(W'\) have the same exact remote syndrome, their
remote-check commutators cancel, so \(WW'\) is a compact completed-wall
centralizer.  Straight-wall local exactness makes it, up to scalar phase, a
completed-wall stabilizer.  The uniqueness clause is for equal exact remote syndromes,
not for crossings agreeing only in anyon class.
\end{proof}

\subsubsection{Without Normalization}\label{sec:without-normalization}

Normalizing the ribbon is not a gauge choice: two solutions of the period
equation differ by a syzygy, whose checks inside the removed block the wall
has merged away (Sec.~\ref{sec:junctions}), and by the proposition below an
unnormalized realization can transmit a label pair the duality forbids while
normalized walls of the same code do not.  The anomaly lies in the
realization, not the code.  No necessary-and-sufficient criterion for an
unnormalized realization to be anomalous is proved.

\begin{proposition}[An unnormalized wall realization]
\label{rem:shallow-necessity}
There is a code with $0<Q<\infty$, an un-normalized fixed longitudinal-ribbon
wall realization of it with a patch basis and lifts, and a finite Pauli
of weight $21$ that is an admissible crossing in the sense of
Definition~\ref{def:separated-crossing-relation} with this realization in
place of the transverse-normalized wall and whose remote label pair lies
outside $\operatorname{Graph}(D_{\rm lab})$.  So the remote-label relation
this realization defines is not contained in
\(\operatorname{Graph}(D_{\rm lab})\).
\end{proposition}

\begin{proof}
Fix $\BBc{2,2,1,4}$, so that
$f=1+x+x^{-2}y^4$ and $g=1+y+xy^{-2}$, with $l_0=m_0=3$, $Q=2$ and
$\kappa=x^2y^2$, and fix the un-normalized realization whose period ribbons
are
\begin{equation}\label{eq:shallow-realization}
\begin{aligned}
 h={}&y^3+x^2y^3+y^4+xy^4+y^5+x^2y^5,\\
 v={}&x^2+x^4+y+x^4y+y^2+x^2y^2+y^3+xy^3+x^2y^3+x^3y^3,\\
 c={}&xy^2+y^3+y^4+xy^4,\\
 d={}&x^3y^{-1}+x^2+x^3+y+y^2+x^2y^2,
\end{aligned}
\end{equation}
which satisfy $f^\star h+g^\star v=1+x^3$ and $f^\star c+g^\star d=1+y^3$
exactly.  Its comparison polynomial reduces to
\(\overline\Delta=y^4+x^2y^4+xy^5+x^2y^5+y^6+xy^6\),
whose transverse support $\{4,5,6\}$ lies outside the range
$\{0,1,2\}$ of Eq.~\eqref{eq:normalized-transverse-support}.  For this realization there
is a finite Pauli of weight $21$ whose designated
remote excitations are retained checks, of minimum distance $1$ from the wall
block.  It commutes with every composite and interblock row of a
buffered window, but its remote label pair is
$(a_L,a_R)=(1,\,e_ym_1)\notin\operatorname{Graph}(D_{\rm lab})$, the $-$ and
$+$ banks being the left and right sides as fixed above.  In the label bases
$V_e=V_m=\operatorname{span}(1,y)$ and the component
order~\eqref{eq:pauli-component-order}, one realizing Pauli is
\begin{equation}\label{eq:shallow-witness}
\begin{aligned}
 u_1={}&x^{-2}y^{-3}+xy^{-3}+y^{-2}
        +x^{-2}y^{-1}+x^{-1}y^{-1}+y^{-1}+x^{-2}\\
      &+x^{-3}y+x^{-2}y+x^{-1}y+x^{-2}y^2+x^{-2}y^3
        +x^{-3}y^4+x^{-3}y^5,\\
 v_1={}&x^{-3}y^{-1}+x^{-2}y^{-1}+x^{-2}+x^{-3}y+x^{-3}y^4,\\
 u_2={}&x^{-3}y^{-2}+x^{-3}y^2,
 \qquad v_2=0,
\end{aligned}
\end{equation}
with support bounds $[-3,1]\times[-3,5]$.  Direct Laurent multiplication
gives its exact syndromes: all non-wall syndrome sites are designated
strict-bank checks, and the wall-block violations are paired by the
composites.  These data and its commutators with the composites and interblock
rows follow from Eq.~\eqref{eq:shallow-realization}.  For the patch
family of Eq.~\eqref{eq:endpoint-constant-3} built from these ribbons, exact
$\mathbb F_2$ arithmetic confirms that the basis centralizes the scaffold with
ledgers spanning $\mathcal I_e$, and that the Pauli commutes with every
translate of every patch row meeting its support.  It therefore meets
every clause of admissibility.
\end{proof}


\subsection{Finite Walls}\label{sec:finite-walls}

\begin{definition}[Crossing window]
\label{def:qualifying-crossing-window}
For a transverse-normalized wall, take the pivot
basis $\{\xi_\nu\}$ of $\mathcal I_e$ and the patch rows $M_\nu$ of
Eq.~\eqref{eq:endpoint-constant-3} as the fixed patch basis and family of
Definition~\ref{def:natural-termination}, fix a crossing datum
(Definition~\ref{def:crossing-constants}) of depth $N$ and radius $r_\times$,
and put
\begin{equation}\label{eq:finite-window-constants}
 \rho_{\rm win}
 =\rho_{\rm end}+r_{\rm inc}
  +\max\{2r_{\rm chk},\,d_T,\,r_{\rm pkg}\},
 \qquad
 N_{\rm win}=2\rho_{\rm win}+3,
 \qquad
 L_\times^{\rm room}=\max\{r_\times+2r_{\rm chk},\,r_{\rm pkg}\}.
\end{equation}
Write $N_{\rm wall}$ for the number of blocks of the finite wall, reserving
$N$ for the transverse crossing depth.  Let the wall be naturally terminated,
with blocks carrying exponents $0,\ldots,N_{\rm wall}-1$ and
$N_{\rm wall}\geq N_{\rm win}$.  A \emph{crossing window} at a wall block
$\tau$ is a pair
\(\Omega_{\rm in}=N_{L_\times^{\rm room}}(\{\tau\})\subset
\Omega_{\rm out}=N_{\rho_{\rm win}}(\Omega_{\rm in})\) such that
\begin{enumerate}
\item[\textup{(W1)}] \emph{Bulk window.}
 $\Omega_{\rm out}=N_{\rho_{\rm win}}(\Omega_{\rm in})$ lifts to the plane
 without wrapping and lies in the straight-wall interior; in particular the
 longitudinal projection of its lift lies in the exponent interval
 $\{0,\ldots,N_{\rm wall}-1\}$;
\item[\textup{(W2)}] \emph{Endcaps far.}
 Both endcaps are at least $N_{\rm win}$ complete blocks from
 $\Omega_{\rm out}$.
\end{enumerate}
By the choice of $L_\times^{\rm room}$ it then has two further properties.
\begin{enumerate}
\item[\textup{(W3)}] \emph{Room for one package per direction.}
 For each patch direction $\nu$ the block $b_\nu=\tau$ satisfies
 $N_{r_{\rm pkg}}(b_\nu)\subseteq\Omega_{\rm in}$.
\item[\textup{(W4)}] \emph{Room for the basis crossings.}
 $\Omega_{\rm in}$ contains, with a $2r_{\rm chk}$-collar, the supports of the
 datum's admissible crossings at $\tau$ realizing every label of the fixed
 bases of $V_e$ and $V_m$, in both crossing directions and both CSS tones:
 such crossings exist by Theorem~\ref{thm:transmission}, and their supports lie in
 $N_{r_\times}(\{\tau\})$ (Definition~\ref{def:crossing-constants}).
\end{enumerate}
\end{definition}

\begin{lemma}
\label{lem:finite-window-straight-wall-agreement}
On a naturally terminated wall
(Definition~\ref{def:natural-termination}),
let $\Omega_{\rm in}\subset\Omega_{\rm out}$ satisfy \textup{(W1)}--\textup{(W2)}
and let $W$ be a finite crossing supported in $\Omega_{\rm in}$.  Then every
composite, interblock or installed patch row whose support can meet $W$ is,
after lifting, the corresponding translate of the infinite straight-wall row,
and no endpoint row is encountered.  Every pristine check whose commutator with
$W$ can be nonzero has support in $\Omega_{\rm out}$ and is the corresponding
lifted check, and likewise for a designated remote check and for a removed bank
check used as bookkeeping.  Every interblock row that can meet $W$ has its entire
support in $\Omega_{\rm out}$.  Consequently the finite commutator table of $W$
with everything that can contribute to its syndrome, ledger or interblock
obstruction coincides with the table computed from the lifted infinite
straight-wall scaffold.
\end{lemma}
\begin{proof}
By \textup{(W2)} the finite wall has at least $N_{\rm win}$ blocks of
straight-wall interior on both sides of $\Omega_{\rm out}$, so natural
termination (Definition~\ref{def:natural-termination}) installs every
composite, interblock and patch row meeting $\Omega_{\rm out}$ as the
translate of the straight-wall row, the whole package of such a patch
translate lying in the wall since \(N_{\rm win}>2r_{\rm pkg}\geq\max A_a-\min A_a\); and by \textup{(W1)} the lift is injective
there.  A pristine check $q$ with nonzero commutator against $W$ meets
$\operatorname{supp}W\subseteq\Omega_{\rm in}$, so
$\operatorname{supp}q\subseteq N_{2r_{\rm chk}}(\Omega_{\rm in})
\subseteq\Omega_{\rm out}$ by Eq.~\eqref{eq:finite-window-constants}, and
likewise for a designated remote check and a removed bank check; an interblock row
$T$ meeting $W$ has diameter at most $d_T$, so
$\operatorname{supp}T\subseteq N_{d_T}(\Omega_{\rm in})\subseteq\Omega_{\rm out}$.
\end{proof}

\begin{corollary}
\label{cor:finite-transmission}
This is the finite-wall form of Theorem~\ref{thm:transmission}.
Let $\Omega_{\rm in}$ be a crossing window
(Definition~\ref{def:qualifying-crossing-window}).

Restrict $\operatorname{Rel}_{\rm wall}$ to admissible crossings $W$ with
$\operatorname{supp}W\subseteq\Omega_{\rm in}$.  Then:

\emph{(i)} every electric and magnetic label is realized in both crossing
directions by an admissible crossing;

\emph{(ii)} the restricted relation satisfies
$\operatorname{Rel}_{\rm wall}=\operatorname{Graph}(D_{\rm lab})$;

\emph{(iii)} if $W$ and $W'$ are two such crossings with the same exact remote
syndrome, then $WW'$ is, up to scalar phase, a completed-wall stabilizer, and admits a
factorization into installed rows (retained checks, composites, interblock rows and installed patch rows) with \emph{every individual factor}
supported in $\Omega_{\rm out}$.  Retained checks are named explicitly here
because the scaffold $\mathcal R_0$ contains them, so they occur among the
factors even though they are not completed-wall rows in the sense of
Definition~\ref{def:separated-crossing-relation}.
\end{corollary}
\begin{proof}
By \textup{(W4)}, \(\Omega_{\rm in}\) contains a basis crossing for every label, in both directions and CSS tones, at the datum's depth and at one common anchor, while \textup{(W1)}--\textup{(W2)} and the prescribed collars make them admissible for the finite wall.
Products of these crossings realize every electric and magnetic label, remain supported in \(\Omega_{\rm in}\), and remain admissible at that anchor.
Since \(D_{\rm lab}\) is additive, this proves \textup{(i)} and the inclusion
\(\operatorname{Graph}(D_{\rm lab})\subseteq\operatorname{Rel}_{\rm wall}\).
Conversely, by Lemma~\ref{lem:finite-window-straight-wall-agreement} every wall row meeting a crossing supported in \(\Omega_{\rm in}\) is the corresponding installed straight-wall row.
Thus every restricted admissible crossing is covered by Theorem~\ref{thm:transmission}, giving
\(\operatorname{Rel}_{\rm wall}\subseteq\operatorname{Graph}(D_{\rm lab})\) and proving \textup{(ii)}.
For \textup{(iii)}, put \(P=WW'\).
Equality of the exact remote syndromes and admissibility imply that \(P\) is supported in \(\Omega_{\rm in}\) and centralizes every scaffold row; hence \(h:=\ell(P)\in\mathcal I_e\) by Lemma~\ref{lem:charge-ledger}.
Since \(r_{\rm inc}=r_{\rm chk}+r_T+1\), Eqs.~\eqref{eq:finite-window-constants} and~\eqref{eq:effective-R0} give \(\rho_{\rm win}\ge R_0\).
Write \(h=\sum_\nu h_\nu\xi_\nu\); by \textup{(W3)}, Theorem~\ref{thm:completion-lift}, Lemma~\ref{lem:endpoint-completion-retention}\textup{(ii)}, and Definition~\ref{def:natural-termination}, the corresponding translates \(M_{\nu,b_\nu}\) are installed, supported in \(\Omega_{\rm in}\), and have ledger \(\xi_\nu\).
For \(M=\prod_\nu M_{\nu,b_\nu}^{h_\nu}\), therefore, \(PM\) is supported in \(\Omega_{\rm in}\), centralizes the scaffold, and has zero ledger.
Corollary~\ref{cor:buffered-injectivity} yields \(PM\in\mathcal S_{\rm in,out}\), so it is a product of retained checks, composites, and interblock rows, each individually supported in \(\Omega_{\rm out}\).
Multiplying by the installed patch factors of \(M\), themselves supported in \(\Omega_{\rm in}\subseteq\Omega_{\rm out}\), gives the required completed-wall-stabilizer factorization of \(P=WW'\), up to
scalar phase.
\end{proof}

\begin{corollary}[Local exactness up to the room]\label{cor:distance}
Let $\Lambda\ge0$ and work in a standard configuration at room
$\rho_0+\Lambda$.  Every Pauli that commutes with
every row and is supported in a non-wrapping box of block diameter at most
$\Lambda$ is, up to phase, a product of rows, each within a fixed buffer of the
box.  With $\delta_H$ the largest block diameter of a row, the code distance
therefore exceeds $1+\Lambda/\delta_H$; for $n\geq2$ the pair loops $Z_{t,b}$, $t<n$, $b\ne0$, are
logical (Theorem~\ref{thm:pca-register}) and bound it by their weight, at most
$2s_{\rm rib}(L_t+W_t)$ (Lemma~\ref{lem:period-rectangle-corridors}).  The same
factorization gives the conditions TQO-1 and TQO-2 of
Ref.~\cite{bravyiTopologicalStability2010} at every scale up to $\Lambda$, for
coarse cells larger than twice the buffer; for a family with the code and its
realization fixed and the walls, separations and room growing in proportion to
a square torus, the gap is then stable under weak local perturbations.
\end{corollary}
\begin{proof}
Walls are at least $\rho_0+\Lambda$ long and apart, also from their own
periodic images, and as in the proof of Lemma~\ref{lem:room} both torus periods
exceed $2(\rho_0+\Lambda)$.  Put
\[
 a=r_{\rm chk}+r_{\rm pkg}+2,
 \qquad
 e=a+\rho_{\rm win}+N_{\rm win}+1,
 \qquad
 b=e+\rho_{\rm end}+2 ;
\]
\(\rho_0\) is taken with \(\rho_0\geq e+\rho_{\rm end}+3\) and
\(\rho_0\geq b+r_{\rm chk}+\max\{r_{\rm chk},r_{\rm row}\}+2\).  Let $A$ be a box of
diameter at most $\Lambda$.  Then \(N_b(A)\) lifts without wrapping, and if
$A$ is within \(r_{\rm chk}\) of the carrier of a wall $s$, \(N_b(A)\) contains
no removed check of another wall.  If $A$ is farther than \(r_{\rm chk}\) from
every carrier, Lemma~\ref{lem:compact-filling} factors the Pauli into retained
checks within \(r_{\rm chk}\) of $A$.  If $A$ is within \(r_{\rm chk}\) of the
carrier of wall $s$ and its longitudinal projection lies in the wall at least
$e$ blocks from both end blocks, then \(\Omega_{\rm in}=N_a(A)\) and
\(\Omega_{\rm out}=N_{\rho_{\rm win}}(\Omega_{\rm in})\) satisfy
\textup{(W1)}--\textup{(W3)}, with $b_\nu$ the carrier block nearest $A$, and
the proof of Corollary~\ref{cor:finite-transmission}\textup{(iii)}, which uses
of $WW'$ only its support and its commutation with the rows, and of the window
only \textup{(W1)}--\textup{(W3)}, factors the Pauli in
\(\Omega_{\rm out}\subseteq N_b(A)\).  Otherwise $A$ lies within $e$ blocks of
the end block of an endpoint $s$; the least block rectangle containing $A$ and
that block lies in \(N_{e+1}(A)\) and at least
\(\rho_0+\Lambda-(\Lambda+e+3)\geq\rho_{\rm end}\) blocks from $s_{\rm opp}$,
so it is an endpoint window with \(\Omega_{\rm out}\subseteq N_b(A)\), and
Theorem~\ref{thm:endpoint-exactness} gives the factorization.  A
commuting Pauli of weight at most $1+\Lambda/\delta_H$ splits into pieces
linked through shared rows, each commuting with every row by itself and lying
in such a box, so it is a stabilizer.  The factorization is TQO-1, and TQO-2
once each row is assigned to a $2\times2$ square of cells containing it,
chosen in each coordinate toward the side of the row's midpoint;
Theorem~1 of Ref.~\cite{bravyiTopologicalStability2010} then gives the gap.
\end{proof}

\section{The \texorpdfstring{$F$}{F} and \texorpdfstring{$R$}{R} Matrices as Categorical Data}\label{sec:FR}

The fusion rules \eqref{eq:fusion-rule} determine which processes can occur;
the $F$- and $R$-matrices determine their amplitudes.  This section matches the
microscopic data --- the channel group, its pairing, and the two
loop families --- to the data of a Tambara--Yamagami category
(Sec.~\ref{sec:TY}), reads the recoupling off that identification twice, once
from the model and once from the category (Sec.~\ref{sec:Fmatrix}), and reads
off the exchange, separating what the code fixes from the choice of quadratic
refinement (Sec.~\ref{sec:Rmatrix}).

\paragraph*{Setting for this section.}
Sec.~\ref{sec:FR} uses the period ribbons of
Definition~\ref{def:standard-layout}\textup{(S1)} and the label section of
Lemma~\ref{lem:endpoint-caps} in a common non-wrapping lift, so that for
$b\in V_m$ one may write $B_b=\widehat{s_m(b)}$ and $B_b^\iota=\kappa B_b^\star$ and quantify over
$b$ freely.  Results in this section assume this Setting and state only the
hypotheses that go beyond it.

\subsection{Microscopic Data for the Tambara--Yamagami Category}\label{sec:TY}

We first separate a pairing fixed directly by BB ribbon algebra from the
abstract bicharacter of a Tambara--Yamagami (TY) category.  Recall that TY
associativity data are classified by a non-degenerate symmetric bicharacter
$\chi$ on $A$ together with a sign $\nu=\pm1$, giving
$[F^{\sigma\sigma\sigma}_\sigma]_{a,b}=\nu|A|^{-1/2}\chi(a,b)^{-1}$
\cite[Def.~3.1 and Thm.~3.2]{tambaraTensor1998}, while the ordinary braidings of
$\mathrm{TY}(A,\chi,\nu)$ exist precisely when every element of $A$ squares
to the identity \cite[Thm.~1.2]{siehlerBraided2000}, which $A\cong
\mathbb Z_2^Q$ satisfies; the modular data of the Drinfeld centre
$Z(\mathrm{TY})$ are given in Ref.~\cite[\S4.3]{gelakiCenters2009}.  What the code fixes is not the existence
of such data but \emph{which} bicharacter occurs: the identification
$\chi=\lambda$ below makes it the code's pairing.  The exchange phases of the twists are
crossed $R$-symbols of the $\mathbb Z_2$-crossed extension describing
extrinsic defects~\cite[\S V and \S VI.B]{barkeshliSymmetry2019}, the
$G$-graded categorical extension of~\cite{etingofFusionHomotopy2010}, not an
ordinary braiding of the TY fusion category.

The twist and the invariant channels form a Tambara--Yamagami category
$\mathrm{TY}(A,\chi,\nu)$: $A$ is computed from the code below, $\chi=\lambda$ is
Theorem~\ref{prop:chi-equals-lambda}, and of the fusion-category data only the
sign $\nu$ is left undetermined (Sec.~\ref{sec:Rmatrix}).  For fixed $(A,\chi)$ the two
TY fusion categories differ precisely in $\nu$.  Nor can the charge sectors supply it --- the squared
vacuum-channel overlap is $|A|^{-1}$ identically, which fixes
$\lvert[F^{\sigma\sigma\sigma}_\sigma]_{0,0}\rvert$ and erases its sign.
No further $\mathbb F_2$ datum computed here determines $\nu$.

\paragraph*{The code-derived pairing.}
Let
\begin{equation}
 A=\{\epsilon_a=(a,a^\iota):a\in V_m\}\cong V_m
\end{equation}
be the additive group of defect-invariant channels, indexed by their
magnetic components; we use that indexing throughout, writing $\epsilon_a$ for
the channel and $a\in V_m$ for its index.  The pairing $\lambda$ of
Eq.~\eqref{eq:code-channel-pairing}, which Theorem~\ref{thm:pca-weyl} makes the
commutation phase of the two loop families, is read on channels through their
indices, $\lambda(\epsilon_a,\epsilon_b):=\lambda(a,b)$; the arguments fed to
$\mathcal B$ are always the indices, never the pairs themselves.

\begin{lemma}[Channel spin and trivial parent monodromy]\label{lem:code-channel-pairing}
The diagonal of the symmetric non-degenerate bicharacter $\lambda$
(Lemma~\ref{lem:pca-perfect}) is the spin of the channel,
\begin{equation}
 \lambda(a,a)=\theta_{\epsilon_a}=\B{a}{a^\iota}.
\end{equation}
The full parent-theory monodromy on $V_e\oplus V_m$, for $r,r'\in V_e$ and
$s,s'\in V_m$,
\begin{equation}\label{eq:full-anyon-monodromy}
 \mathfrak M(e_rm_s,e_{r'}m_{s'})
 =\B{s}{r'}\B{s'}{r},
\end{equation}
is nevertheless trivial on two invariant channels:
\begin{equation}\label{eq:invariant-full-monodromy}
 \mathfrak M(\epsilon_a,\epsilon_b)
 =\lambda(a,b)\lambda(b,a)=1.
\end{equation}
\end{lemma}
\begin{proof}
The diagonal formula is the spin of the composite of bosonic electric and
magnetic constituents.  Eq.~\eqref{eq:full-anyon-monodromy} gives the first
equality in Eq.~\eqref{eq:invariant-full-monodromy}, and symmetry of $\lambda$
and $\lambda(a,b)^2=1$ give the second.
\end{proof}

\paragraph*{The abstract TY category.}
The abstract TY category associated with the fusion rules
Eq.~\eqref{eq:fusion-rule} is specified by a symmetric non-degenerate
bicharacter $\chi:A\times A\to U(1)$ and
\begin{equation}
 \tau_{\rm TY}=\frac{\nu}{\sqrt{|A|}},\qquad \nu\in\{\pm1\}.
\end{equation}
In the normal basis of Ref.~\cite{siehlerBraided2000},
$\chi(a,b)=\sigma_0(a,b)$ is the scalar for the single braiding isomorphism
$\epsilon_a\epsilon_b\to\epsilon_b\epsilon_a$.  It is not their full
monodromy, which is $\sigma_0(a,b)\sigma_0(b,a)=1$.

\subsubsection{Commutation within the Transfer Family}

Theorem~\ref{thm:pca-weyl} gives the commutation phase of the two loop
families, and Lemma~\ref{lem:pca-perfect} makes the pairing behind it perfect.  For them to define two \emph{bases}, each family must
also be internally commuting: one cannot simultaneously measure
noncommuting observables.  For the pair-loop family $Z_s$, of pristine boundaries, this is immediate by
Lemma~\ref{lem:closed-loop-interior}.  The transfer family $X_s$ needs a
computation, which we now record.

\begin{lemma}[The transfer family is isotropic]
\label{lem:L23-isotropy}
Under the hypotheses of Lemma~\ref{lem:pca-ledger}, in a
standard configuration (Definition~\ref{def:standard-layout}), and
for any crossing directions $w_u$ in \textup{(S3)},
\begin{equation}\label{eq:L23-isotropy}
 \left\langle X_{s,a},X_{s,b}\right\rangle=0
 \qquad\text{for all }a,b\in V_m .
\end{equation}
\end{lemma}

\begin{proof}
Write $X_{s,b}=U(b)D(b)$, where $D(b)=D_s(b)D_{n,s}(b)$ collects the two
removed-check transmission dressings and $U(b)$ is the undressed product of the two
crossings and the two corridors.  Substituting the displayed crossing and
corridor formulas gives the exact binary identity
\begin{equation}\label{eq:L23-seed-factorization}
 U(b)=Z\!\left[B_b^\iota(E_1,E_2)\right]X\!\left[B_b(M_1,M_2)\right],
\end{equation}
in which the four finite Laurent seeds $E_1,E_2,M_1,M_2$ collect both
crossing contributions and every corridor edge and, crucially, do not depend
on the label $b$.

Expand the symplectic form using
Eq.~\eqref{eq:L23-seed-factorization} and put
$F_{a,b}=\kappa^\star AB\left(E_1^\star M_1+E_2^\star M_2\right)$, where
$A=\widehat{s_m(a)}$ and $B=\widehat{s_m(b)}$.  Since the Laurent ring is
commutative and $A^\iota=\kappa A^\star$, the two opposite-tone
contributions are $F_{a,b}$ and its antipode $F_{a,b}^\star$.  The
coefficient trace is antipode invariant, so
\[
 \left\langle U(a),U(b)\right\rangle
 =\operatorname{tr}\!\left(F_{a,b}+F_{a,b}^\star\right)
 =2\operatorname{tr}F_{a,b}=0
\]
over $\mathbb F_2$.  Every same-tone pairing vanishes by the CSS form.

For the dressings, products of pristine checks commute, so
$\langle D(a),D(b)\rangle=0$.  By boundary--syndrome adjointness $D_u(a)$ reads only the
exact syndrome of $U(b)$ on its removed checks, which by
Lemma~\ref{lem:pca-ledger} and the isolation \textup{(L5)} is the ledger
$\Lambda_u(b)$ at its own anchor.  Label the blocks of that wall as in
Sec.~\ref{sec:composites}, block $j$ carrying $z^{j-1}$ relative to the anchor;
write $B=\sum_vB_v\widehat v$; and let $c_{j,v}(a)$ be the parity of the
coefficients of $S^X_{j,\iota(v)}$ and $S^Z_{j,v}$ in the prescribed compact
straight-wall dressing, carried into the window with the same factors, each
within \(r_\times\) of the anchor by Definition~\ref{def:crossing-constants}
(Lemma~\ref{lem:finite-window-straight-wall-agreement}).  By
Eq.~\eqref{eq:affine-anchor}, $B^\iota=\sum_vB_v\widehat{\iota(v)}$, so the
anchor contributes $\sum_vB_vc_{1,v}(a)$.  The dressed crossing commutes with
every interblock row, so $c_{j,v}+c_{j+1,v}=t_{j,v}$, the pattern of
Eq.~\eqref{eq:crossing-junction-pattern} with $\widehat r=A^\iota$ and
$\Delta_u$ in place of $\Delta$, and compactness gives
$c_{1,v}=\sum_{j\le0}t_{j,v}$.  Antipode invariance of the trace turns the
pattern into $t_{j,v}(a)=\operatorname{tr}\bigl(\kappa^\star A\widehat v
\Delta_u^\star(z^j+z^{1-j})\bigr)$; as $j\le0$ and $1-j\ge1$ exhaust
$\mathbb Z$, the anchor contributes
$\operatorname{Tr}_z(\kappa^\star AB\,\Delta_u^\star)$, symmetric in $a$ and
$b$.  So the mixed terms $\langle D(a),U(b)\rangle$ and $\langle D(b),U(a)\rangle$
cancel, giving Eq.~\eqref{eq:L23-isotropy}.
\end{proof}

\begin{lemma}[Phase normalization of the transfer family]
\label{lem:L23-group-law}
Assume Lemma~\ref{lem:L23-isotropy}, so that the binary Paulis
$X_{s,a}$, $a\in V_m$, pairwise commute, and recall that the construction is
linear in the label.  Then Hermitian lifts may be chosen so that
\begin{equation}\label{eq:L23-group-law}
 X_{s,0}=I,
 \qquad
 X_{s,a}X_{s,b}=X_{s,a+b}
 \qquad\text{for all }a,b\in V_m ,
\end{equation}
as an identity of operators and not merely modulo scalar phase.
\end{lemma}
\begin{proof}
Fix an $\mathbb F_2$-basis $\mu_1,\ldots,\mu_Q$ of $V_m$ and give each
$X_{s,\mu_i}$ its Hermitian lift; a Hermitian Pauli squares to $I$.  By
Lemma~\ref{lem:L23-isotropy} these commute pairwise, so
\[
 X_{s,\sum_i a_i\mu_i}
 :=\prod_i\bigl(X_{s,\mu_i}\bigr)^{a_i},
 \qquad a_i\in\{0,1\},
\]
is independent of the order of the factors and hence well defined, and it
agrees with the original operator up to phase because the construction is
label-linear.  Since $V_m$ is elementary abelian and each generator squares to
$I$, the assignment is a homomorphism, which is
Eq.~\eqref{eq:L23-group-law}.  Commutativity is what makes the product
unambiguous; without it no such choice need exist.
\end{proof}

\begin{corollary}[Two complete measurement bases]
\label{cor:two-bases}
Under the hypotheses of Lemma~\ref{lem:L23-isotropy} and
Theorem~\ref{thm:exhaustion}, each family $\{Z_{s,a}\}$ and $\{X_{s,b}\}$ is
commuting, carries a group law in its label as an operator identity
(Lemmas~\ref{lem:pair-loop-characters}, \ref{lem:L23-isotropy}
and~\ref{lem:L23-group-law}), and is complete on the $|A|$-dimensional
four-twist space at fixed total charge: the loop--transfer algebra is the full
matrix algebra there (Theorem~\ref{thm:exhaustion}), and
Theorem~\ref{thm:pca-weyl} with non-degeneracy of $\lambda$
(Lemma~\ref{lem:pca-perfect}) gives each family $|A|$ distinct
characters, hence rank-one joint eigenspaces.  The rank-one conclusion does
not need the group law; the recoupling argument below does.  With the transfer route realizing the
$(23)(14)$ cut (Sec.~\ref{sec:probes}), the two joint eigenbases are the
$(12)(34)$ and $(23)(14)$ fusion bases.
\end{corollary}

\subsubsection{The Tambara--Yamagami Identification}

\begin{theorem}[Microscopic TY identification]\label{prop:chi-equals-lambda}
Assume Theorems~\ref{thm:fusion-reconstruction}, \ref{thm:exhaustion} and
\ref{thm:pca-weyl}, the last for the Weyl relation, and
Corollary~\ref{cor:two-bases} for the internal commutativity and group law
of the $X_{s}$ family.  Then, with the transfer route realizing the $(23)(14)$ cut, the $(12)(34)$ and
$(23)(14)$ fusion bases diagonalize the two loop families in
Theorem~\ref{thm:pca-weyl}.  The Weyl relation and global rank exhaustion
force the recoupling between them to be the finite Fourier transform with
kernel $\lambda$, up to row and column phases and a simultaneous relabeling of
$A$; this cut comparison is the categorical associator, whose row-and-column-phase invariant determines $\chi$.  With labels fixed operationally by the Wilson-loop signatures, the
TY bicharacter is therefore the code's own braiding form: in the TY normal
basis,
\begin{equation}\label{eq:microscopic-TY-identification}
 \chi(a,b)=\lambda(a,b)=\B{a}{b^\iota}.
\end{equation}
\end{theorem}
\begin{proof}
Non-degeneracy of $\lambda$ makes Theorem~\ref{thm:pca-weyl} the finite Weyl
algebra of $A$.  Theorem~\ref{thm:exhaustion} supplies the \emph{dimension}
$|A|$ of the completed fusion space, and irreducibility is then automatic.
The exhaustion theorem treats all registers; here there is only the
$Z_{s},X_{s}$ register.  Indeed the commuting family $Z_s$ splits the space into joint eigenspaces
labelled by the characters $\lambda(\cdot,c)$, distinct by non-degeneracy; by
the Weyl relation $X_{s,b}$ maps the $c$-eigenspace onto the $(c+b)$-eigenspace,
so all $|A|$ eigenspaces occur with one common dimension, which the dimension
$|A|$ forces to be one.  This is the finite Stone--von~Neumann statement: the
representation is irreducible and unique up to equivalence.
Choose
$Z_{s,a}|c\rangle=\lambda(a,c)|c\rangle$.  The Weyl relation implies that
$X_{s,b}|c\rangle$ has the $Z_{s}$ character labeled by $c+b$, up to a phase ---
this step needs $\lambda$ \emph{symmetric}, by
Lemma~\ref{lem:pca-perfect}, non-degeneracy alone leaving it twisted.  After rephasing $|c\rangle$, the $X_{s}$ eigenvectors are
Fourier sums with kernel $\lambda(c,d)$.  Row and column phases are not
all fusion-vertex gauge, which fixes the $(0,0)$ entry; read the bicharacter
instead off the phase-independent ratio $U_{ab}U_{00}/(U_{a0}U_{0b})$, equal to
$\lambda(a,b)$ here and to $\chi(a,b)^{-1}$ for the TY normal associator
$F_{ab}=\tau_{\rm TY}\chi(a,b)^{-1}$, both signs on an elementary abelian $A$.
A simultaneous group automorphism is fixed once the microscopic Wilson loops
are assigned their operational labels.
\end{proof}

The code-specific formulas below use $\chi=\lambda$; the abstract TY formulas are
independent of this identification.

\subsection{F-Matrix}\label{sec:Fmatrix}

The recoupling matrix is computed from the model and compared with the
Tambara--Yamagami classification.

\paragraph*{From the model.}
Operationally, recoupling is a change of measurement basis between the
groupings $(12)(34)$ and $(23)(14)$, each with a complete pair-charge family
(Corollary~\ref{cor:two-bases}, the transfer route realizing the $(23)(14)$
cut); the matrix relating the two outcome bases is the
recoupling matrix.  By the proof of Theorem~\ref{prop:chi-equals-lambda} it is
the finite Fourier transform
\begin{equation}\label{eq:microscopic-recoupling}
 \left[\,U_{(12)\to(23)}\,\right]_{a,b}
 =\frac{1}{\sqrt{|A|}}\,\mathcal B(a,b^\iota),
\end{equation}
up to the phase convention chosen for each basis vector.  Nothing
categorical enters this kernel, which follows from the commutation relations
of operators built from the code polynomials.

\paragraph*{From the category.}
In the canonical TY normal basis, the nontrivial recoupling matrix is
\begin{equation}\label{eq:TY-F-matrix}
 \left[F_{\sigma}^{\sigma\sigma\sigma}\right]_{a,b}
 =\frac{\nu}{\sqrt{|A|}}\,\chi(a,b)^{-1}.
\end{equation}
For $A\cong\mathbb Z_2^Q$, $\chi$ takes values in $\{\pm1\}$ and the
inverse may be omitted.  Equation~\eqref{eq:TY-F-matrix}, rather than a
general symmetric-Hadamard ansatz, is the complete canonical solution of the
pentagon equations~\cite{tambaraTensor1998,siehlerBraided2000}.  Other
fusion-vertex gauges multiply rows and columns by the corresponding vertex
phases.  A simultaneous permutation can
always be used merely to reorder a displayed table.  A physical relabeling
of the group objects, however, must be induced by a group automorphism of
$A$; an automorphism preserving the same displayed bicharacter lies in
$\operatorname{Aut}(A,\chi)$.

For $\BBc{1,1,3,1}$ we use throughout the binary-compatible magnetic-label
order
\begin{equation}\label{eq:1131-channel-order}
 (0,x^2,x,x^2+x,1,x^2+1,x+1,x^2+x+1),
\end{equation}
corresponding to the ordered basis $(x^2,x,1)$.  Direct evaluation of
Eq.~\eqref{eq:code-channel-pairing} gives exactly the $8\times8$ sign matrix
tabulated in Table~\ref{table:measurement1131} (Sec.~\ref{sec:measurement}),
read as $\lambda(a,b)=\B{a}{b^\iota}$ with both labels in the
order~\eqref{eq:1131-channel-order}.  It is symmetric, Hadamard, and
biadditive in the channel labels, and its diagonal is the spin
$\theta_{\epsilon_a}$.  Under
Eq.~\eqref{eq:microscopic-TY-identification}, this is the matrix $\chi$
entering Eq.~\eqref{eq:TY-F-matrix}.

\paragraph*{The two descriptions agree.}
Since $\chi=\lambda$ by Theorem~\ref{prop:chi-equals-lambda} and
$\chi^{-1}=\chi$, Eqs.~\eqref{eq:microscopic-recoupling}
and~\eqref{eq:TY-F-matrix} have the same entries up to the overall sign
$\nu$.  Neither makes the other redundant:
Eq.~\eqref{eq:microscopic-recoupling} gives the code's spectral overlap but
not coherence for larger twist configurations, which
Eq.~\eqref{eq:TY-F-matrix}, the complete solution of the pentagon equations,
supplies; the latter in turn needs the data
$(A,\chi)$ from outside the pentagon --- here the code's pairing.

One datum is supplied by neither, the sign $\nu$: it is invisible to the
spins $\theta_{\epsilon_a}$, which are unchanged under $\nu\mapsto-\nu$.  The sign remains undetermined here, and Sec.~\ref{sec:clifford} verifies that no later conclusion
depends on it: $\nu$ contributes only an overall phase to a braid word.

\subsection{R-Matrix}\label{sec:Rmatrix}

The bicharacter $\lambda$ is fixed by the microscopic Wilson/transfer
algebra; a quadratic refinement of it is not fixed by that algebra alone.

The TY hexagon equations are most transparently solved by a quadratic
refinement
\begin{equation}\label{eq:quadratic-refinement}
 \varrho(a+b)=\varrho(a)\varrho(b)\chi(a,b),
 \qquad \varrho(0)=1.
\end{equation}
Because every $a\in A$ obeys $2a=0$, \(\varrho(a)^2=\chi(a,a)\).
In the same normal basis as Eq.~\eqref{eq:TY-F-matrix}, the categorical exchange of two
twists is diagonal in their fusion channel and satisfies
\begin{equation}\label{eq:TY-R-matrix}
 R_{\sigma\sigma}^{\epsilon_a}
 =R_{\sigma\sigma}^{\mathbf1}\,\varrho(a)^{-1},
 \qquad
 \left(R_{\sigma\sigma}^{\mathbf1}\right)^2
 =\tau_{\rm TY}\sum_{c\in A}\varrho(c).
\end{equation}
Thus the relative channel phases are not independent.  Once the values of
$\varrho$ on a basis of $A$ are chosen, all remaining values are fixed by
Eq.~\eqref{eq:quadratic-refinement}.

The code fixes every square $\varrho(a)^2$ of the categorical refinement: by
Eq.~\eqref{eq:quadratic-refinement} and Theorem~\ref{prop:chi-equals-lambda},
$\varrho(a)^2=\chi(a,a)=\lambda(a,a)=\B{a}{a^\iota}=\theta_{\epsilon_a}$, the
spin of the channel (Lemma~\ref{lem:code-channel-pairing}).  Hence the relative
exchange phase $R_{\sigma\sigma}^{\epsilon_a}/R_{\sigma\sigma}^{\mathbf1}=\varrho(a)^{-1}$
squares to $\theta_{\epsilon_a}$: it is $\pm1$ on bosonic and $\pm i$ on
fermionic channels.  What is left undetermined is one sign per basis element
of $A$, a character's worth of freedom.

Extrinsic-defect exchange is defined only up to an overall
phase~\cite{barkeshliTwist2013,bombinTopological2010}.  Separately, changing the
refinement by a character multiplies the exchange by a diagonal Pauli, up to
phase, and Eq.~\eqref{eq:TY-R-matrix} fixes the categorical exchange once the refinement,
$\nu$, and the overall root are chosen.

Which refinement is realized therefore matters for the exchange phases of a
given code, but not for the conclusions drawn from them below.  Every
refinement obeys $\varrho(a)^2=\chi(a,a)\in\{\pm1\}$, so its values are
fourth roots of unity and lie in the phase group of the Pauli group.
Theorem~\ref{thm:clifford} uses those fourth-root factors together with
Eq.~\eqref{eq:quadratic-refinement} itself, not the values alone.
Theorem~\ref{prop:sync} is insensitive to it for a different reason: its conclusion is an identity between binary symplectic matrices, and two refinements differ by a character, so the exchanges differ, up to an overall phase, by a Pauli, which has trivial binary action.  Both statements are therefore the same for all
$2^{Q+1}$ braided choices at each sign $\nu$.

The sign $\nu$ is the Frobenius--Schur indicator of $\sigma$ in the canonical pivotal
structure~\cite[Thm.~3.4]{shimizuFrobenius2011}, and the two signs give
inequivalent fusion categories~\cite[Thm.~3.2]{tambaraTensor1998}.

If $\varrho_0$ is one quadratic refinement, every other one is
$\varrho_0\psi$ for a character $\psi\in\widehat A$.  Hence, for each fixed
$\nu$, there are $|A|=2^Q$ channelwise refinements and, for each, two roots for
$R_{\sigma\sigma}^{\mathbf1}$.  These are the $2^{Q+1}$ ordinary braided choices for the TY fusion category in
Ref.~\cite{siehlerBraided2000}.  A physical twist exchange
agrees with the corresponding exchange of Eq.~\eqref{eq:TY-R-matrix} up to an
overall phase and a Pauli.  Collapsing the channel assignments of these choices to
unordered multisets can lose the Pauli factor distinguishing the exchanges,
although their binary symplectic action is unchanged.

The diagonal of a symmetric $\{\pm1\}$-valued bicharacter is a character, and
its sum obeys a strict dichotomy.

\begin{lemma}[Diagonal character sum]\label{thm:diagonal-character-sum}
Let $\chi:A\times A\to\{\pm1\}$ be a symmetric bicharacter on
$A\cong\mathbb Z_2^Q$.  Then
\begin{equation}
 \sum_{a\in A}\chi(a,a)\in\{0,|A|\}.
\end{equation}
\end{lemma}
Symmetry makes $d_\chi(a)=\chi(a,a)$ a character of $A$, and a character sums
to $|A|$ or to $0$.  The two cases are the two kinds of symmetric form over
$\mathbb F_2$, $d_\chi$ being trivial exactly when $\chi$ is alternating; they
separate the boson-only from the half-fermionic levels of the hierarchy of
Ref.~\cite[\S5]{scrubyHierarchy2020}.  A non-degenerate alternating form
exists only in even dimension: for $a\neq0$ choose $b$ with $\chi(a,b)=-1$; the
plane $\langle a,b\rangle$ is non-degenerate, $A$ is its orthogonal sum with its
complement, and induction applies.

\begin{corollary}[Odd index forces fermions]\label{cor:odd-Q-fermions}
If $Q$ is odd, exactly half of the $2^Q$ channels $\epsilon_a$ are fermions.
\end{corollary}
\begin{proof}
The pairing $\lambda$ is symmetric and non-degenerate
(Lemma~\ref{lem:pca-perfect}).  Were every channel bosonic, $\lambda$
would be alternating, which forces $Q$ to be even.  So $d_\lambda$ is a
non-trivial character and equals $-1$ on exactly half of $A$.
\end{proof}

For the code-derived pairing, $d_\lambda(a)=\B{a}{a^\iota}$ is the spin of
$\epsilon_a$.  For $\BBc{1,1,3,1}$ the diagonal in the order
Eq.~\eqref{eq:1131-channel-order} is
$(+1,-1,+1,-1,-1,+1,-1,+1)$ and sums to zero, as
Corollary~\ref{cor:odd-Q-fermions} requires for $Q=3$.

\section{Measurement-Only Braiding}\label{sec:measurement}

The preceding sections define the twists and their categorical data; this
section shows how to \emph{braid} them without physically moving them,
using only pair-charge measurements by Wilson loops.  The twist is self-dual,
so $\bar\sigma=\sigma$ in the notation below.

We use the measurement-only
framework~\cite{bondersonMeasurementOnly2008}, which replaces the physical
braiding of twists ($\sigma$) by an adaptive sequence of topological charge
measurements.  To braid twists 1 and 4 without moving them, introduce an
auxiliary pair of twists, 2 and 3, initialized
from the vacuum $\mathbf{1}$ into the maximally entangled state
$\left| \bar{\sigma},\sigma;\mathbf{1} \right\rangle_{23}$.  At vacuum total
charge, four twists with $(23)$ in vacuum leave $(14)$ in vacuum as well, so a
third pair, 5 and 6, carries the charge that balances $(14)$.

\paragraph*{Setting.}
Work in a standard configuration on $n=3$ walls
(Definition~\ref{def:standard-layout}), built as in its construction, with
wall 3 the reference.  Number the twists so that wall 2 carries twists 2
and 3, wall 3 carries twists 1 and 4, and wall 1 carries twists 5 and 6, with
twists 2 and 1 the facing ends of wall 2 and the reference wall; the route
$\gamma_2$ of \textup{(S3)} then encloses twists 1 and 2.  With the
no-excitation normalization, Theorem~\ref{thm:exhaustion} fixes the total
pair charge to vacuum, so the three pair labels sum to zero, and with $(23)$
at label $0$ the pairs $(14)$ and $(56)$ carry one common label $q\in A$.
These $2^Q$ values form the register of $Q$ qubits on which the braid acts.
The pair $(56)$ is a spectator: none of the operators below meets wall 1.

\paragraph*{The three measured families.}
For $a\in A$ put
\begin{equation}\label{eq:measured-families}
 \mathsf A_a=Z_{2,a},
 \qquad
 \mathsf B_a=X_{2,a},
 \qquad
 \mathsf C_a\propto X_{2,a}Z_{3,a}:
\end{equation}
the pair loop of $(23)$, the transfer loop along $\gamma_2$, and the product
of that transfer loop with the pair loop of the reference wall, whose cycle
$\gamma_2+\bdry D_3$ encloses twists 2 and 4 modulo two.  We call the
measurements of $\mathsf B$ and $\mathsf C$ the $(12)$ and $(24)$
measurements after the twists these cycles enclose; $\mathsf C$ is the
product just defined, not the transfer loop of a second route.  The family
$\mathsf A$ carries
the positive lift of Lemma~\ref{lem:pair-loop-characters}.  The binary
Paulis of $\mathsf B$ and $\mathsf C$ are linear in $a$ and pairwise
commuting: for $\mathsf B$ by Lemma~\ref{lem:L23-isotropy}; for $\mathsf C$
because the two cross terms $\phi(a,b^\iota)$ and $\phi(b,a^\iota)$ of
Lemma~\ref{lem:pca-pairing} at $t=n$ cancel by
Eq.~\eqref{eq:phi-affine-exchange}, while the remaining terms vanish by
Lemmas~\ref{lem:L23-isotropy} and~\ref{lem:pair-loop-characters}.  The proof
of Lemma~\ref{lem:L23-group-law} therefore gives each family Hermitian lifts
with $F_0=I$ and $F_aF_b=F_{a+b}$, which we fix.  Lemma~\ref{lem:pca-pairing},
at $t=s$ and at $t=n$, together with these isotropy statements and the
commutation of pair loops on distinct walls gives
\begin{equation}\label{eq:measured-families-weyl}
 F_bH_a=\lambda(a,b)\,H_aF_b,
 \qquad a,b\in A,
\end{equation}
for any two distinct families $F,H\in\{\mathsf A,\mathsf B,\mathsf C\}$, with
$\lambda$ non-degenerate (Lemma~\ref{lem:pca-perfect}).  Each family is
therefore diagonal on joint eigenspaces labelled by the characters
$\lambda(t,\cdot)$, $t\in A$, with spectral projectors
\begin{equation}\label{eq:family-projector}
 P_F^t=|A|^{-1}\sum_{a\in A}\lambda(t,a)\,F_a,
 \qquad F\in\{\mathsf A,\mathsf B,\mathsf C\}.
\end{equation}
For $\mathsf A$ the label $t$ is the pair label of $(23)$
(Lemma~\ref{lem:pair-loop-characters}); for $\mathsf B$ and $\mathsf C$ it
is defined relative to the fixed lifts.  All three families lie in $\mathcal L_U$ of
Eq.~\eqref{eq:splitting-U} and so commute with the bulk logical factor.

\begin{lemma}[Phase of $\mathsf C_a\mathsf B_a$ relative to the pair loop $Z_{3,a}$]
\label{lem:measured-refinement}
The fixed lifts determine a function $\eta:A\to\{\pm1,\pm i\}$ by
\begin{equation}\label{eq:measured-refinement}
 \mathsf C_a\mathsf B_a=\eta(a)\,Z_{3,a},
 \qquad a\in A,
\end{equation}
and it satisfies
\begin{equation}\label{eq:measured-refinement-law}
 \eta(a+b)=\eta(a)\,\eta(b)\,\lambda(a,b),
 \qquad
 \eta(0)=1,
 \qquad
 \eta(a)^2=\lambda(a,a).
\end{equation}
Conversely, for every function $\eta'$ obeying
Eq.~\eqref{eq:measured-refinement-law} the operators
$\eta'(a)Z_{3,a}\mathsf B_a$ are Hermitian lifts of $\mathsf C$ with the group
law.  With $\mathsf B$ fixed, the Hermitian lifts of $\mathsf C$ with the group
law are therefore the $|A|$ families $\lambda(r,a)\mathsf C_a$, $r\in A$;
passing to $\lambda(r,a)\mathsf C_a$ replaces $\eta$ by
$\lambda(r,\cdot)\,\eta$ and $P_{\mathsf C}^{t}$ by $P_{\mathsf C}^{t+r}$.
\end{lemma}
\begin{proof}
The operators $\mathsf C_a\mathsf B_a$ and $Z_{3,a}$ have the same binary
Pauli by Eq.~\eqref{eq:measured-families}, so they differ by a scalar
$\eta(a)$, and $\mathsf C_a=\eta(a)Z_{3,a}\mathsf B_a$ because
$\mathsf B_a^2=I$.  Lemma~\ref{lem:pca-pairing} at $t=n$ gives
$\mathsf B_aZ_{3,b}=\lambda(a,b)Z_{3,b}\mathsf B_a$.  With the group laws of
$\mathsf B$ and of $Z_3$ (Lemma~\ref{lem:pair-loop-characters}),
\[
 \mathsf C_a\mathsf C_b
 =\eta(a)\eta(b)\,Z_{3,a}\mathsf B_aZ_{3,b}\mathsf B_b
 =\eta(a)\eta(b)\lambda(a,b)\,Z_{3,a+b}\mathsf B_{a+b},
\]
and the group law of $\mathsf C$ equates this with
$\eta(a+b)Z_{3,a+b}\mathsf B_{a+b}$.  The lifts at label $0$ are $I$, so
$\eta(0)=1$, and $b=a$ gives $\eta(a)^2\lambda(a,a)=1$; hence $\eta$ takes
values in $\{\pm1,\pm i\}$.  For the converse, the displayed computation read
backwards is the group law of $\eta'(a)Z_{3,a}\mathsf B_a$, and its adjoint is
$\overline{\eta'(a)}\lambda(a,a)Z_{3,a}\mathsf B_a$, which is the operator
itself because $\overline{\eta'(a)}\,\eta'(a)^2=\eta'(a)$.  Two Hermitian lifts
of $\mathsf C$ with the group law differ by a homomorphism
$A\to\{\pm1\}$, which is $\lambda(r,\cdot)$ for a unique $r$ by
Lemma~\ref{lem:pca-perfect}; the statement about $P_{\mathsf C}^t$ is
Eq.~\eqref{eq:family-projector}.
\end{proof}

\begin{lemma}[Forced-measurement algebra]\label{lem:forced-measurement}
For distinct $F,H\in\{\mathsf A,\mathsf B,\mathsf C\}$ and all $u,v,w\in A$,
\begin{equation}\label{eq:forced-measurement}
 P_F^u P_H^v P_F^w=|A|^{-1}\lambda(v,u+w)\,H_{u+w}P_F^w .
\end{equation}
In particular $P_F^uP_H^vP_F^u=|A|^{-1}P_F^u$.
\end{lemma}
\begin{proof}
By Eq.~\eqref{eq:measured-families-weyl}, $H_a$ maps the $F$-eigenspace of
label $w$ onto that of label $w+a$, so $P_F^uH_aP_F^w$ equals $H_aP_F^w$ for
$a=u+w$ and vanishes otherwise.  Insert Eq.~\eqref{eq:family-projector} for
$P_H^v$.
\end{proof}

\paragraph*{The forced-measurement protocol.}
A measurement of a family is the projective measurement $\{P_F^t\}_{t\in A}$
of its $Q$ basis loops.  The braid has three stages, forcing the outcome
$0$ of $\mathsf B$, then of $\mathsf C$, then of $\mathsf A$; each stage
starts in the label-$0$ eigenspace of the family forced before it, and the first
in the prepared state, in the range of $P_{\mathsf A}^0$.  A non-demolition
measurement repeats its own outcome, so a nonzero outcome is not corrected
by measuring the same family again.  In a stage forcing $H$ after $F$,
measure $H$; while the outcome is not $0$, measure $F$, accepting any
outcome, and measure $H$ again.

\begin{corollary}[Termination and a history-independent map]
\label{cor:forced-retry}
\textup{(i)} Conditional on any earlier history, each measurement of the
forced family returns $0$ with probability $2^{-Q}$.  A stage therefore
succeeds within $m$ attempts with probability $1-(1-2^{-Q})^m$, and the
protocol terminates with probability one.  \textup{(ii)} A completed run in
which $K$ measurements of forced families did not return $0$ acts on the
prepared state as
\begin{equation}\label{eq:forced-run}
 \pm|A|^{-K}\,P_{\mathsf A}^0P_{\mathsf C}^0P_{\mathsf B}^0P_{\mathsf A}^0,
\end{equation}
so after normalization every completed run implements the same map up to
sign.
\end{corollary}
\begin{proof}
Each measurement of the forced family $H$ follows a measurement of $F$, the
end of the preceding stage, or the preparation, so the state $\rho$ lies in
the range of some $P_F^w$, and
$\operatorname{tr}(P_H^0\rho)=\operatorname{tr}(P_F^wP_H^0P_F^w\rho)=2^{-Q}$
by Lemma~\ref{lem:forced-measurement}; this gives \textup{(i)}.  For
\textup{(ii)}, a stage applies a word
$P_H^0P_F^{u_k}P_H^{v_k}\cdots P_F^{u_1}P_H^{v_1}P_F^0$ with every
$v_i\neq0$.  The group law gives $P_H^tH_a=\lambda(t,a)P_H^t$, so
Eq.~\eqref{eq:forced-measurement} at $w=0$ removes one failed attempt at a
time from the right, leaving $\pm|A|^{-k}P_H^0P_F^0$.  The three stages
multiply to Eq.~\eqref{eq:forced-run}.
\end{proof}

\begin{lemma}[The implemented map]\label{lem:measured-gate}
Put $D_a=\mathsf A_aZ_{3,a}$, the product of the pair loops of $(23)$ and
$(14)$.  The family $D$ is Hermitian with the group law and commutes with
$\mathsf A$, $\mathsf B$ and $\mathsf C$; let
$P_D^{r}=|A|^{-1}\sum_{a\in A}\lambda(r,a)D_a$ be its spectral projectors and
\begin{equation}\label{eq:measured-gate}
 \mathsf U_\eta=\sum_{r\in A}\overline{\eta(r)}\,P_D^{r},
 \qquad
 G_\eta=\sum_{a\in A}\eta(a).
\end{equation}
Then $\mathsf U_\eta$ is unitary, $|G_\eta|^2=|A|$, and for all outcomes
$u,w,v\in A$ of $\mathsf B$, of $\mathsf C$ and of a final measurement of
$\mathsf A$, with $s=u+w+v$,
\begin{equation}\label{eq:measured-product}
 P_{\mathsf A}^{v}P_{\mathsf C}^{w}P_{\mathsf B}^{u}P_{\mathsf A}^{0}
 =\frac{G_\eta}{|A|^{2}}\,\lambda(u,v)\,\overline{\eta(s)}\;
 D_{s}\,\mathsf U_\eta\,\mathsf B_v\,P_{\mathsf A}^{0}.
\end{equation}
Each record $(u,w,v)$ therefore occurs with probability $|A|^{-3}$ on every
prepared state, and applying the Pauli $\mathsf B_vD_s$ afterwards returns
the auxiliary pair to the label $0$ and leaves $\mathsf U_\eta$ up to a
phase.  In particular
$P_{\mathsf A}^0P_{\mathsf C}^0P_{\mathsf B}^0P_{\mathsf A}^0
=G_\eta|A|^{-2}\,\mathsf U_\eta P_{\mathsf A}^0$.  On the range of
$P_{\mathsf A}^0$ the loops $D_a$ and $Z_{3,a}$ agree, so on the register,
where the pair $(14)$ has label $q$, $\mathsf U_\eta$ multiplies by
$\overline{\eta(q)}$ and $D_s$ by $\lambda(s,q)$.
\end{lemma}
\begin{proof}
The families $\mathsf A=Z_2$ and $Z_3$ commute and carry the group law
(Lemma~\ref{lem:pair-loop-characters}), so $D$ is Hermitian with the group
law and commutes with $\mathsf A$.  By Theorem~\ref{thm:pca-weyl} and
Lemma~\ref{lem:pca-pairing} at $t=n$, $\mathsf B_a$ picks up the same sign
$\lambda(a,b)$ from $\mathsf A_b$ and from $Z_{3,b}$, so it commutes with
$D_b$; and $\mathsf C_a=\eta(a)Z_{3,a}\mathsf B_a$ then commutes with $D_b$
as well.  The phases $\overline{\eta(r)}$ have modulus one and the $P_D^r$
are orthogonal projectors summing to $I$, so $\mathsf U_\eta$ is unitary.  By
Eq.~\eqref{eq:measured-refinement-law},
$\eta(a)\lambda(t,a)=\eta(a+t)\overline{\eta(t)}$, hence
\begin{equation}\label{eq:measured-gauss}
 \sum_{a\in A}\eta(a)\lambda(t,a)=G_\eta\,\overline{\eta(t)},
 \qquad t\in A,
\end{equation}
and likewise
$|G_\eta|^2=\sum_{a,c}\eta(a)\overline{\eta(a+c)}
=\sum_c\overline{\eta(c)}\sum_a\lambda(a,c)=|A|$, the inner sum vanishing
for $c\neq0$ by Lemma~\ref{lem:pca-perfect}.

Insert Eq.~\eqref{eq:family-projector} for $P_{\mathsf C}^w$ and
$P_{\mathsf B}^u$.  By Eq.~\eqref{eq:measured-refinement} and the group law of
$\mathsf B$, $\mathsf C_b\mathsf B_a=\eta(b)Z_{3,b}\mathsf B_{a+b}$.  The loops
$Z_{3,b}$ commute with $\mathsf A$, and
$P_{\mathsf A}^v\mathsf B_cP_{\mathsf A}^0$ equals
$\mathsf B_vP_{\mathsf A}^0$ for $c=v$ and vanishes otherwise, by
Eq.~\eqref{eq:measured-families-weyl}.  Only $a=b+v$ survives.  Moreover
$Z_{3,b}\mathsf B_vP_{\mathsf A}^0=D_b\mathsf A_b\mathsf B_vP_{\mathsf A}^0
=\lambda(v,b)D_b\mathsf B_vP_{\mathsf A}^0$, because
$\mathsf A_bP_{\mathsf A}^0=P_{\mathsf A}^0$.  Therefore
\[
 P_{\mathsf A}^{v}P_{\mathsf C}^{w}P_{\mathsf B}^{u}P_{\mathsf A}^{0}
 =|A|^{-2}\lambda(u,v)\sum_{b\in A}\lambda(s,b)\,\eta(b)\,D_{b}\;
 \mathsf B_vP_{\mathsf A}^{0}.
\]
On the range of $P_D^r$ the sum is
$\sum_b\eta(b)\lambda(s+r,b)=G_\eta\overline{\eta(r+s)}$ by
Eq.~\eqref{eq:measured-gauss}, and
$\overline{\eta(r+s)}=\overline{\eta(s)}\,\lambda(r,s)\,\overline{\eta(r)}$
by Eq.~\eqref{eq:measured-refinement-law}; summing over $r$ with
$D_{s}=\sum_r\lambda(r,s)P_D^r$ gives
Eq.~\eqref{eq:measured-product}.  The operator
$D_{s}\mathsf U_\eta\mathsf B_v$ is unitary and a prepared state is fixed
by $P_{\mathsf A}^0$, so the squared norm of the image is
$|G_\eta|^2|A|^{-4}=|A|^{-3}$.  The Paulis $\mathsf B_v$ and $D_s$ square to
$I$ and commute with $\mathsf U_\eta$, which gives the statement about
$\mathsf B_vD_s$.  The last statement is $\mathsf A_a=I$ on the range of
$P_{\mathsf A}^0$ and Eq.~\eqref{eq:pair-loop-eigenvalue}.
\end{proof}

By Corollary~\ref{cor:forced-retry} every completed forced run thus acts as
$\mathsf U_\eta$ up to a phase.  Without forcing, the record $(u,w,v)$
determines the Pauli $\mathsf B_vD_{u+w+v}$ to be applied or tracked; forcing
only the last outcome leaves the Pauli $D_{u+w}$.
Interchanging the two middle measurements gives
$P_{\mathsf A}^0P_{\mathsf B}^0P_{\mathsf C}^0P_{\mathsf A}^0
=\overline{G_\eta}|A|^{-2}\,\mathsf U_\eta^{-1}P_{\mathsf A}^0$ by the same
computation.  Since $r\mapsto\lambda(r,r)$ is a homomorphism, equal to
$\lambda(\cdot,d)$ for one $d\in A$, and $\eta(r)^2=\lambda(r,r)$,
\begin{equation}\label{eq:measured-gate-square}
 \mathsf U_\eta^{2}=\sum_{r\in A}\lambda(r,r)P_D^r=D_{d},
 \qquad
 \mathsf U_\eta^{-1}=D_{d}\,\mathsf U_\eta,
 \qquad
 \mathsf U_{\lambda(r,\cdot)\eta}=D_{r}\,\mathsf U_\eta .
\end{equation}
The square is the same for every lift, and the inverse is one of the $|A|$
maps $D_{r}\mathsf U_\eta$ that the lifts of
Lemma~\ref{lem:measured-refinement} produce.

\paragraph*{Normalization and the gate for $\BBc{1,1,3,1}$.}
Let $\mu_1,\ldots,\mu_Q$ be a basis of $A$, write
$\lambda(\mu_j,\mu_k)=(-1)^{\beta_{jk}}$ with $\beta_{jk}\in\{0,1\}$, and
$q=\sum_jq_j\mu_j$.  Equation~\eqref{eq:measured-refinement-law} gives
$\overline{\eta(q)}=\prod_j\overline{\eta(\mu_j)}^{\,q_j}
\prod_{j<k}(-1)^{\beta_{jk}q_jq_k}$, and $\eta(\mu_j)^2=(-1)^{\beta_{jj}}$.  The
characters $\lambda(r,\cdot)$ take every pattern of signs on the basis, so by
Lemma~\ref{lem:measured-refinement} the lift of $\mathsf C$ can be chosen
with $\eta(\mu_j)=(-i)^{\beta_{jj}}$ for every $j$, and then
\begin{equation}\label{eq:measured-gate-normal}
 \overline{\eta(q)}=i^{\,\sum_{j,k}\beta_{jk}q_jq_k},
\end{equation}
the exponent being computed in $\mathbb Z$.  For $\BBc{1,1,3,1}$ and the basis
$(1,x,x^2)$, with coordinates $(q_0,q_1,q_2)$,
Table~\ref{table:measurement1131} gives
$\beta_{00}=\beta_{22}=\beta_{12}=\beta_{21}=1$ and all other entries zero, so
\begin{equation}\label{eq:measured-gate-1131}
 \mathsf U_\eta\lvert q\rangle
 =i^{\,q_0+q_2}(-1)^{q_1q_2}\lvert q\rangle ,
\end{equation}
which on the three qubits $q_0,q_1,q_2$ is $S_0S_2\,\mathrm{CZ}_{12}$ with
$S=\operatorname{diag}(1,i)$: the gate $U_0$ of the main text.  The other
seven lifts multiply it by $D_{r}$, $r\neq0$, and on these qubits
$D_s=Z_0^{s_0}Z_1^{s_2}Z_2^{s_1+s_2}$.  The assignment of qubits
depends on the basis: in the basis $(1,x^2,x+x^2)$, in which $\beta$ is the
identity matrix, the operator of Eq.~\eqref{eq:measured-gate-1131} is
$S\otimes S\otimes S^{\dagger}$.

\begin{figure}[htbp]
    \centering
    \begin{minipage}[t]{0.28\linewidth}
        \centering
        \subfloat[\label{fig:measure_12}]{%
            \includegraphics[trim=0 1 25 23,clip,height=62bp]{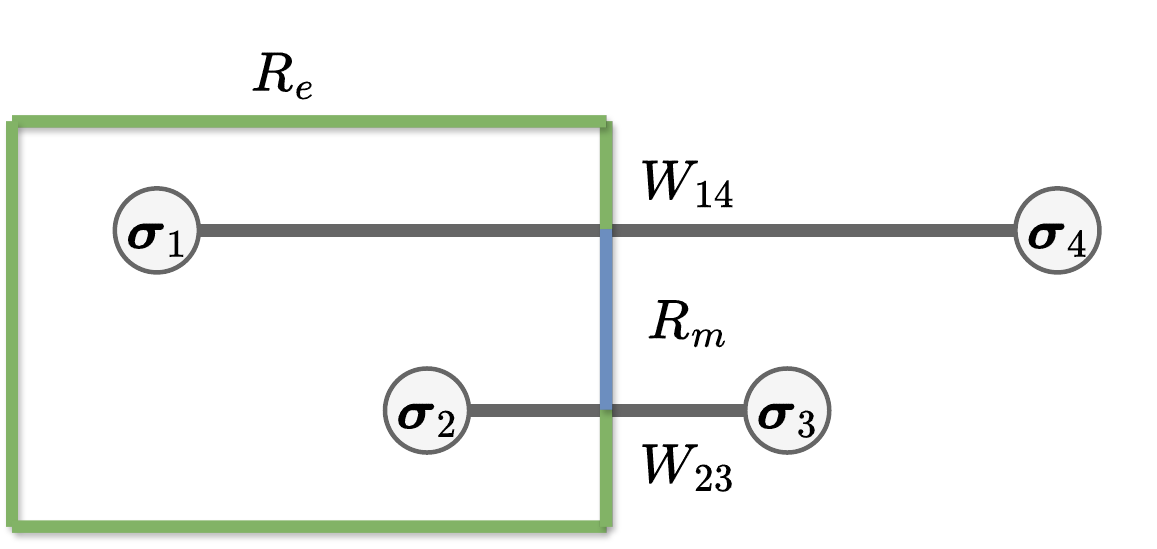}}
        \par\vspace{3pt}
        \footnotesize Measurement of (1,2)
    \end{minipage}
    \hfill
    \begin{minipage}[t]{0.28\linewidth}
        \centering
        \subfloat[\label{fig:measure_24}]{%
            \includegraphics[trim=23 1 0 23,clip,height=62bp]{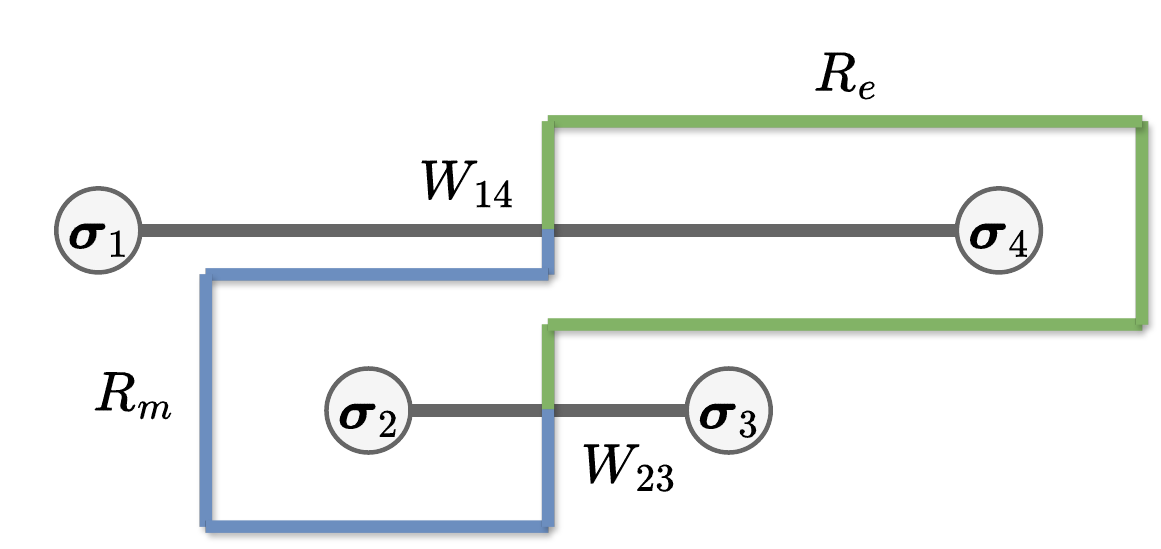}}
        \par\vspace{3pt}
        \footnotesize Measurement of (2,4)
    \end{minipage}
    \hfill
    \begin{minipage}[t]{0.40\linewidth}
        \centering
        \subfloat[\label{fig:measure_23}]{%
            \includegraphics[trim=23 1 25 11,clip,height=62bp]{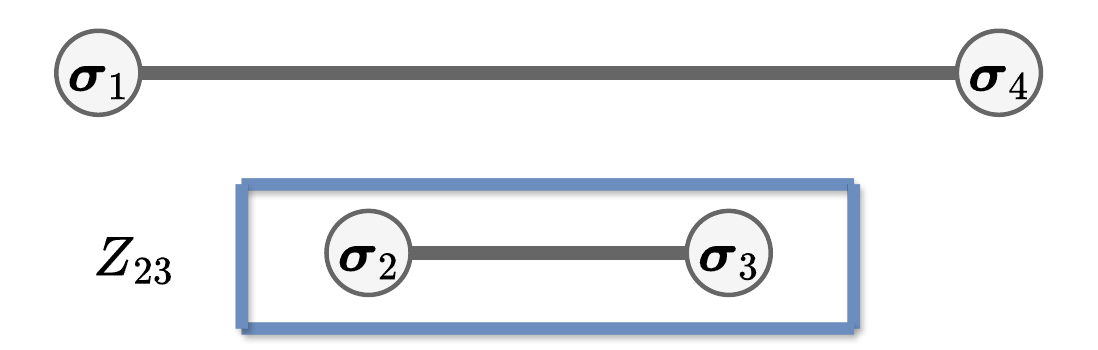}}
        \par\vspace{3pt}
        \footnotesize Measurement of (2,3)
    \end{minipage}

    \caption{Forced measurements by Wilson loops; panels (a), (b) and (c)
    show closed curves round the pairs $(1,2)$, $(2,4)$ and $(2,3)$ that the
    three measurements concern.  The walls of the
    pairs $(1,4)$ and $(2,3)$ are drawn one above the other, whereas the
    Setting places them on one line.  In (a) and (b) the loop passes
    through each wall once: $W_{14}$ is the piece that goes through the
    $(\sigma_1,\sigma_4)$ wall and $W_{23}$ the piece that goes through the
    $(\sigma_2,\sigma_3)$ wall, and each is exactly where the loop changes
    species, since a wall exchanges $e$ and $m$.  Between the two crossings
    the loop is an ordinary bulk ribbon: the electric corridor $R_e$ and the
    magnetic corridor $R_m$.  The drawn loops therefore have the four-piece form
    $W_{14}R_eW_{23}R_m$ of the transfer loops of
    Definition~\ref{def:pca-families}.  The loop of (a) encloses twists 1 and
    2, as the cycle $\gamma_2$ of $\mathsf B$ does; the loop of (b) encloses
    twists 2 and 4, as the cycle $\gamma_2+\bdry D_3$ of $\mathsf C$ does
    modulo two, and stands for that enclosure only: $\mathsf C$ is the product
    of Eq.~\eqref{eq:measured-families}.  In (c) the loop meets no wall at all: it is the closed pristine
    period rectangle $Z_{23}$ around the $(\sigma_2,\sigma_3)$ pair, of a
    single species throughout.  Test-polynomial labels are suppressed.}
    \label{fig:wilson_force_measurements}
\end{figure}

Fig.~\ref{fig:wilson_force_measurements} shows, on the twists 1 to 4, a
closed curve round the pair that each measurement concerns.  For $\mathsf C$
the curve of panel (b) records only that the cycle $\gamma_2+\bdry D_3$
encloses twists 2 and 4 modulo two; the measured operators are the products
of Eq.~\eqref{eq:measured-families}.  The outcome of a family is read from its basis loops: on the
eigenspace of label $t$ the loop $F_X$ returns $\lambda(t,X)=\B{t}{X^\iota}$.
For $\mathsf A$, whose label is the pair label of $(23)$, this is
$\B{A_1}{X^\iota}=\B{X}{A_2}$ in the channel notation $\epsilon=(A_1,A_2)$ of
Sec.~\ref{sec:absorbable}.
Table~\ref{table:measurement1131} lists these phases for the
$\BBc{1,1,3,1}$ code, where a basis of three Wilson loops, for example
$X=1,x,x^2$, gives a complete three-bit signature.

\begin{table}[htbp]
\centering
\renewcommand{\arraystretch}{1.3}
\begin{tabular}{c|cccccccc}
\hline
$(A_1, A_2)$ & $0$ & $x^2$ & $x$ & $x^2+x$ & $1$ & $x^2+1$ & $x+1$ & $x^2+x+1$ \\
\hline
$(0,0)$                 & $+1$ & $+1$ & $+1$ & $+1$ & $+1$ & $+1$ & $+1$ & $+1$ \\
$(x^2,x^2+x+1)$         & $+1$ & $-1$ & $-1$ & $+1$ & $+1$ & $-1$ & $-1$ & $+1$ \\
$(x,x^2+1)$             & $+1$ & $-1$ & $+1$ & $-1$ & $+1$ & $-1$ & $+1$ & $-1$ \\
$(x^2+x,x)$             & $+1$ & $+1$ & $-1$ & $-1$ & $+1$ & $+1$ & $-1$ & $-1$ \\
$(1,1)$                 & $+1$ & $+1$ & $+1$ & $+1$ & $-1$ & $-1$ & $-1$ & $-1$ \\
$(x^2+1,x^2+x)$         & $+1$ & $-1$ & $-1$ & $+1$ & $-1$ & $+1$ & $+1$ & $-1$ \\
$(x+1,x^2)$             & $+1$ & $-1$ & $+1$ & $-1$ & $-1$ & $+1$ & $-1$ & $+1$ \\
$(x^2+x+1,x+1)$         & $+1$ & $+1$ & $-1$ & $-1$ & $-1$ & $-1$ & $+1$ & $+1$ \\
\hline
\end{tabular}
\caption{Measurement phases for the $\BBc{1,1,3,1}$ code.  The first column
lists the possible fusion outcomes $(A_1, A_2)$ of $\sigma \times \sigma$,
and the top row represents the Wilson loop measurement operators defined by
polynomial $X$.  The matrix elements display the expected measurement
values, which geometrically correspond to the braiding phase
$\B{A_1}{X^\iota} = \B{X}{A_2}$.  Its diagonal, $X=A_1$, is the spin
$\theta_{\epsilon_a}=\B{A_1}{A_2}$ of the channel: four bosons and four
fermions.  For this code $[\kappa]=1$ on labels, so
$X^\iota=X^\star$ as labels.  For example, the syndrome vector
$(+1,+1,-1)$ for the basis loops $1,x,x^2$ identifies the channel
$(x,x^2+1)$.  Measuring the three basis loops $1$, $x$, and $x^{2}$
therefore reads out the pair's fusion channel as a three-bit register; this
register is the operational content of the logical-qubit counting quoted in
the main text: in the common-frame, commensurate configurations of
Theorem~\ref{thm:exhaustion}, these registers have one global charge
constraint and give $(n_{\rm pair}-1)Q$ logical qubits.}
\label{table:measurement1131}\label{table:channels1131}
\end{table}

In the standard description of extrinsic defects
(Sec.~\ref{sec:fusiontwists}) the forced sequence teleports the state
information and reproduces the braiding unitary
$R_{\sigma\sigma}^{\left(14\right)}$.  Here $M=(M_1,M_2,M_3)$ records the three outcome histories, $e^{i\varphi_M}$ is their
branch phase, $d_\sigma = \sqrt{2^Q}$ is the
quantum dimension of the twist, and $\breve{\Pi}$ denotes the \emph{normalized}
forced-measurement maps of the diagrammatic conventions of
Bonderson~\cite{bondersonMeasurementOnly2008}, not raw products of outcome
projectors, whose magnitude would carry the branch probability:
\begin{eqnarray}
&& \!\!\!\! \breve{\Pi}_{M_{3}}^{\left(23\leftarrow 24 \right)} \breve{\Pi}_{M_{2}}^{\left(24\leftarrow 12 \right)} \breve{\Pi}_{M_{1}}^{\left(12\leftarrow 23 \right)}
\notag \\
&&= \frac{e^{i \varphi_{M}}}{d_{\sigma}^{3}}
\vcenter{\hbox{\includegraphics{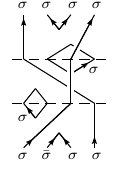}}}
= \frac{e^{i \varphi_{M}}}{d_{\sigma}}
\vcenter{\hbox{\includegraphics{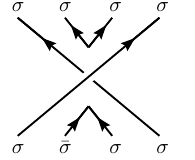}}}
\notag \\
\label{eq:measurement_braid}
&&= e^{i \varphi_{M}} R_{\sigma\sigma}^{\left(14\right)} \otimes  \left| \bar{\sigma},\sigma;\mathbf{1} \right\rangle_{23} \left\langle \bar{\sigma},\sigma;\mathbf{1} \right|_{23}
,
\end{eqnarray}

Eq.~\eqref{eq:measurement_braid}, with the exchange taken at the refinement
$\varrho=\eta$, is the categorical reading of
Corollary~\ref{cor:forced-retry} and Lemma~\ref{lem:measured-gate}.  Every
completed run ends in the range of
$P_{\mathsf A}^0$, so the auxiliary twists 2 and 3 return to the vacuum
channel $\mathbf{1}$ and the resource is available again, and the
probability $2^{-Q}=d_\sigma^{-2}$ of each forced attempt is the
one of Ref.~\cite[text following Eq.~(9)]{bondersonMeasurementOnly2008}.
Equation~\eqref{eq:measured-refinement-law} is
Eq.~\eqref{eq:quadratic-refinement} at $\chi=\lambda$, and on the channel
$\epsilon_q$ the map $\mathsf U_\eta$ has the channel dependence
$\eta(q)^{-1}$ of the exchange of Eq.~\eqref{eq:TY-R-matrix} at
$\varrho=\eta$.  The $|A|$
lifts of $\mathsf C$ realize the $|A|$ refinements, and by
Eq.~\eqref{eq:measured-gate-square} both orders of the middle measurements
lie among them.  The labels of $\mathsf B$ and $\mathsf C$ are defined
relative to the lifts, so the algebra used here does not determine which
refinement belongs to the vacuum channels of the pairs $(12)$ and $(24)$.

\begin{remark}[Braiding versus measurement resources]
Forced-measurement teleportation reproduces braid unitaries, so it stays inside
the synchronized subgroup of Theorem~\ref{prop:sync}.  Arbitrary joint
Wilson--transfer Pauli measurements, stabilizer ancilla pairs and adaptive
feedforward --- capabilities that straight-wall transmission does not supply
and the pure-braiding result does not assume --- generate the full projective
Clifford group, since the Pauli $\pi/4$ rotations generate it and each is a
$P\otimes Y$ measurement with a $\ket{0}$
ancilla~\cite[Figs.~5 and~11(b)]{litinskiGame2019}.
\end{remark}

\section{Computational Power}\label{sec:computation}

Sections~\ref{sec:twists}--\ref{sec:measurement} determine what the twists
\emph{are} and how to braid them; this section determines what those braids
can \emph{compute}.  The answer is two-sided: every braid unitary, in the canonical normal gauge, normalizes
the fusion-register Pauli group and so lies in the Clifford group
(Sec.~\ref{sec:clifford}), yet the braid image is a proper,
label-synchronized subgroup of it whenever $Q>1$ and $N\ge2$
(Sec.~\ref{sec:propersubgroup}); for $N=1$ there are no register qubits.  Both
statements are proved from the categorical data.
Clifford containment is expected from the stacked-code
hierarchy~\cite[\S6]{scrubyHierarchy2020} and from the general bound on braiding
defect-encoded qubits~\cite[Thm.~3]{websterFaulttolerant2020}, whose encoding
hypotheses differ from the register here; the Ising braid image is already a
proper Clifford subgroup from three qubits on~\cite[\S VI]{ahlbrechtImplementation2009},
and the obstruction found here for $Q>1$ is of another kind, a synchronization
of the $Q$ label components.

\subsection{Bounded within the Clifford Group}\label{sec:clifford}

With the register of Theorems~\ref{thm:exhaustion}, \ref{thm:pca-register}, \ref{thm:endpoint-exactness} and~\ref{thm:pca-weyl} and the TY
fusion rules, choose the basis
$\{|c\rangle:c\in A\}$ for one independent twist-pair register.  Define
\begin{equation}\label{eq:abstract-Heisenberg-action}
 Z_a|c\rangle=\chi(a,c)|c\rangle,
 \qquad
 X_b|c\rangle=|c+b\rangle.
\end{equation}
Under the microscopic identification
Eq.~\eqref{eq:microscopic-TY-identification}, these reproduce the label and
adjoint action of the Wilson and transfer operators of
Lemma~\ref{lem:pair-loop-characters}, Theorem~\ref{thm:pca-weyl} and
Corollary~\ref{cor:pca-charge-action}; the identification is of that action,
not of literal operators with fixed scalars, which is all the Clifford
argument below uses.  Since
$A\cong\mathbb F_2^Q$, they generate, with the scalar $iI$, the $Q$-qubit Pauli group.  Writing
$\chi(a,b)=(-1)^{\bar\chi(a,b)}$, the Pauli labels $(a,b)\in A\oplus A$
carry the non-degenerate symplectic form
\begin{equation}
 \Omega((a,b),(a',b'))
 =\bar\chi(a,b')+\bar\chi(a',b).
\end{equation}

The Clifford property follows directly from the canonical TY matrices.
The sign $\nu$ multiplies Eq.~\eqref{eq:TY-F-matrix} as a whole and shifts the
common exchange phase of Eq.~\eqref{eq:TY-R-matrix}, so it cancels in every
conjugation and contributes only an overall scalar phase to a braid word.  The
Clifford-normalization and synchronization conclusions below are conjugation
statements and so do not depend on it, and the register dimension is fixed by
the fusion rules; the amplitudes in the proof of
Lemma~\ref{lem:associator-labels} retain it.  Dropping it, denote the resulting Fourier
transform by $\mathcal F_\chi$.  Bicharacter
orthogonality gives
\begin{equation}\label{eq:F-Pauli-action}
 \mathcal F_\chi X_b\mathcal F_\chi^\dagger=Z_b,
 \qquad
 \mathcal F_\chi Z_a\mathcal F_\chi^\dagger=X_a.
\end{equation}
Let $U_R$ be the diagonal exchange in Eq.~\eqref{eq:TY-R-matrix}.  The
quadratic-refinement identity gives
\begin{equation}\label{eq:R-Pauli-action}
 U_RX_bU_R^\dagger
 =\varrho(b)^{-1}X_bZ_b,
 \qquad
 U_RZ_aU_R^\dagger=Z_a.
\end{equation}
Thus both elementary recoupling and exchange normalize the Pauli group.  So does
every diagonal bicharacter phase $\theta(c)=\chi(\alpha(c),\beta(c))$ with affine
label maps $\alpha,\beta$: bilinearity makes $\theta(c+b)/\theta(c)$ a scalar times
a character of $c$, so conjugation by $\theta$ sends every translation to a Pauli
and fixes every diagonal Pauli.  The
remaining Tambara--Yamagami associators an elementary braid meets are
one-dimensional and cancel between $F$ and $F^{-1}$; the proof of
Lemma~\ref{lem:associator-labels} below shows this and computes every
elementary generator's conjugation action on both halves of the register
Pauli group outright.

\begin{theorem}[Clifford bound]\label{thm:clifford}
For $2N$ twists at fixed total charge, the fusion register consists of
$Q(N-1)$ qubits.  For every braided TY realization with
$A\cong\mathbb Z_2^Q$, in the canonical normal gauge of
Lemma~\ref{lem:associator-labels}, every braid unitary is a composition of bicharacter
Fourier kernels, associator phases, and diagonal exchanges, each of which
normalizes the register Pauli group.  Hence the braid image lies in the
$Q(N-1)$-qubit Clifford group.  This is the Clifford action on the code's Wilson/transfer
register.
\end{theorem}
\begin{proof}
Use the cumulative fusion basis
$|a_1,\ldots,a_{N-1}\rangle$, $a_j\in A$.  Every elementary braid is
obtained by local TY recouplings and one diagonal exchange, and
Lemma~\ref{lem:associator-labels} shows that it conjugates every
$X^{(k)}_b$ and $Z^{(k)}_b$ to a register Pauli up to a scalar, the Fourier
and exchange factors doing so by Eqs.~\eqref{eq:F-Pauli-action}
and~\eqref{eq:R-Pauli-action}.  Therefore every braid word normalizes the
register Pauli group.  The normalizer of the $Q(N-1)$-qubit Pauli group is the
corresponding Clifford group.
\end{proof}

\subsection{Label Synchronization and the Proper Subgroup}\label{sec:propersubgroup}

The properness statement below cannot be had from the lattice alone.  Theorem~\ref{thm:transmission} constrains the
remote labels of a crossing and defines no unitary on the fusion register, so
it cannot by itself force the tensor form: writing $T^{\dagger_\phi}$ for the
adjoint of $T\in GL(A)$ with respect to the register pairing
$(a,b)\mapsto\phi(a,b^\iota)$ of Lemma~\ref{lem:pca-pairing},
$\phi(Ta,b^\iota)=\phi(a,(T^{\dagger_\phi}b)^\iota)$, the map $X_{s,a}\mapsto X_{s,Ta}$,
$Z_{s,b}\mapsto Z_{s,(T^{\dagger_\phi})^{-1}b}$ preserves the Weyl relations
and is a Clifford automorphism of the register, and for $Q>1$ a mixing $T$ --- a
shear, for instance --- mixes the label directions.

For $Q>1$ and $N\ge2$ the Clifford bound of Theorem~\ref{thm:clifford} is not
saturated.  The
obstruction is a synchronization constraint: every braid generator's binary symplectic action
is identical on all $Q$ components of the channel label, so the braid
group can never mix the $A$-label directions.

\begin{lemma}[Which labels the associators carry]\label{lem:associator-labels}
In the canonical Tambara--Yamagami normal basis, write the cumulative path with
intermediate objects $x_{2j-1}=\sigma$ and $x_{2j}=\epsilon_{a_j}$, and put
$a_0=0$ with $a_N$ the fixed total charge, whose value enters the boundary
phases only and not the symplectic action.  Then every elementary braid carries the register labels by the
identity: for each $b\in A$,
\[
 B_{2j-1}X^{(k)}_bB_{2j-1}^{\dagger}
  \doteq X^{(k)}_bZ^{(j-1)}_bZ^{(j)}_b
 \quad\bigl(k\in\{j-1,j\}\bigr),
 \qquad
 B_{2j-1}X^{(k)}_bB_{2j-1}^{\dagger}=X^{(k)}_b
 \quad\bigl(k\notin\{j-1,j\}\bigr),
\]
\[
 B_{2j-1}Z^{(k)}_bB_{2j-1}^{\dagger}=Z^{(k)}_b
 \quad(\text{every }k),
 \qquad
 B_{2j}X^{(j)}_bB_{2j}^{\dagger}=X^{(j)}_b,
 \qquad
 B_{2j}Z^{(j)}_bB_{2j}^{\dagger}\doteq X^{(j)}_bZ^{(j)}_b,
\]
with every index $k$ ranging over $\{1,\dots,N-1\}$, with $\doteq$ suppressing
a scalar independent of the channel labels, and with $Z^{(0)}_b=Z^{(N)}_b=I$, there being no
register at either boundary.  An even generator acts on no register but the
$j$th.  These are the conjugation rules on \emph{both} generating halves of
the register Pauli group, so they determine the induced binary symplectic
matrix of each elementary braid, and no automorphism of $A$ other than the
identity occurs.
\end{lemma}

\begin{proof}
For an odd generator the recoupling is
$F^{\epsilon_{a_{j-1}}\sigma\sigma}_{\epsilon_{a_j}}$, which is
one-dimensional because $\sigma\times\epsilon_a=\sigma$ with multiplicity one,
so its phase is a scalar and cancels between $F$ and $F^{-1}$.  What survives
is the exchange phase at the single $\sigma\sigma$ channel
$q_j=a_{j-1}+a_j$, giving
$B_{2j-1}\ket{\mathbf a}=R^{\mathbf 1}_{\sigma\sigma}\varrho(q_j)^{-1}\ket{\mathbf a}$,
and the displayed conjugation on $X$ follows from
$\varrho(q_j)/\varrho(q_j+b)=\varrho(b)^{-1}\chi(q_j,b)$, which is
Eq.~\eqref{eq:quadratic-refinement}.  That same display is diagonal in the
cumulative basis, with a phase depending only on $q_j$.  Every $Z^{(k)}_b$ is
diagonal in that basis, so it commutes with $B_{2j-1}$; and for
$k\notin\{j-1,j\}$ the shift $X^{(k)}_b$ leaves $q_j$ unchanged, so it commutes
too.  These are the two remaining displayed rules, and neither needs a separate
argument.  With $X$ and $Z$ both accounted for, the symplectic action is
determined; the rules for $X^{(k)}$ alone would not determine it.

For an even generator the only non-scalar associator is
$F^{\sigma\sigma\sigma}_{\sigma}$ of Eq.~\eqref{eq:TY-F-matrix}, whose entries
are $\nu|A|^{-1/2}\chi(c,d)^{-1}$ with $c$ and $d$ the two channel labels
themselves.  Bicharacter orthogonality gives $F^{-1}\doteq F$, so
$B_{2j}=F^{-1}U_RF$ on register $j$ up to an overall phase, and
\[
 \bra{c}B_{2j}\ket{a}
 =\frac{R^{\mathbf 1}_{\sigma\sigma}}{|A|}
  \sum_{d\in A}\chi(c,d)^{-1}\varrho(d)^{-1}\chi(d,a)^{-1}
 =C\,\varrho(a+c),
\]
by shifting $d\mapsto d+a+c$ and applying
Eq.~\eqref{eq:quadratic-refinement}, with $C$ independent of the labels.  The
displayed conjugations follow.

That the arguments of $\chi$ are the channel labels and not images of them is
the whole content: a diagonal phase $\chi(Tc_1,c_2)$ with $T\ne I$ would be
Clifford but would carry $X^{(1)}_b$ to $X^{(1)}_bZ^{(2)}_{Tb}$ and $X^{(2)}_b$
to $X^{(2)}_bZ^{(1)}_{T^{\dagger}b}$, and no such phase occurs, the odd
associator being scalar and the even one being Eq.~\eqref{eq:TY-F-matrix}.  The
boundary conditions do not enter this exclusion, every automorphism fixing $0$;
what excludes $T$ is that those two are the only associators an elementary
braid meets.  A fusion-vertex gauge change rephases path states by a diagonal
$G(\mathbf a)$ and conjugates the displayed matrices --- not always by a
Clifford, so the gauge cannot be left free.  Existence of the canonical normal
gauge is an input from the classification recalled in Sec.~\ref{sec:FR}, not
something established here; in it, multiplicity being one throughout, a
residual change alters phases only and cannot mix distinct simple-object
sectors.
\end{proof}

\begin{theorem}[Label synchronization]\label{prop:sync}
Let $N\ge2$.  For every braided TY realization with $A\cong\mathbb Z_2^Q$, in
the cumulative fusion basis \emph{taken in the canonical normal gauge of
Lemma~\ref{lem:associator-labels}} every braid generator has an
$A$-label-preserving binary symplectic action
\begin{equation}\label{eq:synchronized-braid}
 S_i=M_i\otimes I_Q,
\end{equation}
where $M_i$ is the corresponding Ising-label action on the $N-1$ cumulative
registers.  Consequently, for $Q>1$ pure twist braiding is a \emph{proper}
subgroup of the $Q(N-1)$-qubit Clifford group.  This survives arbitrarily many
ancillary twists, provided each is obtained by repeated vacuum-pair insertion
--- not an arbitrary multi-twist state of total vacuum charge --- and is
returned to the vacuum \emph{cleanly}, as a unitary tensor factor rather than
by postselection.  This is the braid action on the code's Wilson/transfer
register.
\end{theorem}
\begin{proof}
Use the cumulative labels of Lemma~\ref{lem:associator-labels}, with $a_N$
the total charge.  Its rules for both Pauli families, including boundary
generators, preserve the label $b$ and give Eq.~\eqref{eq:synchronized-braid}.

Synchronization survives ancillary twists, including interleaved ones,
provided the return to vacuum is \emph{clean}.  Appending a vacuum pair sends
$(a_1,\dots,a_{N-1})$ to $(a_1,\dots,a_{N-1},a_N)$, the total charge being
unchanged; inserting one between
completed pairs duplicates a cumulative label, a controlled addition; and
inserting one at an odd position adjoins the vacuum Fourier column
$\mathcal F_\chi\ket0$.  Each is realized by a Clifford encoding circuit of controlled additions and
bicharacter Fourier transforms, with binary action of the form $M\otimes I_Q$,
and every enlarged-system braid has binary symplectic action $M_U\otimes I_Q$
by Lemma~\ref{lem:associator-labels}.  Call the return \emph{clean} when
$U(\ket{\psi}_R\otimes\ket{\Omega}_A)
 =(V\ket{\psi}_R)\otimes\ket{\Omega'}_A$ for every $\ket{\psi}_R$, with
$\ket{\Omega'}_A$ one fixed ancillary state; $V$ is then unitary, by norm
preservation in equal finite dimensions.  For a data Pauli $P$ write
$U(P\otimes I)U^{\dagger}=P_R\otimes P_A$, a Pauli since $U$ is Clifford.
Applying it to the displayed state gives
$P_RV\ket{\psi}\otimes P_A\ket{\Omega'}=VP\ket{\psi}\otimes\ket{\Omega'}$ for
every $\ket{\psi}$, hence $P_A\ket{\Omega'}=\lambda_P\ket{\Omega'}$ and
$VPV^{\dagger}=\lambda_PP_R$: $V$ is Clifford.  In the coordinates these circuits fix, the whole Clifford acts as
$M\otimes I_Q$, and the clean-factor identity gives $S_V=M_{RR}\otimes I_Q$,
with $M_{RR}$ its data-to-data block.  The
qualification is essential: postselecting on the vacuum sector,
$P_\Omega UP_\Omega$, gives a Kraus operator rather than a unitary tensor
factor, and the argument does not apply to it.  Neither
$\ket{\Omega'}=\ket{\Omega}$ nor a stabilizer ancillary state is needed here;
the vacuum-pair preparation is what fixes the encoding, not what makes
the return clean.

For $Q>1$, the register relabeling $X_{s,a}\mapsto X_{s,Ta}$,
$Z_{s,b}\mapsto Z_{s,(T^{\dagger_\phi})^{-1}b}$ displayed above, with $T$ a
shear, is Clifford and mixes the label directions; no $M\otimes I_Q$ can do
that, acting as it does identically on every component.  The synchronized
braid image is therefore proper.
\end{proof}

The common phase $R_{\sigma\sigma}^{\mathbf1}$ is irrelevant to Pauli
conjugation and the refinement does not change the binary action; the $M_i$
generate the binary symplectic image of the Ising braid
group~\cite{frankoExtraspecial2006}.

For $Q=1$, $A\cong\mathbb Z_2$ and Corollary~\ref{cor:odd-Q-fermions} makes the
nontrivial channel a fermion, so $\sigma\times\sigma=\mathbf 1+\epsilon$ with
$\epsilon$ fermionic: the Ising fusion rules of the toric-code
twist~\cite{bombinTopological2010}.  The synchronization of
Theorem~\ref{prop:sync} is then empty, and each braid generator has the binary
symplectic action $M_i$ of the Ising twists, so it agrees with theirs up to a
phase and a Pauli.

\end{document}